\documentclass[letterpaper,twocolumn,10pt]{article}
\usepackage{usenix}
    
\usepackage{tikz}
\usepackage{amsmath}

\usepackage{filecontents}

\usepackage{nicefrac}
\usepackage{siunitx}
\usepackage{array,framed}
\usepackage{booktabs}
\usepackage{subcaption}
\usepackage{latexsym,fancyhdr,url}
\usepackage{enumerate}
\usepackage[algo2e]{algorithm2e}
\usepackage{algorithmic}
\usepackage{algorithm}

\newcommand{\algorithmicbreak}{\textbf{break}}
\newcommand{\BREAK}{\STATE \algorithmicbreak}
\usepackage{graphics}
\usepackage{graphicx}
\usepackage{epsfig}
\usepackage{wrapfig}
\usepackage{color}
\usepackage{float}
\usepackage{xparse} 
\usepackage{xspace}
\usepackage{multirow}
\usepackage{csvsimple}
\usepackage{balance}
\usepackage{mathrsfs}
\usepackage{epsfig}
\usepackage{pifont}
\newcommand{\argmin}{\operatornamewithlimits{argmin}}
\newcommand{\argmax}{\operatornamewithlimits{argmax}}

\usepackage{mathtools}
\usepackage{tikz}
\usepackage{pgfplots} 
\usepackage{pgfplotstable}
\pgfplotsset{compat = newest}
\usetikzlibrary{arrows}
\usetikzlibrary{patterns}
\usepackage{stackrel}
\usepackage{pstricks}
\usepackage{wrapfig}
\usepackage[utf8]{inputenc}
\usepackage{booktabs}
\usepackage{expl3}
\usepackage{collcell}
\usepackage{colortbl} 
\usepackage{fp}
\usepackage{xurl}
\usepackage{hyperref}
\usepackage{rotating}
\usepackage{amsthm, amssymb}
\usepackage{soul}
\theoremstyle{definition}
\newtheorem{theorem}{Theorem}

\usetikzlibrary{
  shapes.geometric,
  arrows,
  external,
  pgfplots.groupplots,
  matrix
}

\DeclareMathAlphabet{\mathcal}{OMS}{cmsy}{m}{n}

\newcommand{\fullcircle}{\begin{tikzpicture}[baseline=-0.4ex]
  \draw (0,0) circle [radius=0.2em];
  \filldraw[fill=black] (0,0) circle [radius=0.2em];
\end{tikzpicture}}

\newcommand{\emptycircle}{\begin{tikzpicture}[baseline=-0.4ex]
  \draw (0,0) circle [radius=0.2em];
\end{tikzpicture}}

\DeclareGraphicsExtensions{%
    .png,.PNG,%
    .pdf,.PDF,%
    .jpg,.mps,.jpeg,.jbig2,.jb2,.JPG,.JPEG,.JBIG2,.JB2}

\newcounter{note}[section]

\newcommand{\chkmrk}{\ding{52}\xspace}
\newcommand{\xmark}{\ding{55}\xspace}%

\newcommand{\secref}[1]{\mbox{Sec.~\ref{#1}}\xspace}

\newcommand{\figref}[1]{\mbox{Fig.~\ref{#1}}\xspace}
\newcommand{\figsref}[2]{\mbox{Figs.~\ref{#1}--\ref{#2}}\xspace}
\newcommand{\subfigsref}[2]{\mbox{Figs.~\ref{#1}--\subref{#2}}\xspace}

\newcommand{\tblref}[1]{\mbox{Table~\ref{#1}}\xspace}

\newcommand{\appref}[1]{\mbox{App.~\ref{#1}}\xspace}
\newcommand{\eqnref}[1]{\mbox{Eq.~(\ref{#1})}\xspace}

\newcommand{\algref}[1]{\mbox{Alg.~\ref{#1}}\xspace}

\def\lf{\left\lfloor}   
\def\rf{\right\rfloor}
\newcommand{\floor}[1]{\ensuremath{\lf#1\rf}\xspace}

\newcommand{\defeq}{\stackrel{\mathclap{\tiny\mbox{def}}}{=}}

\newlength{\figureheight}

\newcommand{\Norm}[2]{\ensuremath{\left\Vert{#2}\right\Vert_{#1}}\xspace}

\newcommand{\euclideanNorm}[1]{\ensuremath{\Norm{2}{#1}}\xspace}
\newcommand{\infinityNorm}[1]{\ensuremath{\Norm{\infty}{#1}}\xspace}
\newcommand{\dotProduct}[2]{\ensuremath{{#1} \cdot {#2}}\xspace}
\NewDocumentCommand{\genericNat}{ g }{\ensuremath{z\IfNoValueF{#1}{_{#1}}}\xspace}

\newcommand{\setSize}[1]{\ensuremath{\left|{#1}\right|}\xspace}

\newcommand{\realsPos}{\ensuremath{\mathbb{R}_{>0}}\xspace}
\newcommand{\IntegersPos}{\ensuremath{\mathbb{Z}_{>0}}\xspace}
\newcommand{\getsr}{\ensuremath{\overset{\scriptscriptstyle\$}{\leftarrow}}\xspace}

\newcommand{\prob}[1]{\ensuremath{\mathbb{P}\left({#1}\right)}\xspace}
\newcommand{\probb}[2]{\ensuremath{\mathbb{P}_{#1}\left({#2}\right)}\xspace}
\newcommand{\average}{\ensuremath{\mathsf{avg}}\xspace}

\NewDocumentCommand{\expv}{ o g }{\ensuremath{\mathbb{E}\IfNoValueF{#1}{_{#1}}\left({#2}\right)}\xspace}
\NewDocumentCommand{\gradient}{ g }{\ensuremath{\nabla\IfNoValueF{#1}{_{#1}}}\xspace}
\newcommand{\cset}[3]{\ensuremath{#1\{}{#2}\ensuremath{\;#1|} \ifmmode{\;}\fi {#3}\ensuremath{#1\}}\xspace}
\newcommand{\cprob}[3]{\ensuremath{\mathbb{P}#1(}{#2}\ensuremath{\;#1|} \ifmmode{\;}\fi {#3}\ensuremath{#1)}\xspace}
\newcommand{\cprobb}[2]{\ensuremath{\mathbb{P}(}{#1}\ensuremath{\;|} \ifmmode{\;}\fi {#2}\ensuremath{)}\xspace}
\newcommand{\cexpv}[3]{\ensuremath{\mathbb{E}#1(}{#2}\ensuremath{\;#1|} \ifmmode{\;}\fi {#3}\ensuremath{#1)}\xspace}

\newcommand{\algNotation}[1]{\ensuremath{\mathcal{#1}}\xspace}
\newcommand{\arrComponent}[2]{\ensuremath{\big[{#1}\big]_{#2}}\xspace}

\ExplSyntaxOn
\cs_new:Npn \intensity #1
   {\fp_eval:n {\MinIntensity + (1.0-\MinIntensity) * (((\MaxNumber-{#1})/(\MaxNumber-\MinNumber)))}
   }
\ExplSyntaxOff
\FPeval{\MinIntensity}{0.5}   
\FPeval{\MinNumber}{0.0}
\FPeval{\MaxNumber}{0.0}
\newcommand{\ApplyGradientX}[1]{\cellcolor[gray]{\intensity{#1}}{\raggedleft #1}}
\newcolumntype{X}{>{\collectcell\ApplyGradientX}r<{\endcollectcell}}

\newcommand{\MarkedVersionIdx}{\ensuremath{j}\xspace}
\newcommand{\MarkedVersionIdxAlt}{\ensuremath{j'}\xspace}
\newcommand{\SampleIdx}{\ensuremath{i}\xspace}
\newcommand{\VariableIdx}{\ensuremath{w}\xspace}
\newcommand{\SampleIdxAlt}{\ensuremath{i'}\xspace}
\newcommand{\TrainDataset}{\ensuremath{D}\xspace}
\newcommand{\TrainDataSample}{\ensuremath{v}\xspace}
\NewDocumentCommand{\Data}{ g g o }{\ensuremath{x\IfNoValueF{#1}{\IfNoValueTF{#3}{_{#1}}{_{{#1},{#3}}}}\IfNoValueF{#2}{^{#2}}}\xspace}
\NewDocumentCommand{\Dataset}{ g g o }{\ensuremath{X\IfNoValueF{#1}{\IfNoValueTF{#3}{_{#1}}{_{{#1},{#3}}}}\IfNoValueF{#2}{^{#2}}}\xspace}
\NewDocumentCommand{\MLModel}{ g }{\ensuremath{f\IfNoValueF{#1}{_{#1}}}\xspace}
\newcommand{\FeatureExtractor}{\ensuremath{h}\xspace}
\newcommand{\MLPBit}{\ensuremath{b}\xspace}
\newcommand{\DOBit}{\ensuremath{b'}\xspace}
\NewDocumentCommand{\HiddenInfo}{ g }{\ensuremath{H\IfNoValueF{#1}{_{#1}}}\xspace}
\newcommand{\MLPractitioner}{\ensuremath{\algNotation{A}}\xspace}
\newcommand{\OriginalDataset}{\ensuremath{Z}\xspace}
\NewDocumentCommand{\MarkedData}{ g }{\ensuremath{x'\IfNoValueF{#1}{_{#1}}}\xspace}
\newcommand{\MarkedDataset}{\ensuremath{X'}\xspace}
\newcommand{\MarkAlg}{\ensuremath{\algNotation{M}}\xspace}
\NewDocumentCommand{\DetectAlg}{ g }{\ensuremath{\algNotation{D}\IfNoValueF{#1}{^{#1}}}\xspace}
\newcommand{\FDRBound}{\ensuremath{p}\xspace}
\newcommand{\DataDistribution}{\ensuremath{\algNotation{X}}\xspace}
\newcommand{\experiment}[3]{\ensuremath{\mathsf{Expt}^{{#1}\text{-}{#2}}_{#3}}\xspace}
\newcommand{\AuditExptWParams}[1]{\ensuremath{\experiment{\AuditExpt}{#1}{\DataDistribution, \MarkAlg, \DetectAlg}}\xspace}
\newcommand{\AuditExpt}{\textsc{audit}\xspace}
\newcommand{\DatasetSize}{\ensuremath{q}\xspace}
\newcommand{\OriginalDatasetSize}{\ensuremath{z}\xspace}
\NewDocumentCommand{\ScoreFunction}{ g }{\ensuremath{g\IfNoValueF{#1}{^{#1}}}\xspace}
\NewDocumentCommand{\Mark}{ g g }{\ensuremath{\delta\IfNoValueF{#1}{_{#1}}\IfNoValueF{#2}{^{#2}}}\xspace}
\NewDocumentCommand{\MarkAlt}{ g g }{\ensuremath{\delta'\IfNoValueF{#1}{_{#1}}\IfNoValueF{#2}{^{#2}}}\xspace}
\newcommand{\MarkBound}{\ensuremath{\epsilon}\xspace}

\newcommand{\Distance}[2]{\ensuremath{d({#1},{#2})}\xspace}

\newcommand{\FalseDetectionRate}{\ensuremath{\mathsf{FDR}}\xspace}
\newcommand{\TrueDetectionRate}{\ensuremath{\mathsf{TDR}}\xspace}
\newcommand{\FalseDetectionRateWParams}{\ensuremath{\mathsf{FDR}_{\DataDistribution, \MarkAlg, \DetectAlg}}\xspace}
\newcommand{\TrueDetectionRateWParams}{\ensuremath{\mathsf{TDR}_{\DataDistribution, \MarkAlg, \DetectAlg}}\xspace}
\newcommand{\ReferenceDatasetSize}{\ensuremath{m'}\xspace}
\newcommand{\TrainDatasetSize}{\ensuremath{m}\xspace}
\newcommand{\ParameterBetaDistribution}{\ensuremath{\beta}\xspace}
\newcommand{\NumberMarkedData}{\ensuremath{n}\xspace}

\newcommand{\RankVariable}{\ensuremath{r}\xspace}
\newcommand{\UnitVector}[2]{\ensuremath{u}_{#1}^{#2}\xspace}
\newcommand{\MIAlg}[1]{\ensuremath{\algNotation{I}^{#1}}\xspace}
\newcommand{\Rank}[1]{\ensuremath{\mathsf{Rank}\big(#1\big)}\xspace}
\newcommand{\RankSum}{\ensuremath{\mathsf{RankSum}}\xspace}
\newcommand{\DPnoiseStd}{\ensuremath{\sigma}\xspace}
\newcommand{\GaussianNoiseStd}{\ensuremath{\sigma}\xspace}
\newcommand{\Accuracy}{\ensuremath{\mathsf{Acc}}\xspace}
\newcommand{\AccuracyDiff}{\ensuremath{\triangle\mathsf{Acc}}\xspace}
\newcommand{\NumAugment}{\ensuremath{k}\xspace}
\newcommand{\Loss}{\ensuremath{\ell}\xspace}
\newcommand{\MIAScore}[2]{\ensuremath{\mathsf{MIA}_{#1,#2}}\xspace}
\newcommand{\MIAThreshold}{\ensuremath{\eta}\xspace}
\newcommand{\AuxiliaryData}{\ensuremath{a}\xspace}
\newcommand{\AuxiliaryDataLabel}{\ensuremath{y'}\xspace}
\newcommand{\AuxiliaryDataset}{\ensuremath{A}\xspace}
\newcommand{\ReferenceModel}{\ensuremath{f'}\xspace}
\newcommand{\ReferenceModelset}{\ensuremath{F'}\xspace}
\newcommand{\RMIAParameter}{\ensuremath{\gamma}\xspace}
\newcommand{\RMIAParameterAlt}{\ensuremath{\lambda}\xspace}
\newcommand{\DataLabel}{\ensuremath{y}\xspace}

\newcommand{\UnlearningRate}{\ensuremath{\tau}\xspace}

\newcommand{\UnlearntData}{\ensuremath{u}\xspace}
\newcommand{\UnlearntDataset}{\ensuremath{U}\xspace}
\newcommand{\PerturbedData}{\ensuremath{u'}\xspace}
\newcommand{\PerturbedDataset}{\ensuremath{U'}\xspace}
\newcommand{\UnlearnWeight}{\ensuremath{\rho}\xspace}
\newcommand{\MLModelAlt}{\ensuremath{f'}\xspace}
\newcommand{\OriginalMLModel}{\ensuremath{f''}\xspace}
\newcommand{\IncorrectLabel}{\ensuremath{y''}\xspace}

\NewDocumentCommand{\IndicatorFunction}{ g }{\ensuremath{\mathbb{I}\IfNoValueF{#1}{\big({#1}\big)}}\xspace}
\NewDocumentCommand{\indicatorRVIdx}{ g }{\ensuremath{n\IfNoValueF{#1}{_{#1}}}\xspace}

\NewDocumentCommand{\sumIndicatorVals}{ g }{\ensuremath{\mathbb{N}\IfNoValueF{#1}{_{#1}}}\xspace}
\newcommand{\sumThreshold}{\ensuremath{T}\xspace}

\newcommand{\nullHypothesis}{\ensuremath{H_0}\xspace}
\NewDocumentCommand{\pValue}{ g }{\ensuremath{p\IfNoValueF{#1}{_{#1}}}\xspace}

\newcommand{\ObservationTime}{\ensuremath{t}\xspace}

\newcommand{\TotalSuccess}{\ensuremath{n'}\xspace}
\newcommand{\Confidence}{\ensuremath{\alpha}\xspace}
\newcommand{\ConfidenceInterval}[2]{\ensuremath{C_{#1}({#2})}\xspace}
\newcommand{\ConfidenceIntervalLower}[2]{\ensuremath{L_{#1}({#2})}\xspace}
\newcommand{\ConfidenceIntervalUpper}[2]{\ensuremath{U_{#1}({#2})}\xspace}
\newcommand{\MeasurementSequence}{\ensuremath{S}\xspace}
\newcommand{\PPRMartingale}{\ensuremath{\mathsf{PPRM}}\xspace}
\newcommand{\QueryNumber}{\ensuremath{c}\xspace}
\newcommand{\QueryNumberVariable}{\ensuremath{l}\xspace}
\newcommand{\NumberUsedDataSamples}{\ensuremath{o}\xspace}
\newcommand{\WeightDecay}{\ensuremath{\mathsf{WeightDecay}}\xspace}

\definecolor{steelblue76114176}{RGB}{76,114,176}
\definecolor{peru22113282}{RGB}{221,132,82}
\definecolor{mediumseagreen85168104}{RGB}{85,168,104}
\definecolor{indianred1967882}{RGB}{196,78,82}

\begin{document}

\date{}

\title{\Large \bf Instance-Level Data-Use Auditing of Visual ML Models}

\author{
{\rm Zonghao Huang}\\
Duke University
\and
{\rm Neil Zhenqiang Gong}\\
Duke University
\and
{\rm Michael K. Reiter}\\
Duke University
} 


\maketitle

\begin{abstract}
    The growing trend of legal disputes over the unauthorized use of
    data in machine learning (ML) systems highlights the urgent need
    for reliable data-use auditing mechanisms to ensure accountability
    and transparency in ML. We present the first proactive,
    instance-level, data-use auditing method designed to enable data
    owners to audit the use of their individual data instances in ML
    models, providing more fine-grained auditing results than previous
    work. To do so, our research generalizes previous work integrating
    black-box membership inference and sequential hypothesis testing,
    expanding its scope of application while preserving the
    quantifiable and tunable false-detection rate that is its
    hallmark.  We evaluate our method on three types of visual ML
    models: image classifiers, visual encoders, and vision-language
    models (Contrastive Language-Image Pretraining (CLIP) and
    Bootstrapping Language-Image Pretraining (BLIP) models).  In
    addition, we apply our method to evaluate the performance of two
    state-of-the-art approximate unlearning methods. As a noteworthy
    second contribution, our work reveals that neither method
    successfully removes the influence of the unlearned data instances
    from image classifiers and CLIP models, even if sacrificing model
    utility by $10\%$.
\end{abstract}

\section{Introduction} \label{sec:intro}

The rapid advances of machine learning (ML) depend on the
availability of massive amounts of training data.  However, the
developers of these ML models often do not disclose the origins of
their training data, raising legal disputes over unauthorized data-use
in training ML models. 
For example, in 2020, multiple lawsuits were filed against Clearview AI,
claiming that Clearview AI scrapped millions of photos online to train
its facial recognition models, violating
the rights of the users of those images~\cite{Hill2020}.
Recently, California passed AI legislation~\cite{ab-2013} that
underscores the importance of tracing the origins of data used in
training ML models. In addition, established data regulations, such as
the General Data Protection Regulation (GDPR) in
Europe~\cite{mantelero2013:eu}, the California Consumer Privacy Act
(CCPA) in the United States~\cite{ccpa}, and Canada's PIPEDA privacy
legislation~\cite{cofone2020:right}, grant individuals the right to
know how their data is being used. The growing trend of legal disputes
over unauthorized data-use in ML and the legislation of data
protection regulations highlight an urgent need for reliable data-use
auditing to ensure accountability and transparency in ML
models, addressing both legal and ethical concerns.

Data-use auditing is a technique that a data owner can use to verify
whether her published data has been used in the training of an ML
model. This approach can be broadly categorized into two levels:
dataset-level~\cite{sablayrolles2020:radioactive, maini2021:dataset,
  dziedzic2022:dataset, li2022:untargeted, tang2023:did, li2023:black,
  du2024:orl, guo2024:domain, wenger2024:data, wang2024:diagnosis,
  huang2024:auditdata} and instance-level~\cite{shokri2017:membership,
  salem2019:ml, pan2020:privacy, liu2021:encodermi, ye2022:enhanced,
  carlini2022:membership, ko2023:practical, zarifzadeh2024:low,
  shi2024:detecting} data-use auditing. Dataset-level data-use
auditing is applied in scenarios where a data owner possesses a
substantial dataset. It produces an auditing result for the entire
dataset, by aggregating information across individual
instances~\cite{huang2024:auditdata} or detecting a significant signal
in an ML model that requires learning from multiple data
samples~\cite{sablayrolles2020:radioactive,
  li2022:untargeted}. However, dataset-level auditing is unsuitable
for cases where the data owner has a small dataset or even a single
data instance. Instance-level data-use auditing addresses this
limitation by offering fine-grained auditing results, assessing the
use of just a few data instances in ML models. In this work, we focus
on instance-level data-use auditing in ML.

To our knowledge, all existing instance-level data-use auditing
methods are \emph{passive}, requiring no modification to the audited
data instance before its publication.  They apply techniques
originally developed for membership inference
attacks~\cite{shokri2017:membership, he2024:difficulty}. These
techniques usually require access to auxiliary data sampled from the
same distribution as the training data of the audited ML model and use
them to train reference models (i.e., ML models similar to the audited
model)~\cite{shokri2017:membership, ye2022:enhanced,
  carlini2022:membership, zarifzadeh2024:low}.  In addition, passive
data-use auditing cannot provide guarantees on false-detection rates,
rendering its detection results less reliable and convincing.  These
limit the applications of passive data-use auditing, as discussed by a
concurrent work~\cite{zhang2024:membership}.

In this work, we propose the \emph{first} proactive, instance-level,
data-use auditing method for the image domain.  Our contribution is a
strict generalization of previous work on dataset-level data-use
auditing~\cite{huang2024:auditdata}, adapting it to apply to
instance-level auditing, as well.  Our approach consists of two key
components: a data-marking algorithm and a data-use detection
algorithm.  The data marking algorithm, which the data owner applies
prior to data publication, generates $\NumberMarkedData$ distinct
marked versions of a data instance by adding $\NumberMarkedData$
unique marks (i.e., image pixel alterations).  Each marked version is
carefully designed to preserve the utility of the original data
instance while ensuring that the marked versions are maximally
distinct, where we measure distinction by the distances between their
high-level features prepared by a pretrained feature extractor.  This
marking process is agnostic to the visual ML task in which the marked
data might be used (including, e.g., labels). After generating the
$\NumberMarkedData$ marked data, the data owner publishes only one
version, selected uniformly at random, while keeping the remaining
versions secret.

Once an ML model is accessible---even in only a black-box way---any
``useful'' membership inference method can be applied to measure the
``memorization'' score of each marked version, including the published
one and those kept secret.  If an ML model has not used the published
marked data instance in training, then the rank of its
``memorization'' score relative to the unpublished versions should
follow a uniform distribution over $\{1, 2, \dots,
\NumberMarkedData\}$ since we select it uniformly at random in the
data-marking step. If, instead, the ML model has used the published
marked data item in training, then its rank (based on its score) is
more likely to be high, as the ML model tends to memorize its training
data~\cite{song2017:machine}.

We develop a novel sequential method to estimate the summed ranks of
the published data. This approach allows a data owner to stop querying
the audited ML model earlier to save cost and conclude if the ML model
was trained with her published data, with any desired false-detection
rate.  While previous work~\cite{huang2024:auditdata} also leveraged a
hypothesis test to ensure a target false-detection rate, it did so
only for dataset-level auditing, using only $\NumberMarkedData = 2$
(vs.\ $\NumberMarkedData \gg 2$ here) and a hypothesis test that is a
special case of ours.

We study the performance of our instance-level data-use auditing
method on three types of visual ML models, namely image
classifiers~\cite{deng2009:imagenet, simonyan2015:very, he2016:deep},
visual encoders~\cite{chen2020:simple}, and vision-language models
(Contrastive Language-Image Pretraining
(CLIP)~\cite{radford2021:learning} and Bootstrapping Language-Image
Pre-training (BLIP)~\cite{li2022:blip} models).  In the case of
auditing image classifiers, we conducted experiments to evaluate our
method on visual benchmark datasets under various settings.  We
empirically demonstrate the applicability of our method in scenarios
where a data owner has only a few data instances (or even only one) to
audit.  Additionally, while the state-of-the-art passive
instance-level data-use auditing methods (i.e., membership inference
methods) lack formal guarantees on false-detection rates, we still
performed an empirical comparison with these approaches. Our method
showed comparable true-detection rates when our formal false-detection
rate was set to be at the same level as the empirical rates of passive
auditing methods.  Our advantage is particularly evident in practical
scenarios where the data owner does not have access to reference
models similar to the audited model. We also examine the effectiveness
of our method against several countermeasures (e.g, preprocessing
training data before their use in model training).  While these
countermeasures can degrade the detection performance of our method,
they do so at the cost of substantially diminishing the utility of the
audited model.  Finally, we extend our evaluation to auditing visual
encoders and vision-language models, further highlighting its uses
across diverse visual ML models.

As another example of the utility of a quantified false-detection
rate, we study the application of our data-use auditing method to
verifying machine unlearning~\cite{cao2015:towards, guo2020:certified,
  bourtoule2021:machine, warnecke2021:machine}.  Machine unlearning is
a technique used to remove the information of specific data instances
from an ML model upon the requests of their data owners, fulfilling
the \emph{right to be forgotten} as mandated by data regulations
(e.g., GDPR~\cite{mantelero2013:eu} and CCPA~\cite{ccpa}). We apply
our approach to evaluate the performance of two state-of-the-art
approximate unlearning methods (i.e., Warnecke et al.'s gradient-based
method~\cite{warnecke2021:machine} and a fine-tuning-based
method~\cite{golatkar2020:forgetting, hu2024:duty}), by detecting
data-use in the ML model updated by these two approximate unlearning
methods. In findings of independent interest, our experiments revealed
that these two methods failed to remove the influence of specific data
instances from visual models even if sacrificing model utility by
$10\%$. Specifically, our method continued to detect those data
instances after unlearning, at a rate convincingly higher than the
false-detection-rate bound that our technique enforced.  Our
experiments also demonstrate that our method is a useful tool for a
data owner to obtain evidence that unlearning was successful.  As
highlighted by previous works (e.g.,~\cite{thudi2022:necessity}), such
auditable evidence is important in machine-unlearning systems.

To summarize, our contributions are as follows:
\begin{itemize}
\item We generalize previous work in dataset-level data-use auditing
  to develop the first method for proactive, instance-level, data-use
  auditing for the image domain.  Our method consists of a
  data-marking algorithm generating $\NumberMarkedData$ maximally
  distinct marked data for each raw data instance, and a detection
  algorithm that is built upon membership inference and a new
  sequential hypothesis test that offers a tunable and quantifiable
  false-detection rate.
\item We demonstrate the effectiveness and robustness 
  of the proposed method and its
  applicability across diverse visual ML models by applying it to
  audit the use of data instances in three types of visual ML models,
  namely image classifiers, visual encoders, and
  vision-language models, under various settings.
\item We illustrate the utility of our method for evaluating
  state-of-the-art machine unlearning techniques, showing that two
  such techniques do not work.  Specifically, our method still
  detected the use of data-items in models whose accuracy was degraded
  by up to 10\% by attempts to unlearn those data-items, at a rate
  convincingly larger than the false-detection-rate bound for which
  our method was configured. Thus, we provide a useful tool
  for a data owner to obtain auditable evidence that the unlearning
  procedure was correctly implemented.
\end{itemize}

\section{Related Work} \label{sec:related_work}

\subsection{Data-Use Auditing in ML}
In this section, we review related work on proactive data-use auditing
of ML models and highlight how our approach addresses a critical gap
in the existing literature.  Proactive data-use auditing involves
modifying the to-be-audited data before they are
published~\cite{sablayrolles2020:radioactive,yu2021:deepfake,
  zou2022:anti, li2022:untargeted, tang2023:did, li2023:black,
  guo2024:domain, wenger2024:data, wang2024:diagnosis,
  wei2024:proving, huang2024:auditdata, du2025:sok} and typically
consists of a data-marking algorithm and a data-use detection
algorithm.  When its data-marking algorithm leverages randomness to
modify data to establish a distribution for a test statistic under the
null hypothesis that the data has not been used, its data-use
detection algorithm can provide a statistical guarantee on the
false-detection rate~\cite{sablayrolles2020:radioactive,
  huang2024:auditdata}.  Existing methods address only
\emph{dataset-level} data-use auditing, some that are tailored to a
specific domain (e.g., image
classifiers~\cite{sablayrolles2020:radioactive,li2022:untargeted, chen2025:anonymity},
language models~\cite{wei2024:proving}, or text-to-image generative
models~\cite{wang2024:diagnosis,li2024:towards}) and others that are
general~\cite{huang2024:auditdata}.  The dataset-level proactive
data-use auditing methods are applicable only in scenarios where a
data owner has a substantial dataset with a large number of data
instances. They produce auditing results by aggregating information
across individual instances~\cite{huang2024:auditdata, chen2025:anonymity} or detecting
prominent signals in an ML model that requires learning from multiple
data samples~\cite{sablayrolles2020:radioactive, li2022:untargeted}.

For example, a line of works on dataset-level proactive data-use
auditing~\cite{li2022:untargeted, tang2023:did, li2023:black} is based
on backdoor attacks~\cite{gu2019:badnets, saha2020:hidden}, to enable
a data owner to modify a substantial portion of her dataset and then
detect its use by eliciting predictable classification results from
the model. But these methods do not provide any guarantee on the
false-detection rate.  Radioactive
data~\cite{sablayrolles2020:radioactive} audits the use of a dataset
in image classifiers with a statistical guarantee on false-detection
rate.  It works by embedding class-specific marks into a subset of the
dataset and analyzing correlations between parameters of the final
layer of the audited image classifier and the embedded marks. Huang et
al.~\cite{huang2024:auditdata} proposed a general framework for
proactive data-use auditing applicable across domains, providing a
quantifiable false-detection rate.  This approach audits the use of a
dataset by comparing and aggregating membership inference scores for
published and unpublished data, derived from its data-marking
algorithm.  However, their approach leverages only two marked versions
per raw data instance in a dataset and so relies on aggregating
information across data instances for detection, which renders their
method unsuitable for instance-level data-use auditing.  Our work
addresses this gap by generalizing their approach to accommodate
instance-level data-use auditing (and their dataset-level auditing
technique) as a special case, yielding the first such instance-level
auditing scheme. More recently, Chen and
Pattabiraman~\cite{chen2025:anonymity} proposed a method that can be
applied to audit the use of a small number of data samples in training
image classifiers, but their method does not bound the false-detection
rate.

Also related to our work, Carlini et al.~\cite{carlini2019:secret}
proposed a method to measure unintended memorization in a language
model by inserting a random sequence into a text dataset and applying
a rank-based test on the inserted random sequence (i.e., the added
mark) relative to other possible marks.  A concurrent
work~\cite{zhang2024:membership} adapts this idea to detect data-use
in language models. Our work's data-use detection algorithm also
adopts a rank-based hypothesis test, but it uses the rank of the
marked data rather than of the added mark, and indeed, we show in
\appref{app:results:additional} that the rank of the added
mark cannot provide strong evidence of data-use in ML models in the
image domain.  Our work differs in other ways, as well: Our method
leverages a sequential hypothesis test of our own design to detect
data use with fewer queries to the model.  And, while these works rely
on random patterns as marks, our data-marking algorithm designs
maximally distinct marked images by solving an optimization problem.

\subsection{Verification of Machine Unlearning}

Existing machine unlearning methods can be categorized into exact
unlearning~\cite{cao2015:towards, bourtoule2021:machine} and
approximate unlearning~\cite{guo2020:certified,
  golatkar2020:forgetting, warnecke2021:machine}.  Exact unlearning
retrains the whole model or a constituent model from scratch on a
dataset excluding the data instances to be removed. Approximate
unlearning, instead, only updates the parameters of the model based on
the data instances to be unlearned.  The guaranteed false-detection
rate of our data-use detection algorithm is powerful in that it
permits the verification of approximate machine unlearning: If the
probability of detecting a data item after applying an algorithm to
unlearn it is higher than the bound on the false-detection rate, then
the algorithm quantifiably does not work.  That is, we conclude that
an (approximate) unlearning algorithm is unsuccessful if the
true-detection rate after unlearning exceeds the false-detection rate
bound.

To our knowledge, existing machine unlearning verification methods are
based on backdoor attacks~\cite{sommer2022:towards,
  guo2024:verifying}.  However, these methods do not provide a formal
guarantee on the false-detection rate.  Moreover, the existing methods
are not applicable to verify if an individual data instance has been
unlearned. Our proposed instance-level data-use auditing method
provides a tool to verify machine unlearning, by detecting if the ML
model still exhibits an individual data instance in its training, with
any desired false-detection rate.

\section{Background}  \label{sec:background} 
 
\subsection{Threat Model}  \label{sec:background:threat_model}
 
We consider a data owner and an ML practitioner.  The data owner
possesses some data instances (here, images) she intends to publish
online.  For instance, she may share her photos on platforms like
Instagram.  The ML practitioner aims to train a useful ML model on a
training dataset for a specific task, such as developing an image
classifier for classification tasks or a visual encoder to extract
image features for general computer vision applications.

The ML practitioner assembles a training dataset by collecting
data posted online by data owners, but does so \emph{without
their authorization}.  He then trains an ML model on the collected
dataset by a learning algorithm designed for his specific ML task.
During training, he might apply additional techniques (e.g., image
preprocessing) to reduce the likelihood that his unauthorized data-use
is detectable, while preserving the utility of the trained model.  The
model is subsequently deployed online to provide services to
customers.

The goal of the data owner is to detect if the model was trained on
her data instances. We have the following assumptions on the data
owner's knowledge and capabilities:
\begin{itemize}
  \item \emph{Knowledge}: The data owner is unaware of the specific
    learning algorithm applied by the ML practitioner and does not
    have access to the architecture or the parameters of the deployed
    ML model.  However, after the model is deployed, the data owner
    can ascertain the form of the ML model (e.g., an image classifier)
    and the format of its inputs and outputs (e.g., the input of an
    image classifier is an image while its output is a vector of
    confidence scores).
  \item \emph{Capabilities}: The data owner can have a black-box
    access to the deployed ML model. In other words, she can obtain
    the output of the ML model by providing her data as
    input. Considering a realistic scenario, however, the data owner
    does not have access to a large amount of data sampled from the
    same distribution as the training samples used to train the
    deployed model.  Consequently, the data owner is unable to train
    an ML model similar to the deployed model that could assist her in
    verifying whether the deployed model was trained using her data
    instances.
\end{itemize}

\subsection{Data-Use Auditing in ML}  \label{sec:background:data_auditing}

\paragraph{Problem definition} 
We focus on a problem, namely \emph{(proactive) data-use auditing in
ML}, where a data owner aims to verify if a \emph{useful} ML model
deployed by an ML practitioner uses her data in training.  To make
this concept precise, we define the data-use auditing of ML in a way
that abstracts away the details of the system model. We do so using
the experiment defined in \figref{fig:data_auditing}.  In data-use
auditing, there are three stages, namely \emph{data marking and
publication}, \emph{ML model training}, and \emph{data-use detection}:
\begin{itemize}
  \item \emph{Data marking and publication}: A data owner has a set of
    data instances $\Dataset \sim \DataDistribution$ of size
    $\setSize{\Dataset} = \DatasetSize$, where $\DataDistribution$
    represents the data distribution from which $\Dataset$ is drawn.
    Prior to publishing data, the data owner applies a data-marking
    algorithm $\MarkAlg$ by $(\MarkedDataset, \HiddenInfo) \leftarrow
    \MarkAlg(\Dataset)$, where $\MarkedDataset$ is the marked version
    of $\Dataset$ that the data owner will publish and $\HiddenInfo$
    is the hidden information that she keeps secret and will use to
    detect data-use in the detection stage.
  \item \emph{ML model training}: An ML practitioner $\MLPractitioner$ assembles his training 
    dataset $\TrainDataset$ that might include 
    the data instances published by the data owner, i.e., $\MarkedDataset \subseteq \TrainDataset$.
    Then he trains an ML model $\MLModel$ on the training dataset by $\MLModel \leftarrow \MLPractitioner(\TrainDataset)$. 
  \item \emph{Data-use detection}: Given oracle (black-box) access to an ML 
    model $\MLModel$, the data owner applies a data-use detection 
    algorithm $\DetectAlg{\MLModel}$ by
    $\DOBit \leftarrow \DetectAlg{\MLModel}(\MarkedDataset, \HiddenInfo)$,
    where $\DOBit \in \{0, 1\}$. If $\DOBit=1$, then the data owner detects the
    use of her data in the ML model; otherwise, she fails to detect.
\end{itemize}

While previous works
(e.g.,~\cite{wenger2024:data,huang2024:auditdata}) address
dataset-level data-use auditing where $\DatasetSize \gg 1$ in
\figref{fig:data_auditing}, our work focuses on \emph{instance-level}
data-use auditing, where the data owner audits the use of only a few
data instances---or even just one---in ML model training.  In other
words, we focus on a data-use auditing where \DatasetSize is small.

\begin{figure}[t]
    \begin{tabbing}
    ***\=***\=\kill
    Experiment $\AuditExptWParams{\MLPBit}(\MLPractitioner, \DatasetSize, \OriginalDatasetSize)$ \\
    \> $\Dataset \sim \DataDistribution$ such that $\setSize{\Dataset} = \DatasetSize$ \\
    \> $(\MarkedDataset, \HiddenInfo) \leftarrow \MarkAlg(\Dataset)$ \\
    \> $\OriginalDataset \sim \DataDistribution$ such that $\setSize{\OriginalDataset} = \OriginalDatasetSize$ and $\MarkedDataset \cap \OriginalDataset = \emptyset$\\
    \> if $\MLPBit = 1$ \\
    \> \> then $\TrainDataset \leftarrow \OriginalDataset \cup \MarkedDataset$ \\
    \> \> else $\TrainDataset \leftarrow \OriginalDataset$ \\
    \> $\MLModel \leftarrow \MLPractitioner(\TrainDataset)$ \\
    \> $\DOBit \leftarrow \DetectAlg{\MLModel}(\MarkedDataset, \HiddenInfo)$ \\
    \> return $\DOBit$\\[10pt]
    $\TrueDetectionRateWParams(\MLPractitioner, \DatasetSize, \OriginalDatasetSize) \defeq \prob{\AuditExptWParams{1}(\MLPractitioner, \DatasetSize, \OriginalDatasetSize)=1}$ \\
    $\FalseDetectionRateWParams(\MLPractitioner, \DatasetSize, \OriginalDatasetSize) \defeq \prob{\AuditExptWParams{0}(\MLPractitioner, \DatasetSize, \OriginalDatasetSize)=1}$
    \end{tabbing}
  \caption{Experiment on data-use auditing in ML and measures 
    on true-detection rate and false-detection rate.}
  \label{fig:data_auditing}
  \end{figure}

\paragraph{Two basic requirements}
There are two basic requirements for a useful data-use auditing
method: \emph{utility-preservation} and \emph{quantifiable
false-detection rate}.
\begin{itemize}
  \item \emph{Utility-preservation}: Its data-marking algorithm should
    return a marked dataset $\MarkedDataset$ that preserves the
    utility (e.g., visual quality for images) of the raw dataset
    $\Dataset$ since the data owner originally wishes to publish
    $\Dataset$. If we use
    $\ell_{\infty}$-norm~\cite{sharif2018:suitability} to
    approximately measure visual similarity for images, then for each
    pair of $\Data \in \Dataset$ and $\MarkedData\in\MarkedDataset$
    such that $\MarkedData$ is the marked version of $\Data$, we
    should have:
    \begin{equation*}
      \infinityNorm{\MarkedData-\Data} \leq \MarkBound,
    \end{equation*}
    where $\MarkBound$ is a parameter enforcing utility preservation,
    i.e., a smaller $\MarkBound$ indicates that $\MarkedData$
    preserves more utility of $\Data$.
  \item \emph{Quantifiable false-detection rate}: Its data-use
    detection algorithm should have a small and bounded
    false-detection rate such that it returns ``detected'' with only a
    small quantifiable probability when the ML model does not use the
    audited data instances in its training. In other words,
    \begin{equation*}
      \prob{\AuditExptWParams{0}(\MLPractitioner, \DatasetSize, \OriginalDatasetSize)=1} \leq \FDRBound,
    \end{equation*}
    where \FDRBound is a pre-defined small value, bounding the false-detection rate.
\end{itemize}

\section{The Proposed Method} \label{sec:proposed_method}

In this section, we propose an instance-level data-use auditing method
that allows a data owner to verify if her data instances were used in
training an ML model. Such an instance-level data-use auditing method
consists of a data-marking algorithm $\MarkAlg$ and a data-use
detection algorithm $\DetectAlg$, which will be introduced in
\secref{sec:proposed_method:data_marking} and
\secref{sec:proposed_method:data_detection}, respectively.  Although
we will focus on evaluating our method in the most challenging case,
i.e., $\DatasetSize=1$, we introduce our proposed method in a general
way, allowing a data owner to audit a set of $\Dataset = \{\Data{1},
\Data{2}, \dots, \Data{\DatasetSize}\}$.

\subsection{Data-Marking Algorithm} \label{sec:proposed_method:data_marking}
The data-marking algorithm $\MarkAlg$, applied at the \emph{data
marking and publication} stage, takes as input a set of raw data
instances $\Dataset$ and outputs a marked version $\MarkedDataset$
that the data owner publishes online and a set $\HiddenInfo$ of hidden
information that she keeps secret and will use to verify if an ML
model used her published data instances in training. \MarkAlg works as
follows: Given a raw data instance $\Data{\SampleIdx} \in \Dataset$,
the data owner first generates $\NumberMarkedData$ ($\NumberMarkedData
\gg 2$) marked versions of $\Data{\SampleIdx}$, namely
$\Data{\SampleIdx}{1}, \Data{\SampleIdx}{2}, \dots,
\Data{\SampleIdx}{\NumberMarkedData}$.  Then, she uniformly at random
samples a data instance $\MarkedData{\SampleIdx} \getsr \{\Data{\SampleIdx}{1}, \Data{\SampleIdx}{2}, \dots, \Data{\SampleIdx}{\NumberMarkedData}\}$
and sets $\HiddenInfo{\SampleIdx} \leftarrow \{\Data{\SampleIdx}{1}, \Data{\SampleIdx}{2}, \dots, \Data{\SampleIdx}{\NumberMarkedData}\} \setminus \{\MarkedData{\SampleIdx}\}$.
As such, she publishes
$\MarkedDataset \leftarrow \{\MarkedData{1}, \MarkedData{2}, \dots, \MarkedData{\DatasetSize}\}$
and keeps
$\HiddenInfo=\bigcup_{\SampleIdx=1}^{\DatasetSize}\HiddenInfo{\SampleIdx}$
secret.

To generate a marked datum $\Data{\SampleIdx}{\MarkedVersionIdx}$, the
data owner adds pixel additions into the raw data instance
$\Data{\SampleIdx}$ by
$\Data{\SampleIdx}{\MarkedVersionIdx} = \Data{\SampleIdx} + \Mark{\SampleIdx}{\MarkedVersionIdx}$.

\paragraph{Two requirements for $\NumberMarkedData$ marked versions of
  $\Data{\SampleIdx}$} 
Following previous work (e.g.,~\cite{huang2024:auditdata}), we have
two basic requirements for generating $\Data{\SampleIdx}{1},
\Data{\SampleIdx}{2}, \dots, \Data{\SampleIdx}{\NumberMarkedData}$:
\begin{itemize}
    \item \emph{Utility preservation}: Since the published data
      instance $\MarkedData{\SampleIdx}$ should preserve the utility
      (i.e., visual quality) of the raw data instance
      $\Data{\SampleIdx}$ (defined by one of the basic requirements
      for data-use auditing in ML,
      see~\secref{sec:background:data_auditing}) and
      $\MarkedData{\SampleIdx}$ is selected from
      $\Data{\SampleIdx}{1}, \Data{\SampleIdx}{2}, \dots,
      \Data{\SampleIdx}{\NumberMarkedData}$, each marked datum
      $\Data{\SampleIdx}{\MarkedVersionIdx}$ should preserve the
      utility (i.e., visual quality) of the raw data instance
      $\Data{\SampleIdx}$ as well.  Formally, given a function used to
      measure the visual similarity between a pair of images (e.g., we
      use $\ell_{\infty}$-norm~\cite{sharif2018:suitability} of pixel
      differences), \emph{utility preservation} requires
        \begin{equation}
            \infinityNorm{\Mark{\SampleIdx}{\MarkedVersionIdx}} \leq \MarkBound,
        \end{equation}
        where $\MarkBound$ is a small scalar controlling visual
        similarity.  A smaller $\MarkBound$ implies that
        $\Data{\SampleIdx}{\MarkedVersionIdx}$ is more similar to
        $\Data{\SampleIdx}$, thereby preserving more utility.
    \item \emph{Distinction}: The $\NumberMarkedData$ marked data
      should be different enough such that membership inference,
      applied in our data-use detection algorithm (see
      \secref{sec:proposed_method:data_detection}), can distinguish an
      ML model trained on one but not the others. Formally, given a
      distance function $\Distance{\cdot}{\cdot}$, \emph{distinction}
      requires that the minimum pairwise distance among the
      $\NumberMarkedData$ marked data should be as large as
      possible. In other words,
        \begin{equation} \label{eq:opimize_distinction}
            \max_{\{\Data{\SampleIdx}{1}, \Data{\SampleIdx}{2}, \dots, \Data{\SampleIdx}{\NumberMarkedData}\}} \min_{1\leq \MarkedVersionIdx < \MarkedVersionIdxAlt \leq \NumberMarkedData} \Distance{\Data{\SampleIdx}{\MarkedVersionIdx}}{\Data{\SampleIdx}{\MarkedVersionIdxAlt}}.
        \end{equation}
        When having access to a pretrained feature extractor
        $\FeatureExtractor$ (e.g., pretrained on
        ImageNet~\cite{deng2009:imagenet}) that prepares the
        high-level features of an image, we define the distance
        function $\Distance{\Data{\SampleIdx}}{\Data{\SampleIdxAlt}}$
        as $\Distance{\Data{\SampleIdx}{\MarkedVersionIdx}}{\Data{\SampleIdx}{\MarkedVersionIdxAlt}} \defeq \euclideanNorm{\FeatureExtractor(\Data{\SampleIdx}{\MarkedVersionIdx})-\FeatureExtractor(\Data{\SampleIdx}{\MarkedVersionIdxAlt})}$.
        As such, \eqnref{eq:opimize_distinction} is reformulated as:
        \begin{equation} \label{eq:opimize_distinction_reformulated}
          \max_{\{\Data{\SampleIdx}{1}, \Data{\SampleIdx}{2}, \dots,
            \Data{\SampleIdx}{\NumberMarkedData}\}} \min_{1\leq
            \MarkedVersionIdx < \MarkedVersionIdxAlt \leq \NumberMarkedData}
          \euclideanNorm{\FeatureExtractor(\Data{\SampleIdx}{\MarkedVersionIdx})
            - \FeatureExtractor(\Data{\SampleIdx}{\MarkedVersionIdxAlt})}.
        \end{equation}
\end{itemize}

\paragraph{Formulating an optimization problem} Given the two requirements, we formulate the problem of generating the marks  as the following optimization problem:
\begin{subequations} \label{eq:generate_mark}
    \begin{align}
    \max_{\{\Mark{\SampleIdx}{1}, \Mark{\SampleIdx}{2}, \dots, \Mark{\SampleIdx}{\NumberMarkedData}\}} & \min_{1\leq \MarkedVersionIdx < \MarkedVersionIdxAlt \leq \NumberMarkedData} \euclideanNorm{\FeatureExtractor(\Data{\SampleIdx}+\Mark{\SampleIdx}{\MarkedVersionIdx})-\FeatureExtractor(\Data{\SampleIdx}+\Mark{\SampleIdx}{\MarkedVersionIdxAlt})}, \label{eq:generate_mark:obj} \\ 
    \text{subject to} & \quad \infinityNorm{\Mark{\SampleIdx}{\MarkedVersionIdx}} \leq \MarkBound,  \label{eq:generate_mark:constraint}
    \end{align}
\end{subequations}
where the objective quantifies the distinction requirement and the
constraint quantifies the utility-preservation constraint.

\paragraph{Solving the optimization problem}
It is challenging to solve \eqnref{eq:generate_mark} using, e.g., a
gradient-based method.  Intuitively, at each iteration of a
gradient-based method, we need to find a pair of marked data such that
their distance is the smallest among
all the pairs.  Then, we use gradient ascent to update the marks for
this pair of marked data. However, this method is costly and slow,
since we need to compute all the pairwise distances to find the closet
pair and only update two marks at each iteration. Thus, we instead
apply a two-step method to approximately solve
\eqnref{eq:generate_mark}. First, we generate $\NumberMarkedData$ unit
vectors of the same dimension as the output dimension of the feature
extractor $\FeatureExtractor$ (e.g., $512$ dimensions for a pretrained
ResNet-18) such that the minimum pairwise distance among the
$\NumberMarkedData$ unit vectors is maximized.  Second, we craft the
$\MarkedVersionIdx$-th mark such that the dot product between the
feature vector of the $\MarkedVersionIdx$-th marked version (prepared
by $\FeatureExtractor$) and the $\MarkedVersionIdx$-th unit vector is
maximized while satisfying the constraint (i.e.,
\eqnref{eq:generate_mark:constraint}).  We present the pseudocode of
the algorithm for generating $\Data{\SampleIdx}{1},
\Data{\SampleIdx}{2}, \dots, \Data{\SampleIdx}{\NumberMarkedData}$ in
\algref{alg:mark_generation} in \appref{app:algorithm}.

\subsection{Data-Use Detection Algorithm} \label{sec:proposed_method:data_detection}

The data-use detection algorithm is applied at the \emph{data-use
detection} stage after an ML model is deployed and accessible.  It is
used to detect if the ML model uses the published data instances in
its training.  Given oracle (i.e., black-box) access to a deployed
ML model $\MLModel$, it takes as inputs a set of published data
instances $\MarkedDataset$ and a set of hidden information
$\HiddenInfo$, and outputs a bit $\DOBit \in \{0, 1\}$ that reflects
the detection result, with $\DOBit=1$ indicating a detection.

\paragraph{Obtaining ``memorization'' scores using \emph{any} membership inference method} 
The data-use detection algorithm measures the ``memorization'' of each published instance $\MarkedData{\SampleIdx}$ 
compared with those marked data instances in the hidden set $\HiddenInfo{\SampleIdx}$ (i.e., the rank of $\MarkedData{\SampleIdx}$ 
among the $\NumberMarkedData$ generated marked data according to their ``memorization''). Such ``memorization'' 
can be measured by any black-box membership inference method 
(e.g., negative modified entropy of the output of an image classifier~\cite{song2021:systematic}) 
whose output indicates the likelihood of the input data being a training sample of an ML model 
(i.e., a larger output indicates a higher likelihood). Specifically, we use a membership inference method 
$\MIAlg{\MLModel}$ (where we are allowed black-box access to the ML model $\MLModel$) to obtain the 
``memorization'' score $\MIAlg{\MLModel}(\Data{\SampleIdx}{\MarkedVersionIdx})$ of a marked data instance 
$\Data{\SampleIdx}{\MarkedVersionIdx}$, which is the published data instance from $\MarkedDataset$ or 
a marked data instance from the hidden information set $\HiddenInfo$.

\paragraph{Formulating a rank-based hypothesis}
We rank the $\NumberMarkedData$ marked data instances of each raw data
instance in increasing order with respect to their ``memorization''
scores, and obtain the rank
$\Rank{\MarkedData{\SampleIdx},\HiddenInfo{\SampleIdx},\MIAlg{\MLModel}}
\in \{1, 2, \dots, \NumberMarkedData\}$ of the published data instance
$\MarkedData{\SampleIdx}$ where $\Rank{\MarkedData{\SampleIdx},
  \HiddenInfo{\SampleIdx},\MIAlg{\MLModel}}=\NumberMarkedData$ means
that $\MarkedData{\SampleIdx}$ has the largest ``memorization'' score
and $\Rank{\MarkedData{\SampleIdx},
  \HiddenInfo{\SampleIdx},\MIAlg{\MLModel}}=1$ means it has the
smallest.  Specifically,
\begin{equation*}
    \Rank{\MarkedData{\SampleIdx},\HiddenInfo{\SampleIdx},\MIAlg{\MLModel}} = 1 + \sum_{\Data\in\HiddenInfo{\SampleIdx}} \IndicatorFunction{\MIAlg{\MLModel}(\MarkedData{\SampleIdx}) > \MIAlg{\MLModel}(\Data)},
\end{equation*}
where $\IndicatorFunction{\cdot}$ is an indicator function returning
$1$ if the input statement is true and $0$ otherwise.\footnote{We
assume that $\MIAlg{\MLModel}(\MarkedData{\SampleIdx}) \neq
\MIAlg{\MLModel}(\Data)$.  If not, we break the tie based on their
indices. Specifically, if $\MarkedData{\SampleIdx} =
\Data{\SampleIdx}{\MarkedVersionIdx}$, $\Data =
\Data{\SampleIdx}{\MarkedVersionIdxAlt}$, and
$\MIAlg{\MLModel}(\MarkedData{\SampleIdx}) = \MIAlg{\MLModel}(\Data)$,
then we define
$\IndicatorFunction{\MIAlg{\MLModel}(\MarkedData{\SampleIdx}) >
  \MIAlg{\MLModel}(\Data)} = 1$ if $\MarkedVersionIdx >
\MarkedVersionIdxAlt$, and 0 otherwise.}  In other words, the rank of
$\MarkedData{\SampleIdx}$ is the number of items in
\HiddenInfo{\SampleIdx} on which \MIAlg{\MLModel} judges \MLModel less
likely to have been trained than $\MarkedData{\SampleIdx}$, plus one.
In this way, we get ranks
\Rank{\MarkedData{1},\HiddenInfo{1},\MIAlg{\MLModel}},
\Rank{\MarkedData{2},\HiddenInfo{2},\MIAlg{\MLModel}}, $\dots$,
\Rank{\MarkedData{\DatasetSize},\HiddenInfo{\DatasetSize},\MIAlg{\MLModel}},
and the sum of these $\DatasetSize$ ranks is:
\begin{equation}
  \RankSum = \DatasetSize + \TotalSuccess,
  \label{eqn:ranksum}
\end{equation}
where $\TotalSuccess =
\sum_{\SampleIdx=1}^{\DatasetSize}\sum_{\Data\in\HiddenInfo{\SampleIdx}}
\IndicatorFunction{\MIAlg{\MLModel}(\MarkedData{\SampleIdx}) >
  \MIAlg{\MLModel}(\Data)}$.  When the ML model does not use any
published data instance from $\MarkedDataset$ in training---which is
our null hypothesis---the rank of $\MarkedData{\SampleIdx}$ is
uniformly distributed over $\{1, 2, \dots, \NumberMarkedData\}$. In
other words, under the null hypothesis, we have for all $\SampleIdx
\in \{1, \ldots, \DatasetSize\}$ and $\RankVariable \in \{1, \dots,
\NumberMarkedData\}$:
\begin{equation*}
    \prob{\Rank{\MarkedData{\SampleIdx},\HiddenInfo{\SampleIdx},\MIAlg{\MLModel}} = \RankVariable} =  \frac{1}{\NumberMarkedData}.
\end{equation*}
Furthermore, under the null hypothesis, the distribution of the sum of
ranks is that of a normalized extended binomial
coefficient~\cite{caiado2007:polynomial} and we have its probability
mass function as:
\begin{equation*}
  \prob{\RankSum = \RankVariable} = \frac{1}{\NumberMarkedData^{\DatasetSize}} \sum_{\VariableIdx=0}^{\floor{\frac{\RankVariable-\DatasetSize}{\NumberMarkedData}}} (-1)^{\VariableIdx} {\DatasetSize \choose \VariableIdx} {\RankVariable - \NumberMarkedData \VariableIdx - 1 \choose \DatasetSize - 1},
\end{equation*}
where $\RankVariable \in \{\DatasetSize,\DatasetSize+1, \dots,
\DatasetSize\NumberMarkedData\}$.  However, when the ML model uses
some published data instances from $\MarkedDataset$ in training, it is
more likely that the sum of ranks is high due to the intuition that
the ML model tends to memorize its training
samples~\cite{song2017:machine} (i.e., some ranks follow a non-uniform
distribution over $\{1, 2, \dots, \NumberMarkedData\}$).  As such, we
can detect the use of some published data instances from
$\MarkedDataset$ based on $\RankSum$.

\paragraph{Detecting data-use sequentially}
Obtaining the sum of ranks requires querying the ML model with all
marked data instances, which can be highly costly when
$\NumberMarkedData$ is large (e.g., we set $\NumberMarkedData=1000$ in
\secref{sec:auditing_ML}). To address this, we propose a sequential
method: Initially we obtain the ``memorization'' scores of the
published data instances, i.e., $\MIAlg{\MLModel}(\MarkedData{1}),
\MIAlg{\MLModel}(\MarkedData{2}), \dots,
\MIAlg{\MLModel}(\MarkedData{\DatasetSize})$, and then, at each time
step, we sample an $\Data$ from $\HiddenInfo$ randomly without
replacement (WoR), measure if
$\MIAlg{\MLModel}(\MarkedData{\SampleIdx}) > \MIAlg{\MLModel}(\Data)$
(where $\Data \in \HiddenInfo{\SampleIdx}$), and estimate
$\TotalSuccess$ and so \RankSum (see \eqnref{eqn:ranksum}) using the
measurements so far.  Following the previous works
(e.g.,~\cite{waudby2020:confidence,huang2024:auditdata}), we estimate
$\TotalSuccess$ at each time step by applying a prior-posterior-ratio
martingale (denoted as $\PPRMartingale$)~\cite{waudby2020:confidence}
on the currently obtained measurements. At the time $\ObservationTime
\in \{1, 2, \dots, \DatasetSize(\NumberMarkedData-1)\}$,
\PPRMartingale takes as inputs the measurements obtained so far (i.e.,
a sequence of $\ObservationTime$ binary values, each indicating if
$\MIAlg{\MLModel}(\MarkedData{\SampleIdx}) > \MIAlg{\MLModel}(\Data)$
where $\Data \in \HiddenInfo{\SampleIdx}$), the size
$\DatasetSize(\NumberMarkedData-1)$ of $\HiddenInfo$, and a confidence
level $\Confidence\in[0, 1]$, and returns a confidence interval
$\ConfidenceInterval{\ObservationTime}{\Confidence} =
[\ConfidenceIntervalLower{\ObservationTime}{\Confidence},
  \ConfidenceIntervalUpper{\ObservationTime}{\Confidence}]$ for
$\TotalSuccess$.  Such a sequence of confidence intervals
$\{\ConfidenceInterval{\ObservationTime}{\Confidence}\}_{\ObservationTime
  \in \{1, 2, \dots, \DatasetSize(\NumberMarkedData-1)\}}$ has the
following guarantee~\cite{waudby2020:confidence}:
\begin{equation*}
  \prob{\exists \ObservationTime \in \{1, 2, \dots, \DatasetSize(\NumberMarkedData-1)\}: \TotalSuccess \notin \ConfidenceInterval{\ObservationTime}{\Confidence}} \leq \Confidence.
\end{equation*}
In words, the probability that 
$\TotalSuccess$ falls outside the confidence intervals at any time step is at most 
$\Confidence$.

As such, we design our data-use detection algorithm as follows:
Initially, the data owner tests the ``memorization'' scores
$\MIAlg{\MLModel}(\MarkedData{1}), \MIAlg{\MLModel}(\MarkedData{2}),
\dots, \MIAlg{\MLModel}(\MarkedData{\DatasetSize})$.  Then, at each
time step, she samples an $\Data$ from $\HiddenInfo$ randomly WoR,
tests the ``memorization'' score $\MIAlg{\MLModel}(\Data)$, measures
if $\MIAlg{\MLModel}(\MarkedData{\SampleIdx}) >
\MIAlg{\MLModel}(\Data)$ where $\Data \in \HiddenInfo{\SampleIdx}$,
and applies \PPRMartingale to obtain a confidence interval
$[\ConfidenceIntervalLower{\ObservationTime}{\Confidence},
  \ConfidenceIntervalUpper{\ObservationTime}{\Confidence}]$ for
$\TotalSuccess$. If the lower bound
$\ConfidenceIntervalLower{\ObservationTime}{\Confidence}$ of the
confidence interval is equal to or larger than a preselected threshold
$\sumThreshold$, the data owner stops sampling and rejects the null
hypothesis, i.e., she returns $\DOBit=1$ concluding that she detects
the use of some members of $\MarkedDataset$ in $\MLModel$.  Otherwise,
she continues sampling.  When all the unpublished data instances have
been exhausted and all lower bounds of confidence intervals are
smaller than the threshold $\sumThreshold$, she returns $\DOBit=0$.
Here $\sumThreshold$ is a tunable parameter that controls the upper
bound of \FalseDetectionRate of our method; see
Theorem~\ref{theorem:fdr} below.

Our data-use detection algorithm is a generalization of Huang et
al.'s~\cite{huang2024:auditdata}, in that their test is a special case
of our method, specifically with $\NumberMarkedData = 2$.  Because of
this limitation, their method works only when $\DatasetSize \gg 1$,
i.e., for dataset-level auditing for large datasets.  Our
generalization enables setting $\NumberMarkedData \gg 2$ in order to
accommodate a small \DatasetSize (even $\DatasetSize = 1$).

\subsection{Guarantee on False-Detection Rate}
A false detection happens when a data-use auditing method outputs
$\DOBit=1$ when the ML model did not use any published data instance
to train. \FalseDetectionRate is the probability that false detection
happens. We show that our instance-level data-use auditing method,
which consists of the data-marking algorithm and data-use detection
algorithm described above, allows us to set $\sumThreshold$ so that
the parameter $\FDRBound$ is an upper bound on \FalseDetectionRate.

\begin{theorem}[Bound on $\FalseDetectionRate$] \label{theorem:fdr}
    Given any $\FDRBound \in [0, 1]$ and $\Confidence < \FDRBound$, when we set
    $\sumThreshold$ such that 
    \begin{equation}
      \sum_{\RankVariable=\sumThreshold+\DatasetSize}^{\DatasetSize\NumberMarkedData}\frac{1}{\NumberMarkedData^{\DatasetSize}} \sum_{\VariableIdx=0}^{\floor{\frac{\RankVariable-\DatasetSize}{\NumberMarkedData}}} (-1)^{\VariableIdx} {\DatasetSize \choose \VariableIdx} {\RankVariable - \NumberMarkedData \VariableIdx - 1 \choose \DatasetSize - 1} \leq \FDRBound - \Confidence,  \label{eq:threshold}
      \end{equation}
    our instance-level data-use auditing algorithm has an
   $\FalseDetectionRate$ no larger than $\FDRBound$. In other words:
    \begin{equation*}
        \prob{\AuditExptWParams{0}(\MLPractitioner, 1, \OriginalDatasetSize)=1} \leq \FDRBound.
    \end{equation*}
  \end{theorem}
  
  \begin{proof}
    See \appref{app:proof}.
  \end{proof}

\section{Auditing Data-Use in ML Models} \label{sec:auditing_ML}

In this section, we apply our data-use auditing method to detect the
unauthorized use of data instances in ML models. We consider three
types of visual models, namely image
classifiers~\cite{deng2009:imagenet, simonyan2015:very, he2016:deep},
visual encoders~\cite{chen2020:simple}, and vision-language models 
(CLIP~\cite{radford2021:learning} and BLIP~\cite{li2022:blip} models).

\subsection{Auditing an Image Classifier} \label{sec:auditing_ML:classifier}

An image classifier is an ML model used to assign labels to
images. Image classifiers are widely used in various fundamental
computer-vision tasks, e.g., disease diagnosis~\cite{yadav2019:deep}
and facial recognition~\cite{parkhi2015:deep}. An image classifier
takes as input an image and outputs a multi-dimensional vector, where
each component represents the predicted probability for its associated
class. To train an image classifier, the ML practitioner needs to
label the images that he collects from the data owners, such that each
image used for training is assigned to only one class.

\begin{table}[ht!]
  \centering
  \begin{tabular}{@{}c@{}p{0.85\columnwidth}@{}}
  \toprule
   Symbol & \multicolumn{1}{c}{Meaning} \\ 
  \midrule 
  \NumberUsedDataSamples & No.\ audited data instances used in training \\
  \NumAugment & No.\ augmentations per data instance in data-use detection \\
  \QueryNumberVariable & No.\ marked data instances queried before detection stops \\
  \TrainDatasetSize & No.\ data samples used to train the audited model \\
  \ReferenceDatasetSize & No.\ data samples used to train the reference model (for competitors that use a reference model) \\
  \FDRBound & Bound on $\FalseDetectionRate$ \\
  \Confidence & Confidence level for confidence-interval sequence \\
  \ParameterBetaDistribution & Controls class imbalance of auxiliary dataset to train reference model (for competitors that use one)\\
  \bottomrule 
  \end{tabular}
  \caption{Additional notation for reporting our experiments.}
  \label{tbl:notations}  
\end{table}

\subsubsection{Experimental Setup}  \label{sec:auditing_ML:classifier:setup}

\paragraph{Datasets} We used three visual benchmark datasets, 
namely CIFAR-100~\cite{krizhevsky2009:learning},
TinyImageNet~\cite{Lle2015:TinyIV}, and
ImageNet~\cite{deng2009:imagenet}. 
Please see their descriptions in \appref{app:setup:dataset}.

\paragraph{Data-marking setting}
In each experiment, we simulated how the data owner would mark her
data and how the ML practitioner would assemble his training dataset
by preparing a marked labeled training dataset, as follows: To prepare
a marked labeled training dataset, we first uniformly at random
sampled $500$ training samples for CIFAR-100, $1{,}000$ for
TinyImageNet, or $1{,}000$ for ImageNet, as the data instances that we
audited. 
However, for a fixed value of $\DatasetSize$, each subset of
$\DatasetSize$ audited samples were treated independently (as if from a
different data owner; i.e., we applied our test for each subset of
size $\DatasetSize$ separately).
Taking each subset of $\DatasetSize$ audited samples as $\Dataset$, we applied our
data-marking algorithm to generate its published version
$\MarkedDataset$ and the hidden information $\HiddenInfo$. By default, we set
$\NumberMarkedData=1000$.  When applying the data-marking algorithm on
each data instance from $\Dataset$, 
we set $\MarkBound=10$ for CIFAR-100 or TinyImageNet, or
$\MarkBound=25$ for ImageNet when the pixel range of an image is $[0,
  255]$ and used ResNet-18 pretrained on ImageNet as
$\FeatureExtractor$, as our default. We applied projected gradient
ascent~\cite{lin2007:projected} to solve \eqnref{eq:generate_mark}: At
each step, we updated a mark by gradient ascent and projected it such
that its associated marked data is a valid image (i.e., the pixel
values of the marked data are integers in the range $[0, 255]$).  As
such, we prepared a marked, labeled, training dataset used to train a
classifier, by replacing each audited data instance in the dataset
with its published version and assigning it a correct label (i.e.,
using its original label).  We present some marked image examples in
\figref{fig:cifar100_examples}, \figref{fig:tinyimagenet_examples},
and \figref{fig:imagenet_examples}, in \appref{app:results:examples}.
We evaluated the runtime overhead of our data-marking algorithm, as shown
in~\appref{app:setup:system_config:marking}.

\paragraph{Model training setting}
In each experiment, we simulated how the ML practitioner would develop
an image classifier by training it from scratch on a training dataset
prepared as described.  We trained the classifier using a standard
stochastic gradient descent (SGD)
algorithm~\cite{amari1993:backpropagation}.
The details on SGD algorithm are presented in \appref{app:setup:training:classifier}.
We used ResNet-18~\cite{he2016:deep} as the default model architecture for
CIFAR-100 and TinyImageNet, and used a larger model, ResNet-50, as the
default for ImageNet.

We report the accuracies (\Accuracy) of the classifiers trained on the
marked datasets (i.e., the fraction of test samples correctly
predicted) and differences between the accuracies of classifiers
trained on marked datasets and those of classifiers trained on clean
datasets, in \tblref{tbl:classifier:acc} in
\appref{app:results:classifier_accuracy}.  The classifiers
trained on the marked datasets preserved good utility, i.e., their
$\Accuracy$ values were similar to those trained on clean datasets,
which demonstrated that the marked data instances preserved good
utility of their original versions.

\paragraph{Data-use detection setting}
In each experiment, we simulated how the ML practitioner would deploy
his classifier and how a data owner would detect the use of her marked
image instance, as follows: Unless otherwise specified, we assumed
that the model returned a vector of confidence scores (i.e., a
multi-dimensional vector where each component represents the predicted
probability for the associated class) in response to an input image,
though we also considered settings where the output was a label or a
label with its associated confidence score.
Using the marked image set $\MarkedDataset$ 
and the associated hidden set $\HiddenInfo$ generated
from the data-marking setting, the data owner applied our data-use
detection algorithm to test if an image classifier was trained using
$\MarkedDataset$.
When applying the data-use detection algorithm, we
followed previous works (e.g.,~\cite{choquette2021:label,
  huang2024:auditdata}) to define the black-box membership inference
method. 
In the implementation of membership inference applied in our data-use detection,
we set $\NumAugment=16$ for CIFAR-100 and
TinyImageNet, and $\NumAugment=64$ for ImageNet, as the default.  
We set $\Confidence=0.001$, and considered $\FDRBound=0.05$,
$\FDRBound=0.01$, and $\FDRBound=0.002$ in the data-use detection
algorithm.  Recall that $\FDRBound$ is the bound on
\FalseDetectionRate of the data auditing method.  For example, setting
$\FDRBound=0.002$ guarantees that $\FalseDetectionRate \leq 0.2\%$.
We evaluated the runtime overhead of our data-use detection algorithm,
as shown in~\appref{app:setup:system_config:detection}.

\paragraph{Baselines}
To our knowledge, there is no existing work on instance-level,
proactive, data-use auditing of ML.  Black-box membership inference
can be used to \emph{passively} infer data-use in an ML model but it
does not provide a bounded \FalseDetectionRate.  In addition,
membership inference assumes the availability of auxiliary data from
the same distribution as the training samples and/or at least one
reference model that is trained on a dataset similar to the training
set of the tested model.  We consider the state-of-the-art black-box
membership inference methods, namely Attack-P~\cite{ye2022:enhanced},
Attack-R~\cite{ye2022:enhanced}, LiRA~\cite{carlini2022:membership},
and RMIA~\cite{zarifzadeh2024:low} as our baselines.  We summarize the
limitations of these membership inference methods applied to data-use
auditing in \tblref{table:limitations}.  We provide detailed
descriptions of the baselines and their implementations in
\appref{app:setup:baselines}.

\begin{table}[ht!]
  \centering
  \begin{tabular}{@{}lc@{\hspace{0.5em}}cc@{\hspace{0.5em}}cc@{\hspace{0.5em}}cc@{\hspace{0.5em}}c@{}}
  \toprule
  & \multicolumn{1}{c}{Auxiliary} & \multicolumn{1}{c}{Reference}&  \multicolumn{1}{c}{Bounded} \\ 
  & \multicolumn{1}{c}{data} & \multicolumn{1}{c}{model} &  \multicolumn{1}{c}{FDR}  \\
  \midrule 
  Our method  & \emptycircle{} & \emptycircle{} & \chkmrk  \\ 
  Attack-P~\cite{ye2022:enhanced} & \fullcircle{} & \emptycircle{} & \xmark \\ 
  Attack-R~\cite{ye2022:enhanced} & \fullcircle{} & \fullcircle{}  & \xmark \\
  LiRA~\cite{carlini2022:membership} & \fullcircle{} & \fullcircle{} & \xmark \\
  RMIA~\cite{zarifzadeh2024:low} & \fullcircle{} & \fullcircle{} & \xmark  \\ 
  \bottomrule 
  \end{tabular} 
  \caption[Short caption without TikZ commands]{Limitations of
    membership inference applied in data-use
    auditing. ``\fullcircle{}'' means the information is needed while
    ``\emptycircle{}'' means the information is not
    needed. ``\chkmrk'' means it provides a bounded
    \FalseDetectionRate while ``\xmark'' means it does not.}
  \label{table:limitations}
\end{table}

\paragraph{Metric}
We use $\TrueDetectionRate$ as measured in \figref{fig:data_auditing}
to evaluate the effectiveness of a data-use auditing
method. $\TrueDetectionRate$ is the fraction of experiments where a
data-use auditing method (i.e., its data-use detection algorithm)
returned ``detected'' (i.e., $\DOBit=1$) when the ML model was trained
using $\MarkedDataset$. 
Under a specific bound on \FalseDetectionRate, a higher $\TrueDetectionRate$ means a more
effective data-use auditing method.

We use $\QueryNumberVariable$ to denote the number of marked data
instances (including the published one and those not published) used
to query the audited ML model before our data-use detection algorithm
stopped. That is, the total query cost for our algorithm was
$\QueryNumberVariable \times \NumAugment$.  Therefore, a lower
$\QueryNumberVariable$ indicates a lower query cost of our method.

\subsubsection{Experimental Results} \label{sec:auditing_ML:classifier:results}

\paragraph{Main results}
Our results of auditing data-use in image classifiers are shown in
\figref{fig:classifier:main}. Under the default setting, our
$\TrueDetectionRate$ ($\DatasetSize=1$) for auditing CIFAR-100
(TinyImageNet) instances was $28.21\%$ ($18.09\%$), $11.59\%$
($6.02\%$), and $3.11\%$ ($1.39\%$), when the bounds on
$\FalseDetectionRate$ were set as $5\%$, $1\%$, and $0.2\%$,
respectively. When the output of a classifier included both the
predicted label and its associated confidence score, the
$\TrueDetectionRate$ remained comparable to that under the default
setting.  When only the predicted label was available, our
$\TrueDetectionRate$ ($\DatasetSize = 1$) was only marginally above
the $\FalseDetectionRate$ level. However, when the data owner
possessed more data instances (i.e., $\DatasetSize > 1$) and all of
them were used in training, our method became significantly more
effective, even when the outputs of the audited classifiers were
labels only (``Label''). For ImageNet, our $\TrueDetectionRate$
($\DatasetSize = 1$) was only slightly higher the
$\FalseDetectionRate$ level at $\DatasetSize = 1$, but clearly
improved on it by $\DatasetSize = 16$.

\figref{fig:classifier:main:partial} presents the results of scenarios
where a data owner had $\DatasetSize=10$ data instances and
$\NumberUsedDataSamples \in \{1, \ldots, \DatasetSize\}$ were used in
training.  As in \figref{fig:classifier:main:partial}, increasing
\NumberUsedDataSamples quickly improved $\TrueDetectionRate$,
noticeably improving over the \FalseDetectionRate bound for both
confidence vectors (``All'') and classification confidences
(``Highest''), even when $\NumberUsedDataSamples = 2$ or $3$.

\begin{figure}[ht!]
  \centering

  \begin{subfigure}[b]{0.45\textwidth}
    \centering
    \resizebox{!}{1.7em}{\newenvironment{customlegend}[1][]{%
    \begingroup
    \csname pgfplots@init@cleared@structures\endcsname
    \pgfplotsset{#1}%
}{%
    \csname pgfplots@createlegend\endcsname
    \endgroup
}%

\def\addlegendimage{\csname pgfplots@addlegendimage\endcsname}

\begin{tikzpicture}

\definecolor{darkslategray38}{RGB}{38,38,38}
\definecolor{indianred1967882}{RGB}{196,78,82}
\definecolor{lightgray204}{RGB}{204,204,204}
\definecolor{mediumseagreen85168104}{RGB}{85,168,104}
\definecolor{peru22113282}{RGB}{221,132,82}
\definecolor{steelblue76114176}{RGB}{76,114,176}

\begin{customlegend}[
    legend style={{font={\small}},{draw=none}}, 
    legend columns=4,
    legend cell align={center},
    legend entries={{All}, {Highest}, {Label}, {$\FalseDetectionRate$ bound}}]
\addlegendimage{very thick, color=steelblue76114176,mark=o}
\addlegendimage{very thick, peru22113282,mark=diamond,dashed,mark options={solid}}
\addlegendimage{very thick, mediumseagreen85168104,mark=triangle,dotted,mark options={solid}}
\addlegendimage{very thick, darkgray,dashdotted}

\end{customlegend}

\end{tikzpicture}}
  \end{subfigure}

  \begin{subfigure}[t]{0.47\textwidth}
    \centering
  \begin{minipage}[t]{0.45\textwidth}
    \centering
    \begin{tikzpicture}
      \begin{axis}[
        xlabel={$\DatasetSize$},
        ylabel={$\TrueDetectionRate (\%)$},
        tick label style={font=\footnotesize},
        xtick={1, 2, 3, 4, 5, 6, 7},
        xticklabels={$1$, , $4$, , $16$, , $64$},
        ymode=log,
        log basis y=10, 
        ymin=0.07, ymax=140,
        ytick={0.1, 1, 10, 100},
        grid=major,
        width=1.0\textwidth,
        height=0.8\textwidth,
        ]
        \addplot[color=steelblue76114176,mark=o,line width=1.5] coordinates {
          (1, 28.21)
          (2, 37.84)
          (3, 57.95)
          (4, 83.7)
          (5, 97.91)
          (6, 100)
          (7, 100)
        };
        \addplot[color=peru22113282,mark=diamond,dashed,mark options={solid},line width=1.5] coordinates {
          (1,27.93)
          (2,37.86)
          (3,57.27)
          (4,82.26)
          (5,97.58)
          (6,99.95)
          (7,100.0)
        };
        \addplot[color=mediumseagreen85168104,mark=triangle,dotted,mark options={solid},line width=1.5] coordinates {
          (1,7.62)
          (2,11.09)
          (3,16.24)
          (4,25.97)
          (5,40.21)
          (6,66.29)
          (7,90.87)
        };
        \addplot[color=darkgray,dashdotted,line width=1.5] coordinates {
          (1, 5)
          (2, 5)
          (3, 5)
          (4, 5)
          (5, 5)
          (6, 5)
          (7, 5)
        };
      \end{axis}
    \end{tikzpicture}
  \end{minipage}
  \hfill
  \begin{minipage}[t]{0.25\textwidth}
    \hspace{-0.225in}
    \begin{tikzpicture}
      \begin{axis}[
        xlabel={$\DatasetSize$},
        xtick={1, 2, 3, 4, 5, 6, 7},
        xticklabels={$1$, , $4$, , $16$, , $64$},
        tick label style={font=\footnotesize},
        ymode=log,
        log basis y=10, 
        ymin=0.07, ymax=140,
        ytick={0.1, 1, 10, 100},
        yticklabels=\empty,
        grid=major,
        width=1.8\textwidth,
        height=1.44\textwidth,
        ]
        \addplot[color=steelblue76114176,mark=o,line width=1.5] coordinates {
          (1, 11.59)
          (2, 17.36)
          (3, 30.59)
          (4, 60.81)
          (5, 91.2)
          (6, 99.86)
          (7, 100)
        };
        \addplot[color=peru22113282,mark=diamond,dashed,mark options={solid},line width=1.5] coordinates {
          (1,11.93)
          (2,17.02)
          (3,29.56)
          (4,59.33)
          (5,90.45)
          (6,99.72)
          (7,100.0)
        };
        \addplot[color=mediumseagreen85168104,mark=triangle,dotted,mark options={solid},line width=1.5] coordinates {
          (1,1.1)
          (2,2.43)
          (3,3.86)
          (4,7.94)
          (5,16.29)
          (6,36.83)
          (7,72.12)
        };
        \addplot[color=darkgray,dashdotted,line width=1.5] coordinates {
          (1, 1)
          (2, 1)
          (3, 1)
          (4, 1)
          (5, 1)
          (6, 1)
          (7, 1)
        };
      \end{axis}
    \end{tikzpicture}
  \end{minipage}
  \hfill
  \begin{minipage}[t]{0.25\textwidth}
    \hspace{-0.21in}
    \begin{tikzpicture}
      \begin{axis}[
        xlabel={$\DatasetSize$},
        xtick={1, 2, 3, 4, 5, 6, 7},
        xticklabels={$1$, , $4$, , $16$, , $64$},
        tick label style={font=\footnotesize},
        ymode=log,
        log basis y=10, 
        ymin=0.07, ymax=140,
        ytick={0.1, 1, 10, 100},
        yticklabels=\empty,
        grid=major,
        width=1.8\textwidth,
        height=1.44\textwidth,
        ]
        \addplot[color=steelblue76114176,mark=o,line width=1.5] coordinates {
          (1, 3.11)
          (2, 5.44)
          (3, 11.68)
          (4, 33.1)
          (5, 73.48)
          (6, 98.53)
          (7, 100)
        };
        \addplot[color=peru22113282,mark=diamond,dashed,mark options={solid},line width=1.5] coordinates {
          (1,3.44)
          (2,5.55)
          (3,11.35)
          (4,30.18)
          (5,70.89)
          (6,97.93)
          (7,100.0)
        };
        \addplot[color=mediumseagreen85168104,mark=triangle,dotted,mark options={solid},line width=1.5] coordinates {
          (1,0.15)
          (2,0.18)
          (3,0.62)
          (4,1.4)
          (5,4.26)
          (6,13.23)
          (7,42.3)
        };
        \addplot[color=darkgray,dashdotted,line width=1.5] coordinates {
          (1, 0.2)
          (2, 0.2)
          (3, 0.2)
          (4, 0.2)
          (5, 0.2)
          (6, 0.2)
          (7, 0.2)
        };
      \end{axis}
    \end{tikzpicture}
  \end{minipage}
  \caption[Short Caption]{CIFAR-100: $@\FalseDetectionRate \le 5\%$ (left), $1\%$ (center), $0.2\%$ (right)}
\end{subfigure}
  \\
  \begin{subfigure}[t]{0.47\textwidth}
    \centering
  \begin{minipage}[t]{0.45\textwidth}
    \centering
    \begin{tikzpicture}
      \begin{axis}[
        xlabel={$\DatasetSize$},
        ylabel={$\TrueDetectionRate (\%)$},
        tick label style={font=\footnotesize},
        xtick={1, 2, 3, 4, 5, 6, 7},
        xticklabels={$1$, , $4$, , $16$, , $64$},
        ymode=log,
        log basis y=10, 
        ymin=0.07, ymax=140,
        ytick={0.1, 1, 10, 100},
        grid=major,
        width=1.0\textwidth,
        height=0.8\textwidth,
        ]
        \addplot[color=steelblue76114176,mark=o,line width=1.5] coordinates {
          (1, 18.09)
          (2, 22.96)
          (3, 34.07)
          (4, 55.21)
          (5, 79.99)
          (6, 96.62)
          (7, 99.94)
        };
        \addplot[color=peru22113282,mark=diamond,dashed,mark options={solid},line width=1.5] coordinates {
          (1,17.55)
          (2,22.96)
          (3,33.27)
          (4,54.05)
          (5,79.07)
          (6,96.88)
          (7,99.93)
        };
        \addplot[color=mediumseagreen85168104,mark=triangle,dotted,mark options={solid},line width=1.5] coordinates {
          (1,6.02)
          (2,6.87)
          (3,8.73)
          (4,11.69)
          (5,16.15)
          (6,24.05)
          (7,37.33)
        };
        \addplot[color=darkgray,dashdotted,line width=1.5] coordinates {
          (1, 5)
          (2, 5)
          (3, 5)
          (4, 5)
          (5, 5)
          (6, 5)
          (7, 5)
        };
      \end{axis}
    \end{tikzpicture}
  \end{minipage}
  \hfill
  \begin{minipage}[t]{0.25\textwidth}
    \hspace{-0.225in}
    \begin{tikzpicture}
      \begin{axis}[
        xlabel={$\DatasetSize$},
        xtick={1, 2, 3, 4, 5, 6, 7},
        xticklabels={$1$, , $4$, , $16$, , $64$},
        tick label style={font=\footnotesize},
        ymode=log,
        log basis y=10, 
        ymin=0.07, ymax=140,
        ytick={0.1, 1, 10, 100},
        yticklabels=\empty,
        grid=major,
        width=1.8\textwidth,
        height=1.44\textwidth,
        ]
        \addplot[color=steelblue76114176,mark=o,line width=1.5] coordinates {
          (1, 6.02)
          (2, 7.84)
          (3, 13.25)
          (4, 27.93)
          (5, 55.59)
          (6, 87.37)
          (7, 99.53)
        };
        \addplot[color=peru22113282,mark=diamond,dashed,mark options={solid},line width=1.5] coordinates {
          (1,6.08)
          (2,8.08)
          (3,12.83)
          (4,27.9)
          (5,55.09)
          (6,87.35)
          (7,99.46)
        };
        \addplot[color=mediumseagreen85168104,mark=triangle,dotted,mark options={solid},line width=1.5] coordinates {
          (1,0.97)
          (2,1.38)
          (3,1.87)
          (4,2.77)
          (5,4.19)
          (6,7.7)
          (7,15.0)
        };
        \addplot[color=darkgray,dashdotted,line width=1.5] coordinates {
          (1, 1)
          (2, 1)
          (3, 1)
          (4, 1)
          (5, 1)
          (6, 1)
          (7, 1)
        };
      \end{axis}
    \end{tikzpicture}
  \end{minipage}
  \hfill
  \begin{minipage}[t]{0.25\textwidth}
    \hspace{-0.21in}
    \begin{tikzpicture}
      \begin{axis}[
        xlabel={$\DatasetSize$},
        xtick={1, 2, 3, 4, 5, 6, 7},
        xticklabels={$1$, , $4$, , $16$, , $64$},
        tick label style={font=\footnotesize},
        ymode=log,
        log basis y=10, 
        ymin=0.07, ymax=140,
        ytick={0.1, 1, 10, 100},
        yticklabels=\empty,
        grid=major,
        width=1.8\textwidth,
        height=1.44\textwidth,
        ]
        \addplot[color=steelblue76114176,mark=o,line width=1.5] coordinates {
          (1, 1.39)
          (2, 2.03)
          (3, 3.42)
          (4, 9.39)
          (5, 27.68)
          (6, 66.69)
          (7, 97.15)
        };
        \addplot[color=peru22113282,mark=diamond,dashed,mark options={solid},line width=1.5] coordinates {
          (1,1.24)
          (2,1.86)
          (3,3.44)
          (4,9.55)
          (5,27.39)
          (6,66.53)
          (7,96.52)
        };
        \addplot[color=mediumseagreen85168104,mark=triangle,dotted,mark options={solid},line width=1.5] coordinates {
          (1,0.1)
          (2,0.13)
          (3,0.19)
          (4,0.39)
          (5,0.61)
          (6,1.47)
          (7,3.76)
        };
        \addplot[color=darkgray,dashdotted,line width=1.5] coordinates {
          (1, 0.2)
          (2, 0.2)
          (3, 0.2)
          (4, 0.2)
          (5, 0.2)
          (6, 0.2)
          (7, 0.2)
        };
      \end{axis}
    \end{tikzpicture}
  \end{minipage}
  \caption[Short Caption]{TinyImageNet: $@\FalseDetectionRate \le 5\%$ (left), $1\%$ (center), $0.2\%$ (right)}
\end{subfigure}
  \\
  \begin{subfigure}[t]{0.47\textwidth}
    \centering
  \begin{minipage}[t]{0.45\textwidth}
    \centering
    \begin{tikzpicture}
      \begin{axis}[
        xlabel={$\DatasetSize$},
        ylabel={$\TrueDetectionRate (\%)$},
        tick label style={font=\footnotesize},
        xtick={1, 2, 3, 4, 5, 6, 7},
        xticklabels={$1$, , $4$, , $16$, , $64$},
        ymode=log,
        log basis y=10, 
        ymin=0.07, ymax=140,
        ytick={0.1, 1, 10, 100},
        grid=major,
        width=1.0\textwidth,
        height=0.8\textwidth,
        ]
        \addplot[color=steelblue76114176,mark=o,line width=1.5] coordinates {
          (1, 9.60)
          (2, 13.51)
          (3, 18.64)
          (4, 26.86)
          (5, 42.29)
          (6, 67.17)
          (7, 89.98)
        };
        \addplot[color=peru22113282,mark=diamond,dashed,mark options={solid},line width=1.5] coordinates {
          (1,8.3)
          (2,11.38)
          (3,14.51)
          (4,21.44)
          (5,33.69)
          (6,52.84)
          (7,79.51)
        };
        \addplot[color=mediumseagreen85168104,mark=triangle,dotted,mark options={solid},line width=1.5] coordinates {
          (1,6.73)
          (2,9.9)
          (3,12.56)
          (4,17.5)
          (5,24.81)
          (6,39.16)
          (7,62.17)
        };
        \addplot[color=darkgray,dashdotted,line width=1.5] coordinates {
          (1, 5)
          (2, 5)
          (3, 5)
          (4, 5)
          (5, 5)
          (6, 5)
          (7, 5)
        };
      \end{axis}
    \end{tikzpicture}
  \end{minipage}
  \hfill
  \begin{minipage}[t]{0.25\textwidth}
    \hspace{-0.225in}
    \begin{tikzpicture}
      \begin{axis}[
        xlabel={$\DatasetSize$},
        xtick={1, 2, 3, 4, 5, 6, 7},
        xticklabels={$1$, , $4$, , $16$, , $64$},
        tick label style={font=\footnotesize},
        ymode=log,
        log basis y=10, 
        ymin=0.07, ymax=140,
        ytick={0.1, 1, 10, 100},
        yticklabels=\empty,
        grid=major,
        width=1.8\textwidth,
        height=1.44\textwidth,
        ]
        \addplot[color=steelblue76114176,mark=o,line width=1.5] coordinates {
          (1, 1.40)
          (2, 3.24)
          (3, 5.48)
          (4, 9.48)
          (5, 18.08)
          (6, 39.21)
          (7, 71.26)
        };
        \addplot[color=peru22113282,mark=diamond,dashed,mark options={solid},line width=1.5] coordinates {
          (1,1.34)
          (2,2.57)
          (3,3.6)
          (4,6.65)
          (5,12.83)
          (6,26.21)
          (7,54.22)
        };
        \addplot[color=mediumseagreen85168104,mark=triangle,dotted,mark options={solid},line width=1.5] coordinates {
          (1,1.01)
          (2,2.2)
          (3,3.25)
          (4,5.01)
          (5,8.12)
          (6,16.05)
          (7,33.74)
        };
        \addplot[color=darkgray,dashdotted,line width=1.5] coordinates {
          (1, 1)
          (2, 1)
          (3, 1)
          (4, 1)
          (5, 1)
          (6, 1)
          (7, 1)
        };
      \end{axis}
    \end{tikzpicture}
  \end{minipage}
  \hfill
  \begin{minipage}[t]{0.25\textwidth}
    \hspace{-0.21in}
    \begin{tikzpicture}
      \begin{axis}[
        xlabel={$\DatasetSize$},
        xtick={1, 2, 3, 4, 5, 6, 7},
        xticklabels={$1$, , $4$, , $16$, , $64$},
        tick label style={font=\footnotesize},
        ymode=log,
        log basis y=10, 
        ymin=0.07, ymax=140,
        ytick={0.1, 1, 10, 100},
        yticklabels=\empty,
        grid=major,
        width=1.8\textwidth,
        height=1.44\textwidth,
        ]
        \addplot[color=steelblue76114176,mark=o,line width=1.5] coordinates {
          (1, 0.20)
          (2, 0.44)
          (3, 0.97)
          (4, 2.1)
          (5, 5.16)
          (6, 15.31)
          (7, 43.25)
        };
        \addplot[color=peru22113282,mark=diamond,dashed,mark options={solid},line width=1.5] coordinates {
          (1,0.22)
          (2,0.22)
          (3,0.51)
          (4,1.38)
          (5,2.93)
          (6,8.66)
          (7,25.54)
        };
        \addplot[color=mediumseagreen85168104,mark=triangle,dotted,mark options={solid},line width=1.5] coordinates {
          (1,0.16)
          (2,0.21)
          (3,0.47)
          (4,0.83)
          (5,1.57)
          (6,4.19)
          (7,11.64)
        };
        \addplot[color=darkgray,dashdotted,line width=1.5] coordinates {
          (1, 0.2)
          (2, 0.2)
          (3, 0.2)
          (4, 0.2)
          (5, 0.2)
          (6, 0.2)
          (7, 0.2)
        };
      \end{axis}
    \end{tikzpicture}
  \end{minipage}
  \caption[Short Caption]{ImageNet: $@\FalseDetectionRate \le 5\%$ (left), $1\%$ (center), $0.2\%$ (right)}
\end{subfigure}
  \caption{$\TrueDetectionRate (\%)$ of auditing image classifiers
    where a data owner had $\DatasetSize$ data
    instances to audit and all of them were used in training. ``All''
    means that the output of a classifier is a full vector of
    confidence scores (our default); ``Highest'' means that the output
    is a label and its associated confidence score; ``Label'' means
    that the output is a label only.}
  \label{fig:classifier:main}
\end{figure}

\begin{figure}[ht!]
  \centering

  \begin{subfigure}[b]{0.45\textwidth}
    \centering
    \resizebox{!}{1.7em}{\newenvironment{customlegend}[1][]{%
    \begingroup
    \csname pgfplots@init@cleared@structures\endcsname
    \pgfplotsset{#1}%
}{%
    \csname pgfplots@createlegend\endcsname
    \endgroup
}%

\def\addlegendimage{\csname pgfplots@addlegendimage\endcsname}

\begin{tikzpicture}

\definecolor{darkslategray38}{RGB}{38,38,38}
\definecolor{indianred1967882}{RGB}{196,78,82}
\definecolor{lightgray204}{RGB}{204,204,204}
\definecolor{mediumseagreen85168104}{RGB}{85,168,104}
\definecolor{peru22113282}{RGB}{221,132,82}
\definecolor{steelblue76114176}{RGB}{76,114,176}

\begin{customlegend}[
    legend style={{font={\small}},{draw=none}}, 
    legend columns=4,
    legend cell align={center},
    legend entries={{All}, {Highest}, {Label}, {$\FalseDetectionRate$ bound}}]
\addlegendimage{very thick, color=steelblue76114176,mark=o}
\addlegendimage{very thick, peru22113282,mark=diamond,dashed,mark options={solid}}
\addlegendimage{very thick, mediumseagreen85168104,mark=triangle,dotted,mark options={solid}}
\addlegendimage{very thick, darkgray,dashdotted}

\end{customlegend}

\end{tikzpicture}}
  \end{subfigure}

  \begin{subfigure}[t]{0.47\textwidth}
    \centering
  \begin{minipage}[t]{0.45\textwidth}
    \centering
    \begin{tikzpicture}
      \begin{axis}[
        xlabel={$\NumberUsedDataSamples$},
        ylabel={$\TrueDetectionRate (\%)$},
        tick label style={font=\footnotesize},
        xtick={1, 2, 3, 4, 5, 6, 7, 8, 9, 10},
        xticklabels={, 2, , 4, , 6, , 8, , 10},
        ymode=log,
        log basis y=10, 
        ymin=0.07, ymax=140,
        ytick={0.1, 1, 10, 100},
        grid=major,
        width=1.0\textwidth,
        height=0.8\textwidth,
        ]
        \addplot[color=steelblue76114176,mark=o,line width=1.5] coordinates {
          (1, 8.03)
          (2, 13.24)
          (3, 19.87)
          (4, 29.4 )
          (5, 40.09)
          (6, 51.63)
          (7, 63.91)
          (8, 74.69)
          (9, 82.86)
          (10, 89.82)
        };
        \addplot[color=peru22113282,mark=diamond,dashed,mark options={solid},line width=1.5] coordinates {
          (1,7.81)
          (2,13.11)
          (3,20.06)
          (4,29.26)
          (5,39.69)
          (6,51.38)
          (7,62.92)
          (8,72.84)
          (9,82.42)
          (10,89.28)
        };
        \addplot[color=mediumseagreen85168104,mark=triangle,dotted,mark options={solid},line width=1.5] coordinates {
          (1,6.41)
          (2,7.6)
          (3,8.97)
          (4,10.79)
          (5,12.83)
          (6,15.86)
          (7,18.92)
          (8,22.16)
          (9,25.78)
          (10,29.23)
        };
        \addplot[color=darkgray,dashdotted,line width=1.5] coordinates {
          (1, 5)
          (2, 5)
          (3, 5)
          (4, 5)
          (5, 5)
          (6, 5)
          (7, 5)
          (8, 5)
          (9, 5)
          (10, 5)
        };
      \end{axis}
    \end{tikzpicture}
  \end{minipage}
  \hfill
  \begin{minipage}[t]{0.25\textwidth}
    \hspace{-0.225in}
    \begin{tikzpicture}
      \begin{axis}[
        xlabel={$\NumberUsedDataSamples$},
        xtick={1, 2, 3, 4, 5, 6, 7, 8, 9, 10},
        xticklabels={, 2, , 4, , 6, , 8, , 10},
        tick label style={font=\footnotesize},
        ymode=log,
        log basis y=10, 
        ymin=0.07, ymax=140,
        ytick={0.1, 1, 10, 100},
        yticklabels=\empty,
        grid=major,
        width=1.8\textwidth,
        height=1.44\textwidth,
        ]
        \addplot[color=steelblue76114176,mark=o,line width=1.5] coordinates {
          (1, 1.64)
          (2, 3.35)
          (3, 5.81)
          (4, 10.02)
          (5, 16.04)
          (6, 24.2)
          (7, 35.34)
          (8, 47.71)
          (9, 60.77)
          (10, 71.52)
        };
        \addplot[color=peru22113282,mark=diamond,dashed,mark options={solid},line width=1.5] coordinates {
          (1,1.55)
          (2,3.43)
          (3,5.85)
          (4,10.06)
          (5,15.48)
          (6,23.94)
          (7,35.01)
          (8,45.99)
          (9,59.04)
          (10,70.11)
        };
        \addplot[color=mediumseagreen85168104,mark=triangle,dotted,mark options={solid},line width=1.5] coordinates {
          (1,1.2)
          (2,1.46)
          (3,2.25)
          (4,2.56)
          (5,3.3)
          (6,4.04)
          (7,4.73)
          (8,6.79)
          (9,7.65)
          (10,10.06)
        };
        \addplot[color=darkgray,dashdotted,line width=1.5] coordinates {
          (1, 1)
          (2, 1)
          (3, 1)
          (4, 1)
          (5, 1)
          (6, 1)
          (7, 1)
          (8, 1)
          (9, 1)
          (10, 1)
        };
      \end{axis}
    \end{tikzpicture}
  \end{minipage}
  \hfill
  \begin{minipage}[t]{0.25\textwidth}
    \hspace{-0.21in}
    \begin{tikzpicture}
      \begin{axis}[
        xlabel={$\NumberUsedDataSamples$},
        xtick={1, 2, 3, 4, 5, 6, 7, 8, 9, 10},
        xticklabels={, 2, , 4, , 6, , 8, , 10},
        tick label style={font=\footnotesize},
        ymode=log,
        log basis y=10, 
        ymin=0.07, ymax=140,
        ytick={0.1, 1, 10, 100},
        yticklabels=\empty,
        grid=major,
        width=1.8\textwidth,
        height=1.44\textwidth,
        ]
        \addplot[color=steelblue76114176,mark=o,line width=1.5] coordinates {
          (1, 0.18)
          (2, 0.55)
          (3, 1.14)
          (4, 2.22)
          (5, 4.14)
          (6, 7.24)
          (7, 13.14)
          (8, 21.49)
          (9, 31.71)
          (10, 43.38)
        };
        \addplot[color=peru22113282,mark=diamond,dashed,mark options={solid},line width=1.5] coordinates {
          (1,0.18)
          (2,0.53)
          (3,0.92)
          (4,1.98)
          (5,3.76)
          (6,7.03)
          (7,12.52)
          (8,19.63)
          (9,30.98)
          (10,42.77)
        };
        \addplot[color=mediumseagreen85168104,mark=triangle,dotted,mark options={solid},line width=1.5] coordinates {
          (1,0.22)
          (2,0.26)
          (3,0.38)
          (4,0.47)
          (5,0.55)
          (6,0.76)
          (7,1.02)
          (8,1.23)
          (9,1.39)
          (10,2.04)
        };
        \addplot[color=darkgray,dashdotted,line width=1.5] coordinates {
          (1, 0.2)
          (2, 0.2)
          (3, 0.2)
          (4, 0.2)
          (5, 0.2)
          (6, 0.2)
          (7, 0.2)
          (8, 0.2)
          (9, 0.2)
          (10, 0.2)
        };
      \end{axis}
    \end{tikzpicture}
  \end{minipage}
  \caption[Short Caption]{CIFAR-100: $@\FalseDetectionRate \le 5\%$ (left), $1\%$ (center), $0.2\%$ (right)}
\end{subfigure}
  \\
  \begin{subfigure}[t]{0.47\textwidth}
    \centering
  \begin{minipage}[t]{0.45\textwidth}
    \centering
    \begin{tikzpicture}
      \begin{axis}[
        xlabel={$\NumberUsedDataSamples$},
        ylabel={$\TrueDetectionRate (\%)$},
        tick label style={font=\footnotesize},
        xtick={1, 2, 3, 4, 5, 6, 7, 8, 9, 10},
        xticklabels={, 2, , 4, , 6, , 8, , 10},
        ymode=log,
        log basis y=10, 
        ymin=0.07, ymax=140,
        ytick={0.1, 1, 10, 100},
        grid=major,
        width=1.0\textwidth,
        height=0.8\textwidth,
        ]
        \addplot[color=steelblue76114176,mark=o,line width=1.5] coordinates {
          (1, 6.86)
          (2, 10.04)
          (3, 13.82)
          (4, 18.22)
          (5, 24.37)
          (6, 30.34)
          (7, 38.8)
          (8, 46.15)
          (9, 55.04)
          (10, 62.48)
        };
        \addplot[color=peru22113282,mark=diamond,dashed,mark options={solid},line width=1.5] coordinates {
          (1,7.31)
          (2,10.03)
          (3,13.37)
          (4,18.24)
          (5,23.26)
          (6,30.94)
          (7,38.3)
          (8,46.6)
          (9,53.99)
          (10,61.61)
        };
        \addplot[color=mediumseagreen85168104,mark=triangle,dotted,mark options={solid},line width=1.5] coordinates {
          (1,5.36)
          (2,6.08)
          (3,6.44)
          (4,6.75)
          (5,8.23)
          (6,8.6)
          (7,9.45)
          (8,10.61)
          (9,11.71)
          (10,12.43)
        };
        \addplot[color=darkgray,dashdotted,line width=1.5] coordinates {
          (1, 5)
          (2, 5)
          (3, 5)
          (4, 5)
          (5, 5)
          (6, 5)
          (7, 5)
          (8, 5)
          (9, 5)
          (10, 5)
        };
      \end{axis}
    \end{tikzpicture}
  \end{minipage}
  \hfill
  \begin{minipage}[t]{0.25\textwidth}
    \hspace{-0.225in}
    \begin{tikzpicture}
      \begin{axis}[
        xlabel={$\NumberUsedDataSamples$},
        xtick={1, 2, 3, 4, 5, 6, 7},
        xtick={1, 2, 3, 4, 5, 6, 7, 8, 9, 10},
        xticklabels={, 2, , 4, , 6, , 8, , 10},
        tick label style={font=\footnotesize},
        ymode=log,
        log basis y=10, 
        ymin=0.07, ymax=140,
        ytick={0.1, 1, 10, 100},
        yticklabels=\empty,
        grid=major,
        width=1.8\textwidth,
        height=1.44\textwidth,
        ]
        \addplot[color=steelblue76114176,mark=o,line width=1.5] coordinates {
          (1, 1.47)
          (2, 2.39)
          (3, 3.74)
          (4, 5.18)
          (5, 8.03)
          (6, 10.93)
          (7, 16.43)
          (8, 20.98)
          (9, 28.51)
          (10,34.95)
        };
        \addplot[color=peru22113282,mark=diamond,dashed,mark options={solid},line width=1.5] coordinates {
          (1,1.6)
          (2,2.04)
          (3,3.37)
          (4,5.07)
          (5,7.28)
          (6,11.08)
          (7,15.28)
          (8,20.69)
          (9,27.25)
          (10,33.69)
        };
        \addplot[color=mediumseagreen85168104,mark=triangle,dotted,mark options={solid},line width=1.5] coordinates {
          (1,0.87)
          (2,1.18)
          (3,1.3)
          (4,1.2)
          (5,1.71)
          (6,1.88)
          (7,2.17)
          (8,2.41)
          (9,2.61)
          (10,3.05)
        };
        \addplot[color=darkgray,dashdotted,line width=1.5] coordinates {
          (1, 1)
          (2, 1)
          (3, 1)
          (4, 1)
          (5, 1)
          (6, 1)
          (7, 1)
          (8, 1)
          (9, 1)
          (10, 1)
        };
      \end{axis}
    \end{tikzpicture}
  \end{minipage}
  \hfill
  \begin{minipage}[t]{0.25\textwidth}
    \hspace{-0.21in}
    \begin{tikzpicture}
      \begin{axis}[
        xlabel={$\NumberUsedDataSamples$},
        xtick={1, 2, 3, 4, 5, 6, 7, 8, 9, 10},
        xticklabels={, 2, , 4, , 6, , 8, , 10},
        tick label style={font=\footnotesize},
        ymode=log,
        log basis y=10, 
        ymin=0.07, ymax=140,
        ytick={0.1, 1, 10, 100},
        yticklabels=\empty,
        grid=major,
        width=1.8\textwidth,
        height=1.44\textwidth,
        ]
        \addplot[color=steelblue76114176,mark=o,line width=1.5] coordinates {
          (1, 0.18)
          (2, 0.34)
          (3, 0.66)
          (4, 0.95)
          (5, 1.53)
          (6, 2.41)
          (7, 4.08)
          (8, 6.17)
          (9, 9.55)
          (10, 14.03)
        };
        \addplot[color=peru22113282,mark=diamond,dashed,mark options={solid},line width=1.5] coordinates {
          (1,0.21)
          (2,0.27)
          (3,0.46)
          (4,1.01)
          (5,1.55)
          (6,2.34)
          (7,4.43)
          (8,6.24)
          (9,9.13)
          (10,12.6)
        };
        \addplot[color=mediumseagreen85168104,mark=triangle,dotted,mark options={solid},line width=1.5] coordinates {
          (1,0.11)
          (2,0.15)
          (3,0.17)
          (4,0.14)
          (5,0.2)
          (6,0.23)
          (7,0.26)
          (8,0.28)
          (9,0.39)
          (10,0.48)
        };
        \addplot[color=darkgray,dashdotted,line width=1.5] coordinates {
          (1, 0.2)
          (2, 0.2)
          (3, 0.2)
          (4, 0.2)
          (5, 0.2)
          (6, 0.2)
          (7, 0.2)
          (8, 0.2)
          (9, 0.2)
          (10, 0.2)
        };
      \end{axis}
    \end{tikzpicture}
  \end{minipage}
  \caption[Short Caption]{TinyImageNet: $@\FalseDetectionRate \le 5\%$ (left), $1\%$ (center), $0.2\%$ (right)}
\end{subfigure}
  \\
  \begin{subfigure}[t]{0.47\textwidth}
    \centering
  \begin{minipage}[t]{0.45\textwidth}
    \centering
    \begin{tikzpicture}
      \begin{axis}[
        xlabel={$\NumberUsedDataSamples$},
        ylabel={$\TrueDetectionRate (\%)$},
        tick label style={font=\footnotesize},
        xtick={1, 2, 3, 4, 5, 6, 7, 8, 9, 10},
        xticklabels={, 2, , 4, , 6, , 8, , 10},
        ymode=log,
        log basis y=10, 
        ymin=0.07, ymax=140,
        ytick={0.1, 1, 10, 100},
        grid=major,
        width=1.0\textwidth,
        height=0.8\textwidth,
        ]
        \addplot[color=steelblue76114176,mark=o,line width=1.5] coordinates {
          (1,6.3)
          (2,7.79)
          (3,9.34)
          (4,11.59)
          (5,13.78)
          (6,17.03)
          (7,19.97)
          (8,22.65)
          (9,27.4)
          (10,31.48)
        };
        \addplot[color=peru22113282,mark=diamond,dashed,mark options={solid},line width=1.5] coordinates {
          (1,6.27)
          (2,7.04)
          (3,8.11)
          (4,10.54)
          (5,11.86)
          (6,13.3)
          (7,15.9)
          (8,18.51)
          (9,21.75)
          (10,25.31)
        };
        \addplot[color=mediumseagreen85168104,mark=triangle,dotted,mark options={solid},line width=1.5] coordinates {
          (1,5.74)
          (2,6.53)
          (3,7.86)
          (4,8.85)
          (5,10.12)
          (6,11.25)
          (7,13.41)
          (8,14.99)
          (9,16.24)
          (10,18.69)
        };
        \addplot[color=darkgray,dashdotted,line width=1.5] coordinates {
          (1, 5)
          (2, 5)
          (3, 5)
          (4, 5)
          (5, 5)
          (6, 5)
          (7, 5)
          (8, 5)
          (9, 5)
          (10, 5)
        };
      \end{axis}
    \end{tikzpicture}
  \end{minipage}
  \hfill
  \begin{minipage}[t]{0.25\textwidth}
    \hspace{-0.225in}
    \begin{tikzpicture}
      \begin{axis}[
        xlabel={$\NumberUsedDataSamples$},
        xtick={1, 2, 3, 4, 5, 6, 7, 8, 9, 10},
        xticklabels={, 2, , 4, , 6, , 8, , 10},
        tick label style={font=\footnotesize},
        ymode=log,
        log basis y=10, 
        ymin=0.07, ymax=140,
        ytick={0.1, 1, 10, 100},
        yticklabels=\empty,
        grid=major,
        width=1.8\textwidth,
        height=1.44\textwidth,
        ]
        \addplot[color=steelblue76114176,mark=o,line width=1.5] coordinates {
          (1,1.43)
          (2,1.63)
          (3,2.24)
          (4,2.6)
          (5,3.65)
          (6,4.76)
          (7,6.1)
          (8,6.83)
          (9,9.53)
          (10,11.72)
        };
        \addplot[color=peru22113282,mark=diamond,dashed,mark options={solid},line width=1.5] coordinates {
          (1,1.34)
          (2,1.53)
          (3,1.71)
          (4,2.42)
          (5,2.85)
          (6,3.27)
          (7,4.19)
          (8,5.36)
          (9,6.59)
          (10,8.5)
        };
        \addplot[color=mediumseagreen85168104,mark=triangle,dotted,mark options={solid},line width=1.5] coordinates {
          (1,1.22)
          (2,1.51)
          (3,1.69)
          (4,1.72)
          (5,2.15)
          (6,2.48)
          (7,3.55)
          (8,3.92)
          (9,4.63)
          (10,5.45)
        };
        \addplot[color=darkgray,dashdotted,line width=1.5] coordinates {
          (1, 1)
          (2, 1)
          (3, 1)
          (4, 1)
          (5, 1)
          (6, 1)
          (7, 1)
          (8, 1)
          (9, 1)
          (10, 1)
        };
      \end{axis}
    \end{tikzpicture}
  \end{minipage}
  \hfill
  \begin{minipage}[t]{0.25\textwidth}
    \hspace{-0.21in}
    \begin{tikzpicture}
      \begin{axis}[
        xlabel={$\NumberUsedDataSamples$},
        xtick={1, 2, 3, 4, 5, 6, 7, 8, 9, 10},
        xticklabels={, 2, , 4, , 6, , 8, , 10},
        tick label style={font=\footnotesize},
        ymode=log,
        log basis y=10, 
        ymin=0.07, ymax=140,
        ytick={0.1, 1, 10, 100},
        yticklabels=\empty,
        grid=major,
        width=1.8\textwidth,
        height=1.44\textwidth,
        ]
        \addplot[color=steelblue76114176,mark=o,line width=1.5] coordinates {
          (1,0.16)
          (2,0.23)
          (3,0.35)
          (4,0.38)
          (5,0.59)
          (6,0.89)
          (7,1.15)
          (8,1.47)
          (9,2.04)
          (10,2.86)
        };
        \addplot[color=peru22113282,mark=diamond,dashed,mark options={solid},line width=1.5] coordinates {
          (1,0.12)
          (2,0.17)
          (3,0.24)
          (4,0.4)
          (5,0.4)
          (6,0.55)
          (7,0.68)
          (8,1.09)
          (9,1.06)
          (10,1.57)
        };
        \addplot[color=mediumseagreen85168104,mark=triangle,dotted,mark options={solid},line width=1.5] coordinates {
          (1,0.17)
          (2,0.17)
          (3,0.24)
          (4,0.25)
          (5,0.33)
          (6,0.33)
          (7,0.48)
          (8,0.71)
          (9,0.86)
          (10,0.89)
        };
        \addplot[color=darkgray,dashdotted,line width=1.5] coordinates {
          (1, 0.2)
          (2, 0.2)
          (3, 0.2)
          (4, 0.2)
          (5, 0.2)
          (6, 0.2)
          (7, 0.2)
          (8, 0.2)
          (9, 0.2)
          (10, 0.2)
        };
      \end{axis}
    \end{tikzpicture}
  \end{minipage}
  \caption[Short Caption]{ImageNet: $@\FalseDetectionRate \le 5\%$ (left), $1\%$ (center), $0.2\%$ (right)}
\end{subfigure}
  \caption{$\TrueDetectionRate (\%)$ of auditing image classifiers
    where a data owner had $\DatasetSize=10$ data
    instances to audit and $\NumberUsedDataSamples$ of them were used
    in training.  ``All'' means that the output of a classifier is a
    full vector of confidence scores (our default); ``Highest'' means
    that the output is a label and its associated confidence score;
    ``Label'' means that the output is a label only.}
  \label{fig:classifier:main:partial}
\end{figure}

We plot the cumulative distribution function (CDF) of
$\QueryNumberVariable$ in the case $\DatasetSize=1$ and
$\NumberMarkedData = 1000$, under the condition that data use is
detected, in \figref{fig:main:cost}.  In \figref{fig:main:cost}, the
area under the curve (AUC) represents the average query cost saved
when the audited data instance was detected.  For CIFAR-100, the AUCs
were $510.24$ and $107.78$ when $\FalseDetectionRate\leq5\%$ and
$\FalseDetectionRate\leq1\%$ respectively.  For TinyImageNet, the AUCs
were $472.42$ and $96.81$ when $\FalseDetectionRate\leq5\%$ and
$\FalseDetectionRate\leq1\%$ respectively.  When we set
$\FDRBound=0.2\%$, it needed to query all the marked data instances
(i.e., $\QueryNumberVariable=1000$) in order detection data use and so
we did not have cost savings in this setting.

\begin{figure}[ht!]
  \centering

  \begin{subfigure}[b]{0.47\textwidth}
    \centering
    \resizebox{!}{2em}{\newenvironment{customlegend}[1][]{%
    \begingroup
    \csname pgfplots@init@cleared@structures\endcsname
    \pgfplotsset{#1}%
}{%
    \csname pgfplots@createlegend\endcsname
    \endgroup
}%

\def\addlegendimage{\csname pgfplots@addlegendimage\endcsname}

\begin{tikzpicture}

\definecolor{darkslategray38}{RGB}{38,38,38}
\definecolor{indianred1967882}{RGB}{196,78,82}
\definecolor{lightgray204}{RGB}{204,204,204}
\definecolor{mediumseagreen85168104}{RGB}{85,168,104}
\definecolor{peru22113282}{RGB}{221,132,82}
\definecolor{steelblue76114176}{RGB}{76,114,176}

\begin{customlegend}[
    legend style={{font={\small}},{draw=none}}, 
    legend columns=4,
    legend cell align={center},
    legend entries={{$\FalseDetectionRate\leq5\%$}, {$\FalseDetectionRate\leq1\%$}}]
\addlegendimage{very thick, steelblue76114176, solid}
\addlegendimage{very thick, peru22113282, dashed}

\end{customlegend}

\end{tikzpicture}}
  \end{subfigure}

  \begin{subfigure}[b]{0.235\textwidth}
    \centering

    \caption[Short Caption]{
      \centering CIFAR-100}
  \end{subfigure}
  \hfill
  \begin{subfigure}[b]{0.235\textwidth}
    \centering
    %
    \caption[Short Caption]{
      \centering TinyImageNet}
  \end{subfigure}

  \caption{CDF of $\QueryNumberVariable$ in the case $\DatasetSize=1$
    and $\NumberMarkedData = 1000$, conditioned on data-use being
    detected.}
  \label{fig:main:cost}
\end{figure}

\paragraph{Empirical false-detection}
We empirically evaluate \FalseDetectionRate of our method by training image classifiers on datasets excluding the
audited data samples. The results on our empirical \FalseDetectionRate are shown in \tblref{tbl:classifier:false_detection}.
As shown in \tblref{tbl:classifier:false_detection}, the empirical \FalseDetectionRate
were less than the bound $\FDRBound$ on false-detection. These results empirically
confirm the upper bounds on \FalseDetectionRate of our method.

\begin{table}[ht!]
  \centering
  {\resizebox{0.47\textwidth}{!}{
  \begin{tabular}{@{}lc@{\hspace{0.5em}}cc@{\hspace{0.5em}}cc@{\hspace{0.5em}}cc@{\hspace{0.5em}}c@{}}
  \toprule
  \multicolumn{1}{c}{\multirow{2}{*}{}} & \multicolumn{3}{c}{$\FalseDetectionRate\le$} \\ 
  & \multicolumn{1}{c}{$5\%$ }& \multicolumn{1}{c}{$1\%$ }& \multicolumn{1}{c}{$0.2\%$ } \\
  \midrule 
  CIFAR-100 & $4.57(\pm0.86)$ &$0.86(\pm0.34)$ &$0.10(\pm0.10)$\\ 
  TinyImageNet & $4.83(\pm0.64)$ &$0.89(\pm0.26)$ &$0.07(\pm0.07)$\\
  ImageNet & $4.59$ &$0.5$ &$0.0$ \\
  \bottomrule 
  \end{tabular}
  }}
  \caption{Empirical measures $(\%)$ of $\FalseDetectionRate$ 
    ($\DatasetSize=1$) of our data-use
    auditing method when applied to audit image classifiers
    that were not trained on the audited data instances.
    Results are averaged over $500\times20$ detections for CIFAR-100
    ($1{,}000\times20$ for TinyImageNet or $1{,}000$ for ImageNet). 
    We trained $20$ classifiers for CIFAR-100 or TinyImageNet, 
    and one classifier for ImageNet,
    in each of which $500$ CIFAR-100 ($1{,}000$ TinyImageNet 
    or $1{,}000$ ImageNet) training
    samples were audited. The numbers in the
    parenthesis are standard deviations among the 20 classifiers.}
  \label{tbl:classifier:false_detection}
\end{table}

\paragraph{Comparison with baselines}
We compare our method with the state-of-the-art membership inference
methods, namely Attack-P~\cite{ye2022:enhanced},
Attack-R~\cite{ye2022:enhanced}, LiRA~\cite{carlini2022:membership},
and RMIA~\cite{zarifzadeh2024:low} when $\DatasetSize=1$. 
Attack-R, LiRA, and RMIA assume the ability to train at
least one ML model, known as a \textit{reference model}, from these
samples.  In those works that proposed Attack-R, LiRA, and RMIA, such
reference models are assumed similar to the audited model (i.e., they
are trained on a dataset similar to the training dataset of the
audited model).

In our implementation of Attack-R, LiRA, and RMIA, we considered a
more realistic setting where only one reference model was used in
membership inference.  We also considered settings where the reference
model is not similar enough to the audited model, by constructing a
dataset used to train the reference model differently from the
training dataset of the audited model.  We constructed such a
``different'' dataset by decreasing its size $\ReferenceDatasetSize$
and/or introducing class imbalance (i.e., the number of data samples
per class was unequal).  Specifically, the class proportions were
drawn from a Beta distribution~\cite{johnson1995:continuous} where we
set its two parameters as the same value $\ParameterBetaDistribution$
that controls the degree of imbalance.  A larger value
$\ParameterBetaDistribution$ leads to greater skewness in the class
proportions.

The comparison results are presented in
\figref{fig:baselines:cifar100} and
\figref{fig:baselines:tinyimagenet}.
(We present both \figref{fig:baselines:cifar100} and \figref{fig:baselines:tinyimagenet} in
\appref{app:results:comparison_tinyimagenet} due to the space limit.)
\figref{fig:baselines:cifar100} shows the auditing/inference results
on CIFAR-100: When Attack-R, LiRA, and RMIA used a reference model
similar to the audited model, our method achieved a
$\TrueDetectionRate$ comparable to those of Attack-R, LiRA, and RMIA
under the same $\FalseDetectionRate$ level. However, when the
reference model was not similar enough to the audited one (e.g., by
decreasing $\ReferenceDatasetSize/\TrainDatasetSize$, where
$\TrainDatasetSize$ is the size of the training set of the audited
model, and/or increasing $\ParameterBetaDistribution$), the
performance of these membership inference methods significantly
degraded, which was also confirmed by their
works~\cite{ye2022:enhanced,carlini2022:membership,zarifzadeh2024:low}.
For example, the state-of-the-art membership inference method, namely
RMIA, had \TrueDetectionRate of $15.27\%$, $2.25\%$, and $0\%$ under
$\FalseDetectionRate\leq5\%$, $\FalseDetectionRate\leq1\%$, and
$\FalseDetectionRate\leq0.2\%$, respectively, when we set
$\ReferenceDatasetSize/\TrainDatasetSize=\frac{1}{8}$ and
$\ParameterBetaDistribution=4$, much lower than ours.

The performance of these three membership inference methods were
highly affected by the reference models. Of course, as shown in the
previous works
(e.g.,~\cite{ye2022:enhanced,carlini2022:membership,zarifzadeh2024:low}),
when more reference models can be trained and used in membership
inference, these methods would achieve a better inference result
(i.e., a higher \TrueDetectionRate).  However, in a realistic scenario
of data-use auditing, it is costly to train a reference model and
challenging to collect a dataset used to train the reference model
that is similar to the training dataset of the audited model.
Attack-P does not require a reference model but its \TrueDetectionRate
was much lower than ours under the same \FalseDetectionRate.  We have
similar observation and conclusion from the results on TinyImageNet,
as shown in \figref{fig:baselines:tinyimagenet} in
\appref{app:results:comparison_tinyimagenet}.  More importantly, all these
membership inference methods do not provide a bound on the
\FalseDetectionRate.  This limits the application of membership
inference methods in auditing data-use of ML models, as discussed in
\secref{sec:intro}.

\paragraph{Robustness against countermeasures}

We study the effectiveness of our data-use auditing method when the ML
practitioner applies countermeasures in the training pipeline to
defeat our method.  We consider these countermeasures:
\begin{itemize}
\item \emph{Image perturbation}: The ML practitioner perturbs the
  training samples before their use in training.  We consider two
  types of perturbation: (1) adding Gaussian noise (parameterized by
  its standard deviation $\GaussianNoiseStd$) and (2) applying our
  marking method (using the default hyperparameters) to add additional
  perturbation, denoted as ``remarking''.
\item \emph{Image denoising}: The ML practitioner applies image
  denoising techniques~\cite{haddad1991:class,huang1979:fast} on the
  training samples in order to remove their added marks.  We consider
  three commonly used denoising methods: (1) Gaussian smoothing, (2)
  median smoothing, and (3) general smoothing.
\item \emph{Privacy-preserving training}: The ML practitioner applies
  differentially private stochastic gradient descent (DPSGD) to train
  his model. DPSGD is the state-of-the-art private learning algorithm
  reducing the memorization and privacy leakage of training
  samples~\cite{aerni2024:evaluations}, and thus it can be considered
  as an attack to a data-use auditing
  method~\cite{huang2024:auditdata}.  It works by clipping the norm of
  the gradients and adding Gaussian noise parameterized by a standard
  deviation $\DPnoiseStd$ (i.e., noise multiplier) into gradients
  during training.
\item \emph{Regularization}: The ML practitioner applies
  regularization techniques, e.g., by adding a penalty proportional to
  the squared values of the model parameters, to reduce memorization
  of training samples. He controls the regularization strength by
  tuning the weight decay (denoted as $\WeightDecay$) parameter in the
  SGD optimizer.
\end{itemize}

Due to the space limit, we report our robustness results in
\appref{app:results:robustness}. As shown in \figref{fig:attack:cifar100}
and \tblref{tbl:classifier:remark_smooth}
in \appref{app:results:robustness}, both perturbation methods decreased our
$\TrueDetectionRate$: ``Remarking'' decreased $\TrueDetectionRate$
($\DatasetSize=1$) from $28.21\%$ to $12.60\%$ under
$\FalseDetectionRate\leq5\%$ but the accuracy \Accuracy on average
dropped by $3.90$ percentage points.  Adding Gaussian noise with a
larger $\GaussianNoiseStd$ made our $\TrueDetectionRate$
($\DatasetSize=1$) closer to the $\FalseDetectionRate$ level but it
also sacrificed more model accuracy. For example, adding Gaussian
noise with $\GaussianNoiseStd=25$ decreased $\TrueDetectionRate$
($\DatasetSize=1$) to a level close to $\FalseDetectionRate$ but at a
cost of $10.11$ percentage points to \Accuracy. As shown in
\tblref{tbl:classifier:remark_smooth} in \appref{app:results:robustness}, image
smoothing decreased our $\TrueDetectionRate$ but significantly
decreased the utility of the trained models.  For example, general
smoothing reduced $\TrueDetectionRate$ ($\DatasetSize=1$) to $14.49\%$
under $\FalseDetectionRate\leq5\%$ but \Accuracy dropped by around
$11$ percentage points. As shown in \figref{fig:classifier:dpsgd} in
\appref{app:results:robustness}, when applying DPSGD, when we set a larger
$\DPnoiseStd$, the trained classifier memorized its training samples
less and thus our $\TrueDetectionRate$ ($\DatasetSize=1$) decreased
under the same level of \FalseDetectionRate.  For example, under
$\FalseDetectionRate\leq5\%$, our $\TrueDetectionRate$
($\DatasetSize=1$) decreased from $28.21\%$ to $9.21\%$ when we
increased $\DPnoiseStd$ from $0$ to $2\times10^{-3}$ ($\DPnoiseStd=0$
corresponds to the non-private setting). However, $\Accuracy$ of the
trained classifiers decreased from $75.53\%$ to $65.82\%$ as
$\DPnoiseStd$ grew. We have similar observations from the auditing
results (see \figref{fig:classifier:weightdecay} in
\appref{app:results:robustness}) when applying stronger regularization
techniques in model training.  To summarize, none of these
countermeasures defeated our auditing method without significantly
sacrificing the utility of the trained ML model.

When the data owner had more data instances (i.e., $\DatasetSize > 1$)
and all of them were used in training, our method became more
effective, i.e., our $\TrueDetectionRate$ increased with
$\DatasetSize$, even when the ML practitioner applied a strong
countermeasure.  For example, when DPSGD with
$\DPnoiseStd=2\times10^{-3}$ was used to train the ML model, our
$\TrueDetectionRate$ achieved $24.34\%$ ($\DatasetSize=8$) and
$85.34\%$ ($\DatasetSize=64$) under $\FalseDetectionRate\leq5\%$
(compared with $9.21\%$ for $\DatasetSize=1$).

\paragraph{Additional results}

In \appref{app:results:additional}, we report results
evaluating different classifier architectures, values for the
utility-preservation parameter $\MarkBound$, numbers
$\NumberMarkedData$ of marked data per raw data instance, and number
$\NumAugment$ of data augmentations in data-use detection.  We also
study how the vulnerability of an image to a membership inference attack 
impacts its auditability, and conduct ablation studies on our
data-marking algorithm and data-use detection algorithm.

\subsection{Auditing Visual Encoder, CLIP, and BLIP}
\label{sec:auditing_ML:encoder}

We additionally experimented with our method to detect data use in
visual encoders and in the CLIP and BLIP models, which we now
introduce briefly.  A visual encoder is a deep neural network used to
extract high-level meaningful features of
images~\cite{chen2020:simple}. It takes as input an image and outputs
a vector of features. Visual encoders are widely used as backbones in
various machine learning tasks in computer vision, e.g., image
classification~\cite{he2016:deep, deng2009:imagenet} and object
detection~\cite{zhao2019:object}.

CLIP is a multimodal model developed by OpenAI in 2021, which can
process both image and text data. It consists of a visual encoder
designed by a convolutional network (e.g., ResNet) or a transformer
architecture~\cite{vaswani2017:attention} and a transformer-based text
encoder that are used to transform image and text data into high-level
structured representations, respectively. CLIP is renowned for its
few-shot capability to classify images based on textual
prompts~\cite{radford2021:learning} and is widely used as a backbone
in vision-language machine learning tasks (e.g., text-to-image
generation~\cite{rombach2022:CVPR}).

BLIP is a multimodal model designed by Li et al.~\cite{li2022:blip} to
unify language and image understanding, enabling advanced tasks like
image captioning and cross-modal retrieval.  It consists of a unimodal
encoder, an image-grounded text encoder, and an image-grounded text
decoder.

\subsubsection{Experimental Setup} \label{sec:auditing_ML:encoder:setup}

\paragraph{Datasets} We used CIFAR-100~\cite{krizhevsky2009:learning} 
and TinyImageNet~\cite{Lle2015:TinyIV} in the experiments to test
auditing of visual encoders. We used the Flickr30k
dataset~\cite{young2014:image} in the experiments on CLIP and BLIP.
Please see their descriptions in \appref{app:setup:dataset}.

\paragraph{Settings}
Due to the space limit, we introduce data-marking setting, model
training/fine-tuning setting, and data-use detection setting in
\appref{app:setup}.

\paragraph{Metric}
We use \TrueDetectionRate, defined in \figref{fig:data_auditing}
\secref{sec:background}, to measure the effectiveness of a
data-use auditing method.

\subsubsection{Experimental Results} \label{sec:auditing_ML:encoder:results}

\paragraph{Auditing visual encoders}
Our results on auditing visual encoders are presented in~\figref{fig:visual_encoder:main}.
When $\DatasetSize=1$, our $\TrueDetectionRate$ ranged from $10.51\%$ to $14.02\%$, from $2.40\%$ to $3.78\%$,
and from $0.34\%$ to $0.56\%$ when the bounds on \FalseDetectionRate were set as $5\%$, $1\%$, and $0.2\%$, respectively.
Although our \TrueDetectionRate on auditing visual encoders were significantly better than those
from a ``random guessing'' method, they were lower than those on auditing image classifiers (see \figref{fig:classifier:main}).
Because visual encoders are trained to learn the general representations of images and thus they memorizes
their training samples less. When the data owner had more data instances (i.e., $\DatasetSize>1$) and
all of them were used in training, our method became significantly more effective, 
i.e., our $\TrueDetectionRate$ increased with $\DatasetSize$.

\begin{figure}[ht!]
  \centering

  \begin{subfigure}[b]{0.47\textwidth}
    \centering
    \resizebox{!}{1.7em}{\newenvironment{customlegend}[1][]{%
    \begingroup
    \csname pgfplots@init@cleared@structures\endcsname
    \pgfplotsset{#1}%
}{%
    \csname pgfplots@createlegend\endcsname
    \endgroup
}%

\def\addlegendimage{\csname pgfplots@addlegendimage\endcsname}

\begin{tikzpicture}

\definecolor{darkslategray38}{RGB}{38,38,38}
\definecolor{indianred1967882}{RGB}{196,78,82}
\definecolor{lightgray204}{RGB}{204,204,204}
\definecolor{mediumseagreen85168104}{RGB}{85,168,104}
\definecolor{peru22113282}{RGB}{221,132,82}
\definecolor{steelblue76114176}{RGB}{76,114,176}

\begin{customlegend}[
    legend style={{font={\small}},{draw=none}}, 
    legend columns=3,
    legend cell align={center},
    legend entries={{CIFAR-100}, {TinyImageNet}, {$\FalseDetectionRate$ bound}}]
\addlegendimage{very thick, color=steelblue76114176,mark=o}
\addlegendimage{very thick, peru22113282,mark=diamond,dashed,mark options={solid}}
\addlegendimage{very thick, darkgray,dashdotted}

\end{customlegend}

\end{tikzpicture}}
  \end{subfigure}

  \begin{subfigure}[t]{0.47\textwidth}
    \centering
  \begin{minipage}[t]{0.45\textwidth}
    \centering
    \begin{tikzpicture}
      \begin{axis}[
        xlabel={$\DatasetSize$},
        ylabel={$\TrueDetectionRate (\%)$},
        tick label style={font=\footnotesize},
        xtick={1, 2, 3, 4, 5, 6, 7},
        xticklabels={$1$, , $4$, , $16$, , $64$},
        ymode=log,
        log basis y=10, 
        ymin=0.07, ymax=140,
        ytick={0.1, 1, 10, 100},
        grid=major,
        width=1.0\textwidth,
        height=0.8\textwidth,
        ]
        \addplot[color=steelblue76114176,mark=o,line width=1.5] coordinates {
          (1, 14.02)
          (2, 20.05)
          (3, 31.53)
          (4, 51.56)
          (5, 77.76)
          (6, 96.21)
          (7, 99.95)
        };
        \addplot[color=peru22113282,mark=diamond,dashed,mark options={solid},line width=1.5] coordinates {
          (1, 10.51)
          (2, 14.05)
          (3, 20.79)
          (4, 31.43)
          (5, 50.04)
          (6, 76.31)
          (7, 95.56)
        };
        \addplot[color=darkgray,dashdotted,line width=1.5] coordinates {
          (1, 5)
          (2, 5)
          (3, 5)
          (4, 5)
          (5, 5)
          (6, 5)
          (7, 5)
        };
      \end{axis}
    \end{tikzpicture}
  \end{minipage}
  \hfill
  \begin{minipage}[t]{0.25\textwidth}
    \hspace{-0.225in}
    \begin{tikzpicture}
      \begin{axis}[
        xlabel={$\DatasetSize$},
        xtick={1, 2, 3, 4, 5, 6, 7},
        xticklabels={$1$, , $4$, , $16$, , $64$},
        tick label style={font=\footnotesize},
        ymode=log,
        log basis y=10, 
        ymin=0.07, ymax=140,
        ytick={0.1, 1, 10, 100},
        yticklabels=\empty,
        grid=major,
        width=1.8\textwidth,
        height=1.44\textwidth,
        ]
        \addplot[color=steelblue76114176,mark=o,line width=1.5] coordinates {
          (1, 3.78)
          (2, 5.81)
          (3, 11.51)
          (4, 24.17)
          (5, 51.12)
          (6, 84.3)
          (7, 99.4)
        };
        \addplot[color=peru22113282,mark=diamond,dashed,mark options={solid},line width=1.5] coordinates {
          (1, 2.40)
          (2, 3.55)
          (3, 5.86)
          (4, 11.12)
          (5, 23.11)
          (6, 49.32)
          (7, 83.7)
        };
        \addplot[color=darkgray,dashdotted,line width=1.5] coordinates {
          (1, 1)
          (2, 1)
          (3, 1)
          (4, 1)
          (5, 1)
          (6, 1)
          (7, 1)
        };
      \end{axis}
    \end{tikzpicture}
  \end{minipage}
  \hfill
  \begin{minipage}[t]{0.25\textwidth}
    \hspace{-0.21in}
    \begin{tikzpicture}
      \begin{axis}[
        xlabel={$\DatasetSize$},
        xtick={1, 2, 3, 4, 5, 6, 7},
        xticklabels={$1$, , $4$, , $16$, , $64$},
        tick label style={font=\footnotesize},
        ymode=log,
        log basis y=10, 
        ymin=0.07, ymax=140,
        ytick={0.1, 1, 10, 100},
        yticklabels=\empty,
        grid=major,
        width=1.8\textwidth,
        height=1.44\textwidth,
        ]
        \addplot[color=steelblue76114176,mark=o,line width=1.5] coordinates {
          (1, 0.56)
          (2, 1.01)
          (3, 2.47)
          (4, 7.1)
          (5, 23.23)
          (6, 60.96)
          (7, 95.56)
        };
        \addplot[color=peru22113282,mark=diamond,dashed,mark options={solid},line width=1.5] coordinates {
          (1, 0.34)
          (2, 0.57)
          (3, 0.98)
          (4, 2.51)
          (5, 6.98)
          (6, 22.28)
          (7, 59.54)
        };
        \addplot[color=darkgray,dashdotted,line width=1.5] coordinates {
          (1, 0.2)
          (2, 0.2)
          (3, 0.2)
          (4, 0.2)
          (5, 0.2)
          (6, 0.2)
          (7, 0.2)
        };
      \end{axis}
    \end{tikzpicture}
  \end{minipage}
\end{subfigure}
  \caption{$\TrueDetectionRate (\%)$ $@\FalseDetectionRate \le 5\%$
    (left), $1\%$ (center), $0.2\%$ (right) of auditing visual
    encoders, when the data owner had $\DatasetSize$ data instances to
    audit and all of them were used in training.}
  \label{fig:visual_encoder:main}
\end{figure}

\paragraph{Auditing CLIP}

Our results on auditing the data-use in CLIP, 
fine-tuned by only one epoch, are shown
in \figref{fig:clip_and_blip}. 
With $\DatasetSize=1$, we achieved \TrueDetectionRate of $9.58\%$,
$2.82\%$, and $0.48\%$ under $\FalseDetectionRate\leq5\%$,
$\FalseDetectionRate\leq1\%$, and $\FalseDetectionRate\leq0.2\%$,
respectively, which were significantly better than random guessing.
When the data owner had more data instances (i.e., $\DatasetSize>1$)
and all of them were used in training, our method became much more
effective, i.e., our $\TrueDetectionRate$ increased with
$\DatasetSize$. For example, we achieved \TrueDetectionRate of
$67.35\%$, $39.78\%$, and $16.45\%$ under
$\FalseDetectionRate\leq5\%$, $\FalseDetectionRate\leq1\%$, and
$\FalseDetectionRate\leq0.2\%$, respectively, when the data owner had
$\DatasetSize=64$ data instances.

We also report our auditing results when the CLIP model was fine-tuned
on the marked datasets for more than one epoch, in
\appref{app:results:additional_clip_blip}.

\begin{figure}[ht!]
  \centering

  \begin{subfigure}[b]{0.47\textwidth}
    \centering
    \resizebox{!}{1.7em}{\newenvironment{customlegend}[1][]{%
    \begingroup
    \csname pgfplots@init@cleared@structures\endcsname
    \pgfplotsset{#1}%
}{%
    \csname pgfplots@createlegend\endcsname
    \endgroup
}%

\def\addlegendimage{\csname pgfplots@addlegendimage\endcsname}

\begin{tikzpicture}

\definecolor{darkslategray38}{RGB}{38,38,38}
\definecolor{indianred1967882}{RGB}{196,78,82}
\definecolor{lightgray204}{RGB}{204,204,204}
\definecolor{mediumseagreen85168104}{RGB}{85,168,104}
\definecolor{peru22113282}{RGB}{221,132,82}
\definecolor{steelblue76114176}{RGB}{76,114,176}

\begin{customlegend}[
    legend style={{font={\small}},{draw=none}}, 
    legend columns=3,
    legend cell align={center},
    legend entries={{CLIP}, {BLIP}, {$\FalseDetectionRate$ bound}}]
\addlegendimage{very thick, color=steelblue76114176,mark=o}
\addlegendimage{very thick, peru22113282,mark=diamond,dashed,mark options={solid}}
\addlegendimage{very thick, darkgray,dashdotted}

\end{customlegend}

\end{tikzpicture}}
  \end{subfigure}

  \begin{subfigure}[t]{0.47\textwidth}
    \centering
  \begin{minipage}[t]{0.45\textwidth}
    \centering
    \begin{tikzpicture}
      \begin{axis}[
        xlabel={$\DatasetSize$},
        ylabel={$\TrueDetectionRate (\%)$},
        tick label style={font=\footnotesize},
        xtick={1, 2, 3, 4, 5, 6, 7},
        xticklabels={$1$, , $4$, , $16$, , $64$},
        ymode=log,
        log basis y=10, 
        ymin=0.07, ymax=140,
        ytick={0.1, 1, 10, 100},
        grid=major,
        width=1.0\textwidth,
        height=0.8\textwidth,
        ]
        \addplot[color=steelblue76114176,mark=o,line width=1.5] coordinates {
          (1,9.58)
          (2,10.43)
          (3,13.54)
          (4,18.71)
          (5,27.33)
          (6,43.2)
          (7,67.35)
        };
        \addplot[color=peru22113282,mark=diamond,dashed,mark options={solid},line width=1.5] coordinates {
          (1,13.07)
          (2,14.59)
          (3,20.47)
          (4,32.06)
          (5,49.14)
          (6,74.37)
          (7,94.72)
        };
        \addplot[color=darkgray,dashdotted,line width=1.5] coordinates {
          (1, 5)
          (2, 5)
          (3, 5)
          (4, 5)
          (5, 5)
          (6, 5)
          (7, 5)
        };
      \end{axis}
    \end{tikzpicture}
  \end{minipage}
  \hfill
  \begin{minipage}[t]{0.25\textwidth}
    \hspace{-0.225in}
    \begin{tikzpicture}
      \begin{axis}[
        xlabel={$\DatasetSize$},
        xtick={1, 2, 3, 4, 5, 6, 7},
        xticklabels={$1$, , $4$, , $16$, , $64$},
        tick label style={font=\footnotesize},
        ymode=log,
        log basis y=10, 
        ymin=0.07, ymax=140,
        ytick={0.1, 1, 10, 100},
        yticklabels=\empty,
        grid=major,
        width=1.8\textwidth,
        height=1.44\textwidth,
        ]
        \addplot[color=steelblue76114176,mark=o,line width=1.5] coordinates {
          (1,2.82)
          (2,2.83)
          (3,3.58)
          (4,5.59)
          (5,9.62)
          (6,18.78)
          (7,39.78)
        };
        \addplot[color=peru22113282,mark=diamond,dashed,mark options={solid},line width=1.5] coordinates {
          (1,3.44)
          (2,4.54)
          (3,6.14)
          (4,11.92)
          (5,23.96)
          (6,47.67)
          (7,82.0)
        };
        \addplot[color=darkgray,dashdotted,line width=1.5] coordinates {
          (1, 1)
          (2, 1)
          (3, 1)
          (4, 1)
          (5, 1)
          (6, 1)
          (7, 1)
        };
      \end{axis}
    \end{tikzpicture}
  \end{minipage}
  \hfill
  \begin{minipage}[t]{0.25\textwidth}
    \hspace{-0.21in}
    \begin{tikzpicture}
      \begin{axis}[
        xlabel={$\DatasetSize$},
        xtick={1, 2, 3, 4, 5, 6, 7},
        xticklabels={$1$, , $4$, , $16$, , $64$},
        tick label style={font=\footnotesize},
        ymode=log,
        log basis y=10, 
        ymin=0.07, ymax=140,
        ytick={0.1, 1, 10, 100},
        yticklabels=\empty,
        grid=major,
        width=1.8\textwidth,
        height=1.44\textwidth,
        ]
        \addplot[color=steelblue76114176,mark=o,line width=1.5] coordinates {
          (1,0.48)
          (2,0.58)
          (3,0.64)
          (4,1.22)
          (5,2.04)
          (6,5.51)
          (7,16.45)
        };
        \addplot[color=peru22113282,mark=diamond,dashed,mark options={solid},line width=1.5] coordinates {
          (1,0.55)
          (2,0.74)
          (3,1.34)
          (4,2.76)
          (5,7.25)
          (6,21.49)
          (7,57.15)
        };
        \addplot[color=darkgray,dashdotted,line width=1.5] coordinates {
          (1, 0.2)
          (2, 0.2)
          (3, 0.2)
          (4, 0.2)
          (5, 0.2)
          (6, 0.2)
          (7, 0.2)
        };
      \end{axis}
    \end{tikzpicture}
  \end{minipage}
\end{subfigure}
  \caption{$\TrueDetectionRate (\%)$ $@\FalseDetectionRate \le 5\%$ (left), $1\%$ (center), $0.2\%$ (right) 
    of auditing the fine-tuned CLIP and BLIP models, when a data owner 
    had $\DatasetSize$ data instances to
    audit and all of them were used in training.}
  \label{fig:clip_and_blip}
\end{figure}

\paragraph{Auditing BLIP}

Our results on auditing the data-use in BLIP, fine-tuned by only
  one epoch, are also included in \figref{fig:clip_and_blip}.
With $\DatasetSize=1$, we achieved \TrueDetectionRate of $13.07\%$,
$3.43\%$, and $0.55\%$ under $\FalseDetectionRate\leq5\%$,
$\FalseDetectionRate\leq1\%$, and $\FalseDetectionRate\leq0.2\%$,
respectively. When the data owner had more data instances (i.e.,
$\DatasetSize>1$) and all of them were used in training, our method
became significantly more effective, i.e., our $\TrueDetectionRate$
increased with $\DatasetSize$. For example, we achieved
\TrueDetectionRate of $94.72\%$, $82.00\%$, and $57.15\%$ under
$\FalseDetectionRate\leq5\%$, $\FalseDetectionRate\leq1\%$, and
$\FalseDetectionRate\leq0.2\%$, respectively, when the data owner had
$\DatasetSize=64$ data instances.

We also report our auditing results when the BLIP models were 
fine-tuned on the marked datasets for more than one epochs,
in \appref{app:results:additional_clip_blip}.

\section{Verification of Machine Unlearning} \label{sec:unlearning}

In this section, we apply our data-use auditing method to verify if
individual data instances have been removed from an ML model by an
approximate unlearning method~\cite{guo2020:certified,
  warnecke2021:machine}.  This study reveals the power of a
false-detection rate bound, in that detecting ``unlearned'' data at a
higher rate than the false-detection-rate bound proves that the
approximate unlearning method does not work.  Indeed, that is exactly
what we find for two unlearning methods in this section.

We focus on the case $\DatasetSize=1$. In our tests, the data owner
applied our data-marking algorithm to generate $\MarkedDataset =
\{\MarkedData\}$ and $\HiddenInfo$ from her raw data $\Dataset =
\{\Data\}$, and releases $\MarkedData$.  The ML practitioner added
$\MarkedData$ to his training dataset and trained an ML model on it.
The ML practitioner then applied a machine unlearning algorithm to
remove \MarkedData from the trained ML model.  The data owner than
audited the resulting model \MLModel for use of \MarkedData.  If the
ML practitioner applied an \textit{exact} unlearning method, i.e.,
retraining a new model \MLModel without using \MarkedData, then this
corresponds to the experiment defined in \figref{fig:data_auditing}
where $\MLPBit=0$.  When the ML practitioner used an
\textit{approximate} unlearning method, however, then this corresponds
to $\MLPBit=1$, since the audited data was still used in the
end-to-end model-training pipeline.  A result showing that the data
owner detected use of \MarkedData at a rate convincingly higher when
$\MLPBit=1$ than when $\MLPBit=0$ (i.e., $\TrueDetectionRate >
\FalseDetectionRate$) would illustrate that the approximate
unlearning method did not work.

\subsection{Experimental Setup}  \label{sec:unlearning:setup}

\paragraph{Visual ML models}
We used image classifiers and CLIP~\cite{radford2021:learning}
in our experiments to verify machine unlearning.

\paragraph{Datasets}
We used benchmark datasets CIFAR-100~\cite{krizhevsky2009:learning}
and TinyImageNet~\cite{Lle2015:TinyIV} for image classifiers, and used
Flickr30k~\cite{young2014:image} for CLIP.  Please see the
descriptions of these datasets in
\appref{app:setup:dataset}.

\paragraph{Data-marking setting}
We followed the data-marking setting described in
\secref{sec:auditing_ML:classifier:setup} and
\secref{sec:auditing_ML:encoder:setup} to prepare marked training
datasets. We set $\NumberMarkedData=1000$ and $\MarkBound=10$,
and used ResNet-18 pretrained on ImageNet as $\FeatureExtractor$.

\paragraph{Model training setting}
We followed the model training setting described in
\secref{sec:auditing_ML:classifier:setup} to train image classifiers
from scratch, and the method described in
\secref{sec:auditing_ML:encoder:setup} to fine-tune CLIP by $5$ epochs,
on the marked training datasets.  We used ResNet-18 as the model
architecture of image classifiers.

\paragraph{Machine unlearning setting}
We considered two state-of-the-art approximate unlearning methods:
Warnecke et al.'s gradient-based method~\cite{warnecke2021:machine}
and a fine-tuning-based
method~\cite{golatkar2020:forgetting,hu2024:duty}. We describe these
methods and their implementations in \appref{app:unlearning}.
In each experiment applying approximate unlearning on image
classifiers, we applied an unlearning method to delete only one
audited data instance from the classifier. For experiments on CLIP
models, we used an unlearning method to unlearn a batch of data
instances (e.g., $250$), since the loss and gradient of a CLIP model
are computed on a batch.  For the gradient-based and
fine-tuning-based methods, we use $\UnlearningRate$ to denote their
unlearning rate.

\paragraph{Data-use detection setting}
We followed the data-use detection setting described in
\secref{sec:auditing_ML:classifier:setup} to detect data-use in an
image classifier, and that described in
\secref{sec:auditing_ML:encoder:setup} to detect data-use in CLIP. We set
$\NumAugment=16$ and $\Confidence=0.001$, and considered
$\FDRBound=0.05$, $\FDRBound=0.01$, and $\FDRBound=0.002$.

\subsection{Experimental Results}  \label{sec:unlearning:results}

\figref{fig:unlearning:cifar100} and
\figsref{fig:unlearning:tinyimagenet}{fig:unlearning:clip:finetune}
show our results.
(\figsref{fig:unlearning:tinyimagenet}{fig:unlearning:clip:finetune}
are presented in \appref{app:unlearning:results}, due to the space
limit.)  As shown there, when we set a larger unlearning rate
$\UnlearningRate$, our $\TrueDetectionRate$ decreased, indicating that
more information of the unlearned data instance was removed from the
updated model.  However, a larger $\UnlearningRate$ led to a lower
model utility as measured by $\Accuracy$, the average fraction of test
data samples correctly predicted by the updated model.  For example,
$\Accuracy$ of the CIFAR-100 models updated by the gradient-based
unlearning method with $\UnlearningRate=0.1$ was $64.56\%$, $11$
percentage points lower than that of models before unlearning, where
we achieved $\TrueDetectionRate = 5.91\%$ at $\FalseDetectionRate \leq
5\%$. In contrast, to maintain good model utility, a small
$\UnlearningRate$ could be used, but neither the gradient-based nor
fine-tuning-based unlearning methods with a small $\UnlearningRate$
could sufficiently remove the unlearned data, since our
$\TrueDetectionRate$ was significantly larger than its
$\FalseDetectionRate$ bound.  For example, $\Accuracy$ of the
CIFAR-100 models updated by the fine-tuning-based method was
$75.05\%$, only $0.48$ percentage points lower than that before
unlearning, but our $\TrueDetectionRate$ was $27.81\%$ at
$\FalseDetectionRate \leq 5\%$. From our results, both the
gradient-based and fine-tuning-based unlearning methods failed to
decrease $\TrueDetectionRate$ to the level of $\FalseDetectionRate$
even after diminishing model utility by $10\%$, and both methods
failed to remove the influence of the unlearned data when maintaining
model utility.

Our experiments also demonstrate that our method is a useful tool for a
data owner to obtain evidence that unlearning was successful. 
In other words, a data owner can use our data-use auditing method to detect if
her requested data instances have been removed from an ML model.
As highlighted by previous works (e.g.,~\cite{thudi2022:necessity}), such
auditable evidence is important in machine-unlearning systems.

\begin{figure}[ht!]
    \centering
    
  \begin{subfigure}[b]{0.47\textwidth}
    \centering
    \resizebox{!}{3em}{\newenvironment{customlegend}[1][]{%
    \begingroup
    \csname pgfplots@init@cleared@structures\endcsname
    \pgfplotsset{#1}%
}{%
    \csname pgfplots@createlegend\endcsname
    \endgroup
}%

\def\addlegendimage{\csname pgfplots@addlegendimage\endcsname}

\begin{tikzpicture}

\definecolor{darkslategray38}{RGB}{38,38,38}
\definecolor{indianred1967882}{RGB}{196,78,82}
\definecolor{lightgray204}{RGB}{204,204,204}
\definecolor{mediumseagreen85168104}{RGB}{85,168,104}
\definecolor{peru22113282}{RGB}{221,132,82}
\definecolor{steelblue76114176}{RGB}{76,114,176}

\begin{customlegend}[
    legend style={{font={\small}},{draw=none}}, 
    legend columns=2,
    legend cell align={center},
    legend entries={{Before unlearning}, {Gradient-based}, {Fine-tuning-based}, {$\FalseDetectionRate$ bound}}]
\addlegendimage{very thick, red, dotted}
\addlegendimage{very thick, steelblue76114176, mark=o, solid}
\addlegendimage{very thick, peru22113282,mark=diamond,dashed,mark options={solid}}
\addlegendimage{very thick, darkgray,dashdotted}

\end{customlegend}

\end{tikzpicture}}
  \end{subfigure}

    \begin{subfigure}[b]{0.235\textwidth}
        \centering
        \begin{tikzpicture}
            \begin{axis}[
                xlabel={$\UnlearningRate (\times 0.01)$ },
                xtick={1, 2, 3, 4, 5},
                xticklabels={2, 4, 6, 8, 10},
                ymode=log,
                log basis y=10, 
                ymin=0.07, ymax=140,
                ytick={0.1, 1, 10, 100},
                grid=major,
                width=1.0\textwidth,
                height=0.8\textwidth,
                ]
                \addplot[color=steelblue76114176,mark=o,line width=1.5] coordinates {
                    (1, 20.5877) 
                    (2, 15.077999999999998) 
                    (3, 11.098600000000001) 
                    (4, 8.918600000000003) 
                    (5, 7.9486000000000026) 
                };
                \addplot[color=peru22113282,mark=diamond,dashed,mark options={solid},line width=1.5] coordinates {
                    (1, 27.8177) 
                    (2, 20.087200000000003) 
                    (3, 12.4976) 
                    (4, 7.828900000000002) 
                    (5, 5.908400000000002)
                };
                \addplot[color=red,dotted,line width=1.5] coordinates {
                    (1, 28.21)
                    (2, 28.21)
                    (3, 28.21)
                    (4, 28.21)
                    (5, 28.21)
                };
                \addplot[color=darkgray,dashdotted,line width=1.5] coordinates {
                    (1, 5)
                    (2, 5)
                    (3, 5)
                    (4, 5)
                    (5, 5)
                };
            \end{axis}
        \end{tikzpicture}
        \caption{$\TrueDetectionRate (\%) @ \FalseDetectionRate \leq 5\%$}
        \label{fig:unlearning:cifar100:5fdr}
    \end{subfigure}
    \hfill
    \begin{subfigure}[b]{0.235\textwidth}
        \centering
        \begin{tikzpicture}
            \begin{axis}[
                xlabel={$\UnlearningRate (\times 0.01)$ },
                xtick={1, 2, 3, 4, 5},
                xticklabels={2, 4, 6, 8, 10},
                ymode=log,
                log basis y=10, 
                ymin=0.07, ymax=140,
                ytick={0.1, 1, 10, 100},
                grid=major,
                width=1.0\textwidth,
                height=0.8\textwidth,
                ]
                \addplot[color=steelblue76114176,mark=o,line width=1.5] coordinates {
                    (1, 7.375100000000001) 
                    (2, 4.806600000000001) 
                    (3, 3.1175000000000006) 
                    (4, 2.3481000000000005) 
                    (5, 1.7690000000000006) 
                };
                \addplot[color=peru22113282,mark=diamond,dashed,mark options={solid},line width=1.5] coordinates {
                    (1, 11.364100000000002) 
                    (2, 7.1859) 
                    (3, 4.027900000000002) 
                    (4, 2.0079000000000002) 
                    (5, 1.4293000000000002)
                };
                \addplot[color=red,dotted,line width=1.5] coordinates {
                    (1, 11.60)
                    (2, 11.60)
                    (3, 11.60)
                    (4, 11.60)
                    (5, 11.60)
                };
                \addplot[color=darkgray,dashdotted,line width=1.5] coordinates {
                    (1, 1)
                    (2, 1)
                    (3, 1)
                    (4, 1)
                    (5, 1)
                };
            \end{axis}
        \end{tikzpicture}
        \caption{$\TrueDetectionRate (\%) @ \FalseDetectionRate \leq 1\%$}
        \label{fig:unlearning:cifar100:1fdr}
    \end{subfigure}
    \\
    \begin{subfigure}[b]{0.235\textwidth}
        \centering
        \begin{tikzpicture}
            \begin{axis}[
                xlabel={$\UnlearningRate (\times 0.01)$ },
                xtick={1, 2, 3, 4, 5},
                xticklabels={2, 4, 6, 8, 10},
                ymode=log,
                log basis y=10, 
                ymin=0.07, ymax=140,
                ytick={0.1, 1, 10, 100},
                grid=major,
                width=1.0\textwidth,
                height=0.8\textwidth,
                ]
                \addplot[color=steelblue76114176,mark=o,line width=1.5] coordinates {
                    (1, 1.5300000000000002) 
                    (2, 0.8900000000000001) 
                    (3, 0.5500000000000003) 
                    (4, 0.33000000000000007) 
                    (5, 0.2800000000000001)   
                };
                \addplot[color=peru22113282,mark=diamond,dashed,mark options={solid},line width=1.5] coordinates {
                    (1, 2.7800000000000007) 
                    (2, 1.6600000000000004) 
                    (3, 0.6600000000000001) 
                    (4, 0.28) 
                    (5, 0.17000000000000007) 
                };
                \addplot[color=red,dotted,line width=1.5] coordinates {
                    (1, 3.12)
                    (2, 3.12)
                    (3, 3.12)
                    (4, 3.12)
                    (5, 3.12)
                };
                \addplot[color=darkgray,dashdotted,line width=1.5] coordinates {
                    (1, 0.2)
                    (2, 0.2)
                    (3, 0.2)
                    (4, 0.2)
                    (5, 0.2)
                };
            \end{axis}
        \end{tikzpicture}
        \caption{$\TrueDetectionRate (\%) @ \FalseDetectionRate \leq 0.2\%$}
        \label{fig:unlearning:cifar100:0.2fdr}
    \end{subfigure}
    \hfill
    \begin{subfigure}[b]{0.235\textwidth}
        \centering
        \begin{tikzpicture}
            \begin{axis}[
                xlabel={$\UnlearningRate (\times 0.01)$ },
                xtick={1, 2, 3, 4, 5},
                xticklabels={2, 4, 6, 8, 10},
                ymin=60, ymax=80,
                grid=major,
                width=1.0\textwidth,
                height=0.8\textwidth,
                ]
                \addplot[color=steelblue76114176,mark=o,line width=1.5] coordinates {
                    (1, 73.304222) 
                    (2, 72.33422399999999) 
                    (3, 70.62700600000002) 
                    (4, 68.11063) 
                    (5, 64.560699) 
                };
                \addplot[color=peru22113282,mark=diamond,dashed,mark options={solid},line width=1.5] coordinates {
                    (1, 75.056964) 
                    (2, 74.35623900000002) 
                    (3, 72.79661000000002) 
                    (4, 69.666949) 
                    (5, 64.129923) 
                };
                \addplot[color=red,dotted,line width=1.5] coordinates {
                    (1, 75.61)
                    (2, 75.61)
                    (3, 75.61)
                    (4, 75.61)
                    (5, 75.61)
                };
            \end{axis}
        \end{tikzpicture}
        \caption{Test accuracy $\Accuracy (\%)$}
        \label{fig:unlearning:cifar100:acc}
    \end{subfigure} 
    \caption{$\TrueDetectionRate(\%)$ of our auditing method for
      CIFAR-100 image classifiers after applying approximate
      unlearning
      (\subfigsref{fig:unlearning:cifar100:5fdr}{fig:unlearning:cifar100:0.2fdr})
      and test accuracies of image classifiers
      (\figref{fig:unlearning:cifar100:acc}).}
    \label{fig:unlearning:cifar100}
  \end{figure}

\section{Conclusion} \label{sec:conclusion}

In this paper, we proposed an instance-level data-use auditing method
for the image domain that a data owner can apply to audit a model for
use of her image instances in training.  Our data-auditing method
leverages any membership-inference technique, folding it into a
sequential hypothesis test of our own design for which we can quantify
and bound the false-detection rate.  Our method strictly generalizes
previous such approaches, permitting ours to be applied to audit for
use of many fewer data instances, even a single one.  By evaluating
our method to audit three types of visual ML models, namely image
classifiers, visual encoders, and CLIP and BLIP models, we demonstrated its
utility across diverse visual models and settings.  We further
demonstrated the power of a bounded false-detection rate when auditing
for use of single data instances, by applying our method to evaluate
two state-of-the-art approximate unlearning methods.  By showing that
our methodology detects a model's use of individual data instances at
a rate substantially higher than the false-positive-rate bound, even
after unlearning via these methods, we quantifiably showed that these
unlearning methods do not work, without substantially decaying model
accuracy.  A future work is to generalize our instance-level data-use
auditing method into other domains, e.g., text data.

\bibliographystyle{plain}
\bibliography{full, references}

\appendix

\section{Pseudocodes for Data Marking Algorithm and Data-Use Detection Algorithm} \label{app:algorithm}
We present the pseudocode of the algorithm for generating
$\Data{\SampleIdx}{1}, \Data{\SampleIdx}{2}, \dots,
\Data{\SampleIdx}{\NumberMarkedData}$,
the pseudocode of our data-marking algorithm, and the pseudocode of
our data-use detection algorithm
in \algref{alg:mark_generation}, \algref{alg:data_marking},
and \algref{alg:data_detection}, respectively.

\paragraph{Marked data generation algorithm}
Marked data generation algorithm (i.e.,~\algref{alg:mark_generation}) is used to generate
$\NumberMarkedData$ marked data instances for each raw data item. 
Given a pretrained feature
extractor $\FeatureExtractor$, a bound $\MarkBound \in \realsPos$,
and the number $\NumberMarkedData \in \IntegersPos$ of marked data,
it takes as input a raw data instance $\Data{\SampleIdx}$
and outputs $\Data{\SampleIdx}{1}, \Data{\SampleIdx}{2}, \dots, \Data{\SampleIdx}{\NumberMarkedData}$.

\paragraph{Data-marking algorithm}
Data-marking algorithm (i.e.,~\algref{alg:data_marking}), applied in data-marking and publication stage, 
is used to generate a marked dataset $\MarkedDataset$ to be published and the hidden information $\HiddenInfo$.
Given a pretrained feature extractor $\FeatureExtractor$, a bound $\MarkBound \in \realsPos$,
and the number $\NumberMarkedData \in \IntegersPos$ of marked data per raw data item, it takes as input
a set of raw data instance $\Dataset$ and outputs a marked dataset $\MarkedDataset$ and the hidden information $\HiddenInfo$.

\paragraph{Data-use detection algorithm}
Data-use detection algorithm (i.e.,~\algref{alg:data_detection}), applied in data-use detection stage,
is used to detect if an ML model used the published dataset in training. Given the number $\NumberMarkedData$ 
of generated marked data instances per each raw data item,
the bound $\FDRBound$ of false-detection rate,
the confidence level $\Confidence < \FDRBound$,
and black-box access to the ML model $\MLModel$, it takes as inputs the marked dataset $\MarkedDataset$ and the hidden information $\HiddenInfo$,
and outputs a binary value $\DOBit$. If $\DOBit=1$, it detects data-use in the ML model; Otherwise, it fails to detect.

\begin{algorithm}[!t]
    \caption{Marked data generation
    algorithm} \label{alg:mark_generation}
    \begin{algorithmic}[1]
    \REQUIRE
    A raw data instance $\Data{\SampleIdx}$, a pretrained feature
    extractor $\FeatureExtractor$, a bound $\MarkBound \in \realsPos$,
    and the number $\NumberMarkedData \in \IntegersPos$ of marked data.
    \STATE Generate $\NumberMarkedData$ unit vectors
    $\UnitVector{\SampleIdx}{1}, \UnitVector{\SampleIdx}{2}, \dots, \UnitVector{\SampleIdx}{\NumberMarkedData}$
    such that their minimum pairwise distance is maximized;
    \FOR{$\MarkedVersionIdx \gets 1, 2, \dots, \NumberMarkedData$}
    \STATE Initialize $\Mark{\SampleIdx}{\MarkedVersionIdx}$;
    \STATE Solve
    $\Mark{\SampleIdx}{\MarkedVersionIdx} \leftarrow \argmax_{\infinityNorm{\Mark{\SampleIdx}{\MarkedVersionIdx}} \leq \MarkBound} \dotProduct{\UnitVector{\SampleIdx}{\MarkedVersionIdx}}{\FeatureExtractor(\Data{\SampleIdx}+\Mark{\SampleIdx}{\MarkedVersionIdx})}$
    where ``\dotProduct{}{}'' denotes the dot product, such that
    $\Data{\SampleIdx}+\Mark{\SampleIdx}{\MarkedVersionIdx}$ is a
    valid image;
    \STATE $\Data{\SampleIdx}{\MarkedVersionIdx} \leftarrow \Data{\SampleIdx}+\Mark{\SampleIdx}{\MarkedVersionIdx}$;
    \ENDFOR
    \ENSURE $\Data{\SampleIdx}{1}, \Data{\SampleIdx}{2}, \dots, \Data{\SampleIdx}{\NumberMarkedData}$.
\end{algorithmic}
\end{algorithm}

\begin{algorithm}[!t]
    \caption{Data-marking algorithm $\MarkAlg$}  \label{alg:data_marking}
    \begin{algorithmic}[1]
    \REQUIRE A set of raw data instance $\Dataset =\{\Data{1}, \Data{2}, \dots, \Data{\DatasetSize}\}$, 
        a pretrained feature extractor $\FeatureExtractor$, a bound $\MarkBound \in \realsPos$,
        and the number $\NumberMarkedData \in \IntegersPos$ of marked data per raw data item.
    \FOR{$\Data{\SampleIdx} \in \Dataset$}
    \STATE $\Data{\SampleIdx}{1}, \Data{\SampleIdx}{2}, \dots, \Data{\SampleIdx}{\NumberMarkedData} \leftarrow \algref{alg:mark_generation}$;
    \STATE $\MarkedData{\SampleIdx} \getsr \{\Data{\SampleIdx}{1}, \Data{\SampleIdx}{2}, \dots, \Data{\SampleIdx}{\NumberMarkedData}\}$;
    \STATE $\HiddenInfo{\SampleIdx} \leftarrow \{\Data{\SampleIdx}{1}, \Data{\SampleIdx}{2}, \dots, \Data{\SampleIdx}{\NumberMarkedData}\} \setminus \{\MarkedData{\SampleIdx}\}$;
    \ENDFOR
    \STATE $\MarkedDataset \leftarrow \{\MarkedData{1}, \MarkedData{2}, \dots, \MarkedData{\DatasetSize}\}$;
    \STATE $\HiddenInfo\leftarrow\bigcup_{\SampleIdx=1}^{\DatasetSize}\HiddenInfo{\SampleIdx}$;
    \ENSURE
    $\MarkedDataset, \HiddenInfo$.
    \end{algorithmic}
\end{algorithm}

\begin{algorithm}[!t]
  \caption{Data-use detection algorithm $\DetectAlg$}  \label{alg:data_detection}
  \begin{algorithmic}[1]
  \REQUIRE A set of published data instances $\MarkedDataset$, the hidden set $\HiddenInfo$, 
      the number $\NumberMarkedData$ of generated marked data instances per each raw data item,
      the bound $\FDRBound$ of false-detection rate,
      the confidence level $\Confidence < \FDRBound$,
      and black-box access to the ML model $\MLModel$.
  \STATE Set $\sumThreshold$ s.t. \eqnref{eq:threshold} is satisfied;
  \STATE Initialize the measurement sequence $\MeasurementSequence \gets \emptyset$;
  \STATE Initialize $\DOBit=0$;
  \STATE Apply membership inference for $\MIAlg{\MLModel}(\MarkedData)$;
  \FOR{$\ObservationTime \gets 1, 2, \dots, \DatasetSize(\NumberMarkedData-1)$}
  \STATE Sample an $\Data$ randomly WoR from $\HiddenInfo$;
  \STATE Apply membership inference for $\MIAlg{\MLModel}(\Data)$;
  \STATE $\MeasurementSequence \gets \MeasurementSequence \cup \{\IndicatorFunction{\MIAlg{\MLModel}(\MarkedData{\SampleIdx}) > \MIAlg{\MLModel}(\Data)}\}$, given $\Data\in \HiddenInfo{\SampleIdx}$;
  \STATE $[\ConfidenceIntervalLower{\ObservationTime}{\Confidence}, \ConfidenceIntervalUpper{\ObservationTime}{\Confidence}] \gets \PPRMartingale(\MeasurementSequence, \DatasetSize(\NumberMarkedData-1), \Confidence)$;
  \IF{$\ConfidenceIntervalLower{\ObservationTime}{\Confidence} \geq \sumThreshold$}
    \STATE $\DOBit=1$;
    \BREAK
  \ENDIF
  \ENDFOR
  \ENSURE
  $\DOBit$.
  \end{algorithmic}
\end{algorithm}

\section{Proof of Theorem~\ref{theorem:fdr}} \label{app:proof}

\begin{proof}
  We use $[\DatasetSize(\NumberMarkedData-1)]$ to represent $\{1,2,\dots,\DatasetSize(\NumberMarkedData-1)\}$.
  We study the probability that under \nullHypothesis there exists a confidence interval whose lower bound is 
  no smaller than a preselected threshold $\sumThreshold \in \{0, 1, \dots, \DatasetSize(\NumberMarkedData-1)\}$, 
  i.e., $\ConfidenceIntervalLower{\ObservationTime}{\Confidence} \geq \sumThreshold$. We have:
  \begin{align*}
    \lefteqn{\cprob{\big}{\exists \ObservationTime \in [\DatasetSize(\NumberMarkedData-1)]:\ConfidenceIntervalLower{\ObservationTime}{\Confidence} \geq \sumThreshold}{\nullHypothesis}} \\
    & = \cprob{\big}{\exists \ObservationTime \in [\DatasetSize(\NumberMarkedData-1)]:\ConfidenceIntervalLower{\ObservationTime}{\Confidence} \geq \sumThreshold}{\TotalSuccess \geq \sumThreshold, \nullHypothesis} \\
    &\quad \times \cprob{\big}{\TotalSuccess \geq \sumThreshold}{\nullHypothesis} \\
    & + \cprob{\big}{\exists \ObservationTime \in [\DatasetSize(\NumberMarkedData-1)]:\ConfidenceIntervalLower{\ObservationTime}{\Confidence} \geq \sumThreshold}{\TotalSuccess < \sumThreshold , \nullHypothesis} \\ 
    & \quad \times \cprob{\big}{\TotalSuccess < \sumThreshold}{\nullHypothesis} \\
    & \leq  \cprob{\big}{\TotalSuccess \geq \sumThreshold}{\nullHypothesis} \\ 
    & + \cprob{\big}{\exists \ObservationTime \in [\DatasetSize(\NumberMarkedData-1)]:\ConfidenceIntervalLower{\ObservationTime}{\Confidence} \geq \sumThreshold}{\TotalSuccess < \sumThreshold, \nullHypothesis} 
    \end{align*}
  Under the null hypothesis, the rank of a published data instance is uniformly distributed over $\{1, 2, \dots, \NumberMarkedData\}$
  and thus the sum of independently, uniformly distributed ranks has the following probability mass function~\cite{caiado2007:polynomial}:
  \begin{equation*}
    \prob{\RankSum = \RankVariable} = \frac{1}{\NumberMarkedData^{\DatasetSize}} \sum_{\VariableIdx=0}^{\floor{\frac{\RankVariable-\DatasetSize}{\NumberMarkedData}}} (-1)^{\VariableIdx} {\DatasetSize \choose \VariableIdx} {\RankVariable - \NumberMarkedData \VariableIdx - 1 \choose \DatasetSize - 1}.
  \end{equation*}
  Because $\RankSum = \TotalSuccess + \DatasetSize$, we have:
  \begin{equation*}
    \cprob{\big}{\TotalSuccess \geq \sumThreshold}{\nullHypothesis} = \sum_{\RankVariable=\sumThreshold+\DatasetSize}^{\DatasetSize\NumberMarkedData}\frac{1}{\NumberMarkedData^{\DatasetSize}} \sum_{\VariableIdx=0}^{\floor{\frac{\RankVariable-\DatasetSize}{\NumberMarkedData}}} (-1)^{\VariableIdx} {\DatasetSize \choose \VariableIdx} {\RankVariable - \NumberMarkedData \VariableIdx - 1 \choose \DatasetSize - 1}.
  \end{equation*}
  Since the confidence sequence is independent of the null hypothesis, we have:
  \begin{align*}
    \lefteqn{\cprob{\big}{\exists \ObservationTime \in [\DatasetSize(\NumberMarkedData-1)]:\ConfidenceIntervalLower{\ObservationTime}{\Confidence} \geq \sumThreshold}{\TotalSuccess < \sumThreshold, \nullHypothesis}} \\
    & = \cprob{\big}{\exists \ObservationTime \in [\DatasetSize(\NumberMarkedData-1)]:\ConfidenceIntervalLower{\ObservationTime}{\Confidence} \geq \sumThreshold}{\TotalSuccess < \sumThreshold}.
  \end{align*}
  Also, according to the guarantee of the confidence sequence~\cite{waudby2020:confidence}, we have:
  \begin{equation*}
    \cprob{\big}{\exists \ObservationTime \in [\DatasetSize(\NumberMarkedData-1)]:\ConfidenceIntervalLower{\ObservationTime}{\Confidence} \geq \sumThreshold}{\TotalSuccess < \sumThreshold} \leq \Confidence.
  \end{equation*}
  Therefore, we have:
  \begin{align*}
    \lefteqn{\cprob{\big}{\exists \ObservationTime \in [\DatasetSize(\NumberMarkedData-1)]:\ConfidenceIntervalLower{\ObservationTime}{\Confidence} \geq \sumThreshold}{\nullHypothesis}} \\
    & \leq  \cprob{\big}{\TotalSuccess \geq \sumThreshold}{\nullHypothesis} \\ 
    & + \cprob{\big}{\exists \ObservationTime \in [\DatasetSize(\NumberMarkedData-1)]:\ConfidenceIntervalLower{\ObservationTime}{\Confidence} \geq \sumThreshold}{\TotalSuccess < \sumThreshold, \nullHypothesis} \\
    & \leq \cprob{\big}{\TotalSuccess \geq \sumThreshold}{\nullHypothesis} + \Confidence
    \end{align*}
  If we set $\sumThreshold$ such that
  \begin{equation*}
    \cprob{\big}{\TotalSuccess \geq \sumThreshold}{\nullHypothesis} \leq \FDRBound - \Confidence.
    \end{equation*}
  In other words,
  \begin{equation}
  \sum_{\RankVariable=\sumThreshold+\DatasetSize}^{\DatasetSize\NumberMarkedData}\frac{1}{\NumberMarkedData^{\DatasetSize}} \sum_{\VariableIdx=0}^{\floor{\frac{\RankVariable-\DatasetSize}{\NumberMarkedData}}} (-1)^{\VariableIdx} {\DatasetSize \choose \VariableIdx} {\RankVariable - \NumberMarkedData \VariableIdx - 1 \choose \DatasetSize - 1} \leq \FDRBound - \Confidence, \label{eq:sumthreshold}
  \end{equation} 
  we have:
  \begin{equation*}
    \cprob{\big}{\exists \ObservationTime \in [\DatasetSize(\NumberMarkedData-1)]:\ConfidenceIntervalLower{\ObservationTime}{\Confidence} \geq \sumThreshold}{\nullHypothesis} \leq \FDRBound.
  \end{equation*}
  Since our detection algorithm rejects \nullHypothesis if it finds a 
  confidence interval whose lower bound satisfies $\ConfidenceIntervalLower{\ObservationTime}{\Confidence} \geq \sumThreshold$,
  $\cprob{\big}{\exists \ObservationTime \in [\DatasetSize(\NumberMarkedData-1)]: \ConfidenceIntervalLower{\ObservationTime}{\Confidence} \geq \sumThreshold}{\nullHypothesis}$ 
  is its false-detection probability. Given any $\FDRBound \in [0, 1]$ and $\Confidence < \FDRBound$, when we set
  $\sumThreshold$ to satisfy \eqnref{eq:sumthreshold}, our detection algorithm has a
\FalseDetectionRate no larger than \FDRBound.
\end{proof}

\section{Experimental Setup} \label{app:setup}

In this section, we describe the experimental setup for our experiments in \secref{sec:auditing_ML}.

\subsection{Datasets} \label{app:setup:dataset}
In \secref{sec:auditing_ML}, we used four image datasets, 
namely CIFAR-100~\cite{krizhevsky2009:learning},
TinyImageNet~\cite{Lle2015:TinyIV},
ImageNet~\cite{deng2009:imagenet}, and Flickr30k~\cite{young2014:image}:
\begin{itemize}
  \item \textbf{CIFAR-100}: CIFAR-100 is a dataset containing
    $60{,}000$ images of $3\times32\times32$ dimensions partitioned 
    into $100$ classes. In CIFAR-100, there are
    $50{,}000$ training samples and $10{,}000$ test samples.
  \item \textbf{TinyImageNet}: TinyImageNet is a dataset containing
    images of $3\times64\times64$ dimensions partitioned into $200$
    classes. In TinyImageNet, there are $100{,}000$ training samples
    and $10{,}000$ validation samples that we used as test samples.
  \item \textbf{ImageNet}: ImageNet is a large-scale dataset
    containing images partitioned into $1{,}000$ classes. There are
    $1{,}281{,}167$ training samples and $50{,}000$ validation
    samples that we used as test samples.
  \item \textbf{Flickr30k}: Flickr30k comprises $31{,}014$ images of varying sizes, 
    each paired with multiple textual descriptions.
\end{itemize}

\subsection{Data-Marking Settings} \label{app:setup:marking}

\paragraph{Visual Encoder}  \label{app:setup:marking:encoder}
We followed the default data-marking setup described in \secref{sec:auditing_ML:classifier:setup}
to prepare marked training datasets, without labels needed.

\paragraph{CLIP}  \label{app:setup:marking:clip}
We first randomly sampled $2{,}500$ images with textual descriptions
as training samples and others as test samples. From $2{,}500$
training samples, we randomly selected $250$ samples as the audited
samples. However, for a fixed value of \DatasetSize, each subset of
\DatasetSize audited samples were treated independently (as if from a
different data owner; i.e., we applied our test for each subset of
size \DatasetSize separately).  We applied our data-marking algorithm
to generate its published version \MarkedDataset and the associated
hidden information set $\HiddenInfo$, where we set $\MarkBound=10$,
$\NumberMarkedData=1000$, and used ResNet-18 pretrained on ImageNet as
$\FeatureExtractor$.  As such, we prepared a marked training dataset
by replacing each audited image with its published version and
assigning it to its original textual description. 
Some examples of marked Flickr30k images are presented in
\figref{fig:flickr30k_examples} in \appref{app:results:examples}.

\paragraph{BLIP}  \label{app:setup:marking:blip}
We followed the data-marking setup of CLIP
to prepare marked training datasets.

\subsection{Model Training Settings} \label{app:setup:training}

\paragraph{Image Classifier}  \label{app:setup:training:classifier}
We trained the classifier using a standard
stochastic gradient descent (SGD)
algorithm~\cite{amari1993:backpropagation}: At each step, the
parameters of the image classifier were updated on a mini-batch of
image-label training pairs (e.g., $128$ for CIFAR-100 or TinyImageNet,
or $256$ for ImageNet) with data normalization and default data
augmentation applied using an SGD optimizer with initial learning rate
of $0.1$ and a weight decay of $5\times10^{-4}$. We trained the
classifier for $100$ epochs.  During the training process, we decayed
the learning rate by a factor of $0.1$ when the number of epochs
reached $37$, $62$, or $87$, following previous work
(e.g.,~\cite{geiping2020:witches}).

\paragraph{Visual Encoder}  \label{app:setup:training:encoder}
We trained the visual encoder by SimCLR algorithm~\cite{chen2020:simple} where a visual encoder and a 
projection head (i.e., a multilayer perceptron with one hidden
layer) are trained simultaneously. The SimCLR algorithm works as follows: At each iteration, 
a mini-batch (i.e., of size 512) of training image samples are selected and two augmented versions
of each sample from the mini-batch are generated by random cropping and resizing, random
color distortion, and random Gaussian blur. Then, the parameters of the visual encoder
and the projection head are updated by minimizing the NT-Xent loss among the generated augmented images, i.e.,
maximizing the cosine similarity between any positive pair (i.e.,
two augmented images generated from the same training sample)
and minimizing the cosine similarity between any negative pair (i.e.,
two augmented images generated from different training samples). To update the parameters of the models,
we used SGD with Nesterov Momentum~\cite{sutskever2013:importance} of $0.9$, a weight decay of $10^{-6}$, and an initial learning rate of $0.6$ 
as the optimizer. We trained the models by $1{,}000$ epochs and applied a cosine annealing schedule~\cite{loshchilov2017:sgdr}
to update the learning rate during the training process.

\paragraph{CLIP}  \label{app:setup:training:clip}
We fine-tuned the pretrained CLIP (\texttt{ViT-B/32}) released from
OpenAI on marked datasets prepared from the data-marking setting. We
followed the pretraining algorithm used by
OpenAI~\cite{radford2021:learning}, applying the Adam
optimizer~\cite{kingma2014:adam} with a learning rate of $10^{-5}$ to
fine-tune the CLIP model on a mini-batch of $256$ training samples at
each iteration.

\paragraph{BLIP}  \label{app:setup:training:blip}
We fine-tuned the pretrained BLIP image captioning base model 
on marked datasets prepared from the data-marking setting. We
applied the AdamW optimizer~\cite{loshchilov2019:decoupled} with a learning rate of $5\times10^{-5}$ to
fine-tune the BLIP model on a mini-batch of $8$ training samples at
each iteration.

\subsection{Data-Use Detection}  \label{app:setup:detection}

\paragraph{Image Classifier}  \label{app:setup:detection:classifier}
We followed previous works (e.g.,~\cite{choquette2021:label, huang2024:auditdata}) to 
define the black-box membership inference method used in data-use 
detection for image classifiers: Given an input (marked) image, 
we first generated $\NumAugment-1$ perturbed versions by randomly cropping and flipping the image, and
then obtained $\NumAugment$ outputs by using the input image and its
$\NumAugment-1$ perturbed versions as inputs to the classifier.  Next,
we averaged the $\NumAugment$ confidence vectors or confidences and
returned the negative modified entropy~\cite{song2021:systematic} as
the ``memorization'' score of the input image. In tests where only the
label was returned, we transformed the $\NumAugment$ returned labels
into $\NumAugment$ one-hot vectors, averaged them, and returned the
negative modified entropy as the ``memorization'' score.

\paragraph{Visual Encoder}  \label{app:setup:detection:encoder}
We assumed that the data owner can obtain a vector of features by providing her marked image or 
its augmented version as input to the visual encoder. Using the marked images $\MarkedDataset$ and 
the associated hidden set $\HiddenInfo$ generated from the data-marking setting, we applied 
our data-use detection algorithm to test if a visual encoder
was trained using $\MarkedDataset$. When applying the data-use detection algorithm,
we followed the previous works (e.g.,~\cite{liu2021:encodermi,huang2024:auditdata,zhu2024:unified}) 
to define the black-box membership inference 
method: Given an input image, we first randomly generated its $\NumAugment$ perturbed 
versions by randomly cropping and flipping the image and then 
obtained $\NumAugment$ feature vectors by using the $\NumAugment$ perturbed versions 
as inputs to the visual encoder. Next, we computed the cosine similarity of each
pair of feature vectors and returned the sum of cosine similarities as
the ``memorization'' score of the input image.
In the implementation of membership inference applied in our data-use detection, 
we set $\NumAugment=64$ as the default. 
We considered $\FDRBound=0.05$, $\FDRBound=0.01$, 
and $\FDRBound=0.002$ in the data-use detection algorithm.

\paragraph{CLIP}  \label{app:setup:detection:clip}
We assumed that the data owner could obtain feature vectors by
providing her marked image and its corresponding textual description
as inputs to the CLIP model. Using the marked images \MarkedDataset and
the associated hidden set $\HiddenInfo$ generated from the data
marking, she applied our data-use detection algorithm to test if a
CLIP was trained/fine-tuned using 
\MarkedDataset.  When applying the
data-use detection algorithm, we followed the previous works
(e.g.,~\cite{ko2023:practical,huang2024:auditdata}) to define the
black-box membership inference method: 
Given an input image and its textual description, 
we obtained their corresponding feature vectors
(one for image and the other for its textual description) by CLIP; 
Then we used the cosine similarity between the two feature vectors 
as the ``memorization'' score of the input image.
We considered $\FDRBound=0.05$, $\FDRBound=0.01$, and
$\FDRBound=0.002$ in the data-use detection algorithm.

\paragraph{BLIP}  \label{app:setup:detection:blip}
We assumed that the data owner could obtain the probability distributions of the output
by providing her marked image. Using the marked images \MarkedDataset and
the associated hidden set $\HiddenInfo$ generated from the data
marking, she applied our data-use detection algorithm to test if a
BLIP image captioning model was trained/fine-tuned using 
\MarkedDataset. When applying the data-use detection algorithm,
we define the membership inference method by the negative cross-entropy
loss score. We considered $\FDRBound=0.05$, $\FDRBound=0.01$, and
$\FDRBound=0.002$ in the data-use detection algorithm.

\subsection{System Configuration and Runtime Overhead} \label{app:setup:system_config}

We conducted all experiments on a server equipped with 
Intel(R) Xeon(R) Silver 4214R CPUs @ 2.40GHz and NVIDIA RTX A5000
GPUs with 24GB memory. We used Python 3.9.15 with PyTorch 
2.2.0 and CUDA 12.1 on Ubuntu 22.04.

\paragraph{Runtime Overhead of Data Marking} \label{app:setup:system_config:marking}
We report GPU time, CPU time, and peak memory usage of running our
data marking algorithm using $\Dataset$ of $\DatasetSize$ size
as input in \figref{fig:overhead:marking}. Its GPU and CPU runtimes 
scale linearly with $\DatasetSize$.

\begin{figure}[ht!]
  \centering

  \begin{subfigure}[b]{0.45\textwidth}
    \centering
    \resizebox{!}{1.7em}{\newenvironment{customlegend}[1][]{%
    \begingroup
    \csname pgfplots@init@cleared@structures\endcsname
    \pgfplotsset{#1}%
}{%
    \csname pgfplots@createlegend\endcsname
    \endgroup
}%

\def\addlegendimage{\csname pgfplots@addlegendimage\endcsname}

\begin{tikzpicture}

\definecolor{darkslategray38}{RGB}{38,38,38}
\definecolor{indianred1967882}{RGB}{196,78,82}
\definecolor{lightgray204}{RGB}{204,204,204}
\definecolor{mediumseagreen85168104}{RGB}{85,168,104}
\definecolor{peru22113282}{RGB}{221,132,82}
\definecolor{steelblue76114176}{RGB}{76,114,176}

\begin{customlegend}[
    legend style={{font={\small}},{draw=none}}, 
    legend columns=4,
    legend cell align={center},
    legend entries={{CIFAR-100}, {TinyImageNet}, {ImageNet}}]
\addlegendimage{very thick, color=steelblue76114176,mark=o}
\addlegendimage{very thick, peru22113282,mark=diamond,dashed,mark options={solid}}
\addlegendimage{very thick, mediumseagreen85168104,mark=triangle,dotted,mark options={solid}}

\end{customlegend}

\end{tikzpicture}}
  \end{subfigure}

  \begin{subfigure}[t]{0.32\textwidth}
    \centering
    \begin{tikzpicture}
      \begin{axis}[
        xlabel={$\DatasetSize$},
        ylabel={GPU time (min)},
        xtick={1, 2, 3, 4, 5, 6, 7},
        xticklabels={1, 2, 4, 8, 16, 32, 64},
        ymin=-20, ymax=620,
        grid=major,
        width=1.0\textwidth,
        height=0.5\textwidth,
        ]
        \addplot[color=steelblue76114176,mark=o,line width=1.5] coordinates {
          (1,7.897980729166666)
          (2,15.721589583333333)
          (3,31.441881249999998)
          (4,62.5205)
          (5,125.190775)
          (6,237.19438333333335)
          (7,453.91253333333333)
        };
        \addplot[color=peru22113282,mark=diamond,dashed,mark options={solid},line width=1.5] coordinates {
          (1,8.1597109375)
          (2,15.452884375)
          (3,30.821337500000002)
          (4,61.483779166666665)
          (5,122.97574166666666)
          (6,241.7246)
          (7,456.56656666666663)
        };
        \addplot[color=mediumseagreen85168104,mark=triangle,dotted,mark options={solid},line width=1.5] coordinates {
          (1,3.4790018229166666)
          (2,6.862492708333334)
          (3,13.678570833333334)
          (4,27.292683333333333)
          (5,54.43249166666667)
          (6,107.425825)
          (7,212.36213333333333)
        };
      \end{axis}
    \end{tikzpicture}
  \end{subfigure}
  \hfill
  \begin{subfigure}[t]{0.32\textwidth}
    \centering
    \begin{tikzpicture}
      \begin{axis}[
        xlabel={$\DatasetSize$},
        ylabel={CPU time (min)},
        xtick={1, 2, 3, 4, 5, 6, 7},
        xticklabels={1, 2, 4, 8, 16, 32, 64},
        ymin=-20, ymax=620,
        grid=major,
        width=1.0\textwidth,
        height=0.5\textwidth,
        ]
        \addplot[color=steelblue76114176,mark=o,line width=1.5] coordinates {
          (1,8.423586496433334)
          (2,16.547314095966666)
          (3,33.070676250383336)
          (4,65.31269091161667)
          (5,131.05003190741667)
          (6,247.04316766413334)
          (7,473.10765510833335)
        };
        \addplot[color=peru22113282,mark=diamond,dashed,mark options={solid},line width=1.5] coordinates {
          (1,8.591233913)
          (2,16.18808911475)
          (3,32.19601136683333)
          (4,63.89584652163333)
          (5,127.63560778543334)
          (6,251.34396805988334)
          (7,476.2745974034833)
        };
        \addplot[color=mediumseagreen85168104,mark=triangle,dotted,mark options={solid},line width=1.5] coordinates {
          (1,3.508756742566667)
          (2,6.9273700195000005)
          (3,13.811613307766667)
          (4,27.569689612983332)
          (5,54.9880038075)
          (6,108.13450430990001)
          (7,214.54853337793335)
        };
      \end{axis}
    \end{tikzpicture}
  \end{subfigure}
  \hfill
  \begin{subfigure}[t]{0.32\textwidth}
    \centering
    \begin{tikzpicture}
      \begin{axis}[
        xlabel={$\DatasetSize$},
        ylabel={Peak memory usage (MB)},
        xtick={1, 2, 3, 4, 5, 6, 7},
        xticklabels={1, 2, 4, 8, 16, 32, 64},
        ymin=-100, ymax=3100,
        grid=major,
        width=1.0\textwidth,
        height=0.5\textwidth,
        ]
        \addplot[color=steelblue76114176,mark=o,line width=1.5] coordinates {
          (1,1937.793024)
          (2,2026.815488)
          (3,1983.46752)
          (4,2024.144896)
          (5,1938.06336)
          (6,1738.87488)
          (7,1734.049792)
        };
        \addplot[color=peru22113282,mark=diamond,dashed,mark options={solid},line width=1.5] coordinates {
          (1,1772.044288)
          (2,1450.115072)
          (3,1530.974208)
          (4,1472.34816)
          (5,1468.338176)
          (6,1531.006976)
          (7,1486.680064)
        };
        \addplot[color=mediumseagreen85168104,mark=triangle,dotted,mark options={solid},line width=1.5] coordinates {
          (1,2229.47328)
          (2,2292.092928)
          (3,2209.73056)
          (4,2370.10944)
          (5,2378.09664)
          (6,2215.489536)
          (7,2309.636096)
        };
      \end{axis}
    \end{tikzpicture}
  \end{subfigure}
  \caption{Overhead of running our data-marking algorithm 
    ($\NumberMarkedData=1000$) for each $\Dataset$ of
    varying $\DatasetSize$ size.}
  \label{fig:overhead:marking}
\end{figure}

\paragraph{Runtime Overhead of Data-Use Detection} \label{app:setup:system_config:detection}
Our data-use detection algorithm involves querying the audited ML model, 
computing a memorization score, and applying a prior-posterior-ratio 
martingale to estimate a confidence interval. We assume that the runtimes 
overhead of querying the ML model occurs on the server side rather than 
on the data owner's side. Since the memorization score is computed 
using a simple operation (e.g., the negative entropy of the classifier output), 
the primary computational burden on the data owner lies in applying 
the prior-posterior-ratio martingale, which runs on CPU only. 
Therefore, we report the CPU runtimes and peak memory usage of applying 
prior-posterior-ratio martingale, in \figref{fig:overhead:detection}.
Its CPU runtimes cost $O(\DatasetSize\log(\DatasetSize))$.

\begin{figure}[ht!]
  \centering

  \begin{subfigure}[t]{0.32\textwidth}
    \centering
    \begin{tikzpicture}
      \begin{axis}[
        xlabel={$\DatasetSize$},
        ylabel={CPU time (min)},
        xtick={1, 2, 3, 4, 5, 6, 7},
        xticklabels={1, 2, 4, 8, 16, 32, 64},
        ymin=-1, ymax=31,
        grid=major,
        width=1.0\textwidth,
        height=0.5\textwidth,
        ]
        \addplot[color=steelblue76114176,mark=o,line width=1.5] coordinates {
          (1,0.04211058581666667)
          (2,0.09485169463333333)
          (3,0.24148303003333332)
          (4,0.5543565155)
          (5,1.672200537533333)
          (6,6.569345813083333)
          (7,23.930749574966665)
        };
      \end{axis}
    \end{tikzpicture}
  \end{subfigure}
  \hfill
  \begin{subfigure}[t]{0.32\textwidth}
    \centering
    \begin{tikzpicture}
      \begin{axis}[
        xlabel={$\DatasetSize$},
        ylabel={Peak memory usage (MB)},
        xtick={1, 2, 3, 4, 5, 6, 7},
        xticklabels={1, 2, 4, 8, 16, 32, 64},
        ymin=-2, ymax=60,
        grid=major,
        width=1.0\textwidth,
        height=0.5\textwidth,
        ]
        \addplot[color=steelblue76114176,mark=o,line width=1.5] coordinates {
          (1,48.10752)
          (2,47.849472)
          (3,48.128)
          (4,47.771648)
          (5,47.665152)
          (6,49.987584)
          (7,51.007488)
        };
      \end{axis}
    \end{tikzpicture}
  \end{subfigure}
  \caption{Overhead of applying prior-posterior-ratio martingale
    for confidence interval estimation in
    data-use detection ($\NumberMarkedData=1000$) for each $\Dataset$ of
    varying $\DatasetSize$ size.}
  \label{fig:overhead:detection}
\end{figure}

\subsection{Baselines} \label{app:setup:baselines}

We introduce the baselines considered in \secref{sec:auditing_ML:classifier:setup},
namely Attack-P~\cite{ye2022:enhanced}, Attack-R~\cite{ye2022:enhanced},
LiRA~\cite{carlini2022:membership}, and RMIA~\cite{zarifzadeh2024:low}.

\paragraph{Basic idea}
Given a data instance $\Data$ and its associated label $\DataLabel$, an ML model $\MLModel$ (i.e., a classifier),
and access to a set of labeled auxiliary data $\AuxiliaryDataset$ and some reference models $\ReferenceModelset$,
these membership inference attacks compute a score $\MIAScore{\AuxiliaryDataset}{\ReferenceModelset}(\Data, \DataLabel, \MLModel)$
and then compare it to a threshold $\MIAThreshold$.
Membership inference attacks infer that $\Data$ is used in training $\MLModel$
if $\MIAScore{\AuxiliaryDataset}{\ReferenceModelset}(\Data, \DataLabel, \MLModel)\geq \MIAThreshold$.
Here $\MIAThreshold$ controls the empirical $\FalseDetectionRate$
of a membership inference attack method. The computation of $\MIAScore{\AuxiliaryDataset}{\ReferenceModelset}(\Data, \DataLabel, \MLModel)$ by
these membership inference attacks is presented in \tblref{table:membership_inference}~\cite[Table 1]{zarifzadeh2024:low}.

\begin{table*}[ht!]
    \centering
    \begin{tabular}{@{}cc@{\hspace{0.5em}}cc@{\hspace{0.5em}}c@{}}
    \toprule
    Membership inference attack & $\MIAScore{\AuxiliaryDataset}{\ReferenceModelset}(\Data, \DataLabel, \MLModel)$ \\ 
    \midrule 
    Attack-P~\cite{ye2022:enhanced} & $\probb{(\AuxiliaryData,\AuxiliaryDataLabel)\in\AuxiliaryDataset}{\frac{\arrComponent{\MLModel(\Data)}{\DataLabel}}{\arrComponent{\MLModel(\AuxiliaryData)}{\AuxiliaryDataLabel}}\geq 1}$ \\ 
    Attack-R~\cite{ye2022:enhanced} & $\probb{\ReferenceModel\in\ReferenceModelset}{\frac{\arrComponent{\MLModel(\Data)}{\DataLabel}}{\arrComponent{\ReferenceModel(\Data)}{\DataLabel}}\geq 1}$ \\
    LiRA~\cite{carlini2022:membership} (offline) & $1-\probb{\ReferenceModel\in\ReferenceModelset}{\log(\frac{\arrComponent{\ReferenceModel(\Data)}{\DataLabel}}{1-\arrComponent{\ReferenceModel(\Data)}{\DataLabel}}) > \log(\frac{\arrComponent{\MLModel(\Data)}{\DataLabel}}{1-\arrComponent{\MLModel(\Data)}{\DataLabel}})}$ \\
    RMIA~\cite{zarifzadeh2024:low} (offline) & $\probb{(\AuxiliaryData,\AuxiliaryDataLabel)\in\AuxiliaryDataset}{\bigg(\frac{\arrComponent{\MLModel(\Data)}{\DataLabel}}{\frac{1}{2}\average_{\ReferenceModel\in\ReferenceModelset}((1+\RMIAParameterAlt)\arrComponent{\MLModel(\Data)}{\DataLabel}+(1-\RMIAParameterAlt))}\bigg)\bigg(\frac{\arrComponent{\MLModel(\AuxiliaryData)}{\AuxiliaryDataLabel}}{\frac{1}{2}\average_{\ReferenceModel\in\ReferenceModelset}((1+\RMIAParameterAlt)\arrComponent{\ReferenceModel(\AuxiliaryData)}{\AuxiliaryDataLabel}+(1-\RMIAParameterAlt))}\bigg)^{-1}\geq \RMIAParameter}$  \\ 
    \bottomrule 
    \end{tabular} 
    \caption[Short caption without TikZ commands]{The computation of $\MIAScore{\AuxiliaryDataset}{\ReferenceModelset}(\Data, \DataLabel,\MLModel)$ by
    membership inference attacks. In RMIA, $\RMIAParameter$ is a hyperparameter that we set $\RMIAParameter=2$ following the previous work~\cite{zarifzadeh2024:low}.
    We set $\RMIAParameterAlt=0.3$ for CIFAR-100 and $\RMIAParameterAlt=0.9$ for TinyImageNet. For each $(\AuxiliaryData,\AuxiliaryDataLabel)\in\AuxiliaryDataset$,
    $\AuxiliaryData$ denotes an auxiliary data instance and $\AuxiliaryDataLabel$ is its associated label. $\arrComponent{\MLModel(\Data)}{\DataLabel}$ denotes
    the confidence score of $\MLModel(\Data)$ associated with $\DataLabel$.
    }
    \label{table:membership_inference}
\end{table*}

\paragraph{Implementation}
For each experiment involving a comparison with baselines, we randomly
divided a dataset's training samples into two equal halves.  One half
(e.g., $25{,}000$ CIFAR-100 training samples or $50{,}000$
TinyImageNet training samples) was used to train the classifier that
we audited, while the other half was used to train reference models
required for membership inference.  The test samples of each dataset
were randomly split into two equal halves: One half was designated as
auxiliary data, while the other half was used to empirically measure
\FalseDetectionRate of a black-box membership inference method.  For
membership inference methods that incorporate data augmentation (e.g.,
LiRA and RMIA), we set the number of augmentations to $2$ in these
methods, following their default
settings~\cite{carlini2022:membership, zarifzadeh2024:low}.  To ensure
a fair comparison, we also set $\NumAugment = 2$ for our method when
benchmarking against these baselines, and limited the baselines'
queries to the audited ML model to be
$\NumAugment\times\NumberMarkedData$ (for $\DatasetSize = 1$).  Since
our data-use detection method allows a data owner to stop querying the
audited ML model early, our method poses $\leq \NumAugment \times
\NumberMarkedData \times \DatasetSize$ queries to each model.
Therefore, under these settings, our method had an equal or lower
query cost than the baselines.

We implemented these membership inference attacks in PyTorch~\cite{paszke2019:pytorch} based on their respective papers and available 
open-source codes.\footnote{https://github.com/tensorflow/privacy/tree/master/research/mi\_lira\_2021}\footnote{https://github.com/privacytrustlab/ml\_privacy\_meter/}
We considered a setting where at most one reference model was accessible, i.e., $\setSize{\ReferenceModelset}=1$.
To simulate the access to an auxiliary dataset $\AuxiliaryDataset$ and 
a reference model $\ReferenceModelset$ ($\setSize{\ReferenceModelset}=1$),
we randomly split the training set of a dataset into two non-overlapping halves, e.g., each half included 
$25{,}000$ CIFAR-100 training samples or $50{,}000$ TinyImageNet training samples.
The first half was used to train an audited ML model $\MLModel$ and the second half was used to train a reference model $\ReferenceModel$.
We also randomly split the test set of a dataset into two non-overlapping halves.
The first half was used as $\AuxiliaryDataset$ and the second half was used as ``a non-member set'' to empirically measure $\FalseDetectionRate$.
In \tblref{table:membership_inference}, the original versions of these methods replace $\arrComponent{\MLModel(\Data)}{\DataLabel}$,
$\arrComponent{\ReferenceModel(\Data)}{\DataLabel}$, $\arrComponent{\MLModel(\AuxiliaryData)}{\AuxiliaryDataLabel}$, 
and $\arrComponent{\ReferenceModel(\AuxiliaryData)}{\AuxiliaryDataLabel}$ 
by a metric (e.g., SM-Taylor-Softmax~\cite{de2015:exploration}) based on model logits (i.e., the outputs before a Softmax layer).
However, in our data-use auditing setting, the model outputs are vectors of confidence scores (i.e., the outputs after the Softmax layer), 
as described in \secref{sec:auditing_ML:classifier}.
Therefore, such a metric is not applicable in a data-use auditing setting. 
We searched for an threshold $\MIAThreshold$ to ensure that the empirical 
$\FalseDetectionRate$ of a membership inference attack was merely below a specified level 
(e.g., $\FalseDetectionRate\leq1\%$). Using the identified $\MIAThreshold$, we then 
calculated $\TrueDetectionRate$ of the membership inference method for the 
specific level of $\FalseDetectionRate$.

\section{Experimental Results} \label{app:results}

\subsection{Examples of Marked Images} \label{app:results:examples}

\begin{figure}[t]
    \centering
    \resizebox{0.46\textwidth}{0.23\textwidth}{%
        \begin{tikzpicture}

            \definecolor{darkgray176}{RGB}{176,176,176}

            \begin{axis}[
                    hide axis,
                    tick align=outside,
                    tick pos=left,
                    x grid style={darkgray176},
                    xmin=-0.5, xmax=127.5,
                    xtick style={color=black},
                    y dir=reverse,
                    y grid style={darkgray176},
                    ymin=-0.5, ymax=63.5,
                    ytick style={color=black}
                ]
                \addplot graphics [includegraphics cmd=\pgfimage,xmin=-0.5, xmax=127.5, ymin=63.5, ymax=-0.5] {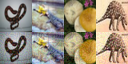};
            \end{axis}

        \end{tikzpicture}
    }
    \caption{Examples of marked CIFAR-100 images ($\MarkBound=10$). First row: raw images; Second row: marked images.}
    \label{fig:cifar100_examples}
\end{figure}

\begin{figure}[t]
    \centering
    \resizebox{0.46\textwidth}{0.23\textwidth}{%
        \begin{tikzpicture}

            \definecolor{darkgray176}{RGB}{176,176,176}

            \begin{axis}[
                    hide axis,
                    tick align=outside,
                    tick pos=left,
                    x grid style={darkgray176},
                    xmin=-0.5, xmax=127.5,
                    xtick style={color=black},
                    y dir=reverse,
                    y grid style={darkgray176},
                    ymin=-0.5, ymax=63.5,
                    ytick style={color=black}
                ]
                \addplot graphics [includegraphics cmd=\pgfimage,xmin=-0.5, xmax=127.5, ymin=63.5, ymax=-0.5] {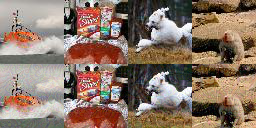};
            \end{axis}

        \end{tikzpicture}
    }
    \caption{Examples of marked TinyImageNet images ($\MarkBound=10$). First row: raw images; Second row: marked images.}
    \label{fig:tinyimagenet_examples}
\end{figure}

\begin{figure}[t]
    \centering
    \resizebox{0.46\textwidth}{0.23\textwidth}{%
        \begin{tikzpicture}

            \definecolor{darkgray176}{RGB}{176,176,176}

            \begin{axis}[
                    hide axis,
                    tick align=outside,
                    tick pos=left,
                    x grid style={darkgray176},
                    xmin=-0.5, xmax=127.5,
                    xtick style={color=black},
                    y dir=reverse,
                    y grid style={darkgray176},
                    ymin=-0.5, ymax=63.5,
                    ytick style={color=black}
                ]
                \addplot graphics [includegraphics cmd=\pgfimage,xmin=-0.5, xmax=127.5, ymin=63.5, ymax=-0.5] {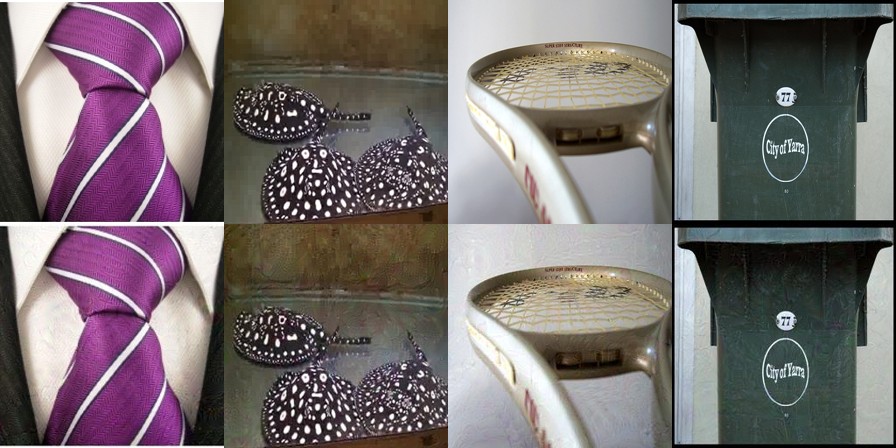};
            \end{axis}

        \end{tikzpicture}
    }
    \caption{Examples of marked ImageNet images ($\MarkBound=25$). First row: raw images; Second row: marked images.}
    \label{fig:imagenet_examples}
\end{figure}

\begin{figure}[t]
    \centering
    \resizebox{0.46\textwidth}{0.575\textwidth}{%
        \begin{tikzpicture}

            \definecolor{darkgray176}{RGB}{176,176,176}

            \begin{axis}[
                    hide axis,
                    tick align=outside,
                    tick pos=left,
                    x grid style={darkgray176},
                    xmin=-0.5, xmax=127.5,
                    xtick style={color=black},
                    y dir=reverse,
                    y grid style={darkgray176},
                    ymin=-0.5, ymax=63.5,
                    ytick style={color=black}
                ]
                \addplot graphics [includegraphics cmd=\pgfimage,xmin=-0.5, xmax=127.5, ymin=63.5, ymax=-0.5] {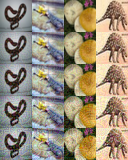};
            \end{axis}

        \end{tikzpicture}
    }
    \caption{Examples of marked CIFAR-100 images with varying $\MarkBound$. First row: raw images; Second row: marked images ($\MarkBound=6$);
        Third row: marked images ($\MarkBound=10$); Fourth row: marked images ($\MarkBound=16$); Last row: marked images ($\MarkBound=20$).}
    \label{fig:eps}
\end{figure}

\begin{figure}[t]
    \centering
    \resizebox{0.46\textwidth}{0.23\textwidth}{%
        \begin{tikzpicture}

            \definecolor{darkgray176}{RGB}{176,176,176}

            \begin{axis}[
                    hide axis,
                    tick align=outside,
                    tick pos=left,
                    x grid style={darkgray176},
                    xmin=-0.5, xmax=127.5,
                    xtick style={color=black},
                    y dir=reverse,
                    y grid style={darkgray176},
                    ymin=-0.5, ymax=63.5,
                    ytick style={color=black}
                ]
                \addplot graphics [includegraphics cmd=\pgfimage,xmin=-0.5, xmax=127.5, ymin=63.5, ymax=-0.5] {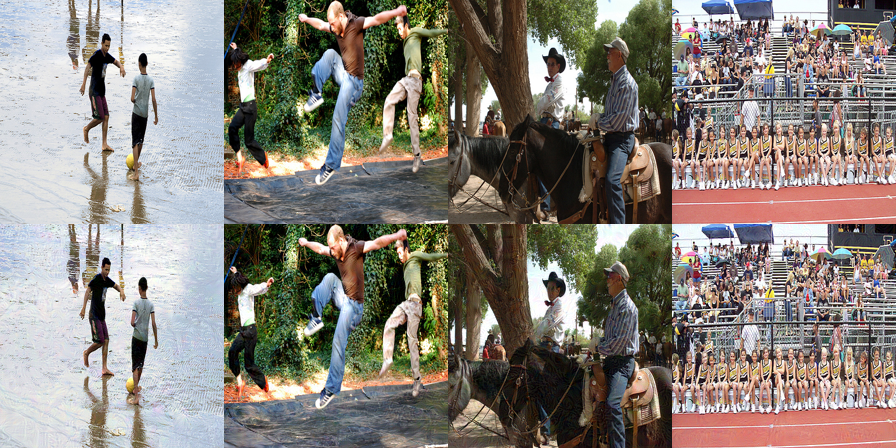};
            \end{axis}

        \end{tikzpicture}
    }
    \caption{Examples of marked Flickr30k images ($\MarkBound=10$). First row: raw images; Second row: marked images.}
    \label{fig:flickr30k_examples}
\end{figure}

We present examples of marked CIFAR-100 images, marked TinyImageNet images, 
and marked Flickr30k images in \figref{fig:cifar100_examples},
\figref{fig:tinyimagenet_examples}, \figref{fig:imagenet_examples}, 
and \figref{fig:flickr30k_examples}, respectively. 
We present examples of marked CIFAR-100 with varying $\MarkBound$
in \figref{fig:eps}.

\subsection{Accuracies of Classifiers Trained on Marked Datasets} \label{app:results:classifier_accuracy}

We report the accuracies of the classifiers trained on the marked
datasets (i.e., the fraction of test samples correctly predicted) and
differences between the accuracies of classifiers trained on marked
datasets and those of classifiers trained on clean datasets, in
\tblref{tbl:classifier:acc}.  The classifiers trained on the marked
datasets preserved good utility, i.e., their $\Accuracy$ values were
similar to those trained on clean datasets, which demonstrated that
the marked data instances preserved good utility of their original
versions.

\begin{table}[ht!]
  \centering
  {\resizebox{0.47\textwidth}{!}{
  \begin{tabular}{@{}lc@{\hspace{0.5em}}cc@{\hspace{0.5em}}cc@{\hspace{0.5em}}c@{}}
  \toprule
   & $\Accuracy (\%)$ & $\AccuracyDiff$ \\ 
  \midrule 
  CIFAR-100 (ResNet-18) & $75.61(\pm0.23)$ & $0.08(\pm0.34)\%$ \\
  TinyImageNet (ResNet-18) & $59.86(\pm0.41)$ & $-0.16(\pm0.60)\%$ \\
  ImageNet (ResNet-50) & $69.08$ & $-0.05\%$\\
  \bottomrule 
  \end{tabular}
  }}
  \caption{Test accuracies of the audited image classifiers and
    differences between accuracies of classifiers trained on marked
    and clean datasets.  For CIFAR-100 and TinyImageNet, results are
    average and standard deviation over $20$ classifiers.  ImageNet
    results are for only one model, due to training cost.}
  \label{tbl:classifier:acc}  
\end{table}

\subsection{Comparison Between Our Method and Baselines}  \label{app:results:comparison_tinyimagenet}

\figref{fig:baselines:cifar100} shows the auditing/inference results
on CIFAR-100: When Attack-R, LiRA, and RMIA used a reference model
similar to the audited model, our method achieved a
$\TrueDetectionRate$ comparable to those of Attack-R, LiRA, and RMIA
under the same $\FalseDetectionRate$ level. However, when the
reference model was not similar enough to the audited one (e.g., by
decreasing $\ReferenceDatasetSize/\TrainDatasetSize$, where
$\TrainDatasetSize$ is the size of the training set of the audited
model, and/or increasing $\ParameterBetaDistribution$), the
performance of these membership inference methods significantly
degraded, which was also confirmed by their
works~\cite{ye2022:enhanced,carlini2022:membership,zarifzadeh2024:low}.
For example, the state-of-the-art membership inference method, namely
RMIA, had \TrueDetectionRate of $15.27\%$, $2.25\%$, and $0\%$ under
$\FalseDetectionRate\leq5\%$, $\FalseDetectionRate\leq1\%$, and
$\FalseDetectionRate\leq0.2\%$, respectively, when we set
$\ReferenceDatasetSize/\TrainDatasetSize=\frac{1}{8}$ and
$\ParameterBetaDistribution=4$, much lower than ours.

The results of comparing our method with baselines on auditing 
TinyImageNet are presented in \figref{fig:baselines:tinyimagenet}.
When Attack-R, LiRA, and RMIA used a reference model
similar to the audited model, our method achieved a
$\TrueDetectionRate$ comparable to those of Attack-R, LiRA, and RMIA
under the same $\FalseDetectionRate$ level. However, when the
reference model was not similar enough to the audited one (e.g., by
decreasing $\ReferenceDatasetSize/\TrainDatasetSize$, where
$\TrainDatasetSize$ is the size of the training set of the audited
model, and/or increasing $\ParameterBetaDistribution$), the
performance of these membership inference methods significantly
degraded, which was also confirmed by their
works~\cite{ye2022:enhanced,carlini2022:membership,zarifzadeh2024:low}.

The performance of these three membership inference methods were
highly affected by the reference models. Of course, as shown in the
previous works
(e.g.,~\cite{ye2022:enhanced,carlini2022:membership,zarifzadeh2024:low}),
when more reference models can be trained and used in membership
inference, these methods would achieve a better inference result
(i.e., a higher \TrueDetectionRate).  However, in a realistic scenario
of data-use auditing, it is costly to train a reference model and
challenging to collect a dataset used to train the reference model
that is similar to the training dataset of the audited model.
Attack-P does not require a reference model but its \TrueDetectionRate
was much lower than ours under the same \FalseDetectionRate. 
More importantly, all these
membership inference methods do not provide a bound on the
\FalseDetectionRate.  This limits the application of membership
inference methods in auditing data-use of ML models, as discussed in
\secref{sec:intro}.

\begin{figure*}[t]
  \begin{center}
  \begin{tabular}{rrccc}
    \toprule
    & $\FalseDetectionRate \le$ & $5\%$ & $1\%$ & $0.2\%$ \\
    \midrule
    \rotatebox{90}{\hspace{-0.5em}Ours} && 30.25 & 12.20 & 3.25 \\[6pt]
    \midrule
    \rotatebox{90}{\hspace{-1.0em}Attack-P} && 7.37 & 0.92 & 0.22 \\[6pt]
    \midrule
    \rotatebox{90}{\hspace{-1.35em}Attack-R}
    &
    \begin{minipage}[b]{3.15em}
    \begin{tabular}{@{}r@{}}
      $\ReferenceDatasetSize/\TrainDatasetSize =$ \\[4pt]
      $\ParameterBetaDistribution = 1$ \\
      $2$ \\
      $3$ \\
      $4$ 
    \end{tabular}
    \end{minipage}
    &
    \begin{minipage}[b]{0.275\textwidth}
      \FPeval{\MinNumber}{13.0}
      \FPeval{\MaxNumber}{34.0}
      \begin{tabular}{*{4}{X}}
        \multicolumn{1}{c}{$1/1$}
        & \multicolumn{1}{c}{$1/2$}
        & \multicolumn{1}{c}{$1/4$}
        & \multicolumn{1}{c}{$1/8$} \\ \midrule
        28.82 & 30.87 & 26.40 & 19.95 \\ 
        27.75 & 28.12 & 24.17 & 16.75 \\ 
        23.27 & 23.47 & 20.10 & 16.87 \\ 
        19.55 & 20.47 & 17.55 & 14.14 \\ 
      \end{tabular}
    \end{minipage}
    &
    \begin{minipage}[b]{0.26\textwidth}
      \FPeval{\MinNumber}{2.0}
      \FPeval{\MaxNumber}{17.0}
      \begin{tabular}{*{4}{X}}
        \multicolumn{1}{c}{$1/1$}
        & \multicolumn{1}{c}{$1/2$}
        & \multicolumn{1}{c}{$1/4$}
        & \multicolumn{1}{c}{$1/8$} \\ \midrule
        14.32 & 13.05 & 6.57 & 3.52 \\ 
        10.82 & 7.65 & 5.02 & 2.72 \\ 
        3.35 & 3.54 & 3.52 & 3.05 \\ 
        2.52 & 2.97 & 3.27 & 2.62 \\
      \end{tabular}
    \end{minipage}
    &
    \begin{minipage}[b]{0.24\textwidth}
      \FPeval{\MinNumber}{0.0}
      \FPeval{\MaxNumber}{7.0}
      \begin{tabular}{*{4}{X}}
        \multicolumn{1}{c}{$1/1$}
        & \multicolumn{1}{c}{$1/2$}
        & \multicolumn{1}{c}{$1/4$}
        & \multicolumn{1}{c}{$1/8$} \\ \midrule
        5.72 & 3.95 & 1.37 & 0.75 \\ 
        2.60 & 1.65 & 0.82 & 0.45 \\ 
        0.52 & 0.89 & 0.62 & 0.72 \\ 
        0.50 & 0.45 & 0.67 & 0.52 \\
      \end{tabular}
    \end{minipage}
    \\
    \midrule
    \rotatebox{90}{\hspace{-0.75em}LiRA}
    &
    \begin{minipage}[b]{3.15em}
    \begin{tabular}{@{}r@{}}
      $\ReferenceDatasetSize/\TrainDatasetSize =$ \\[4pt]
      $\ParameterBetaDistribution = 1$ \\
      $2$ \\
      $3$ \\
      $4$ 
    \end{tabular}
    \end{minipage}
    &
    \begin{minipage}[b]{0.275\textwidth}
      \FPeval{\MinNumber}{13.0}
      \FPeval{\MaxNumber}{34.0}
      \begin{tabular}{*{4}{X}}
        \multicolumn{1}{c}{$1/1$}
        & \multicolumn{1}{c}{$1/2$}
        & \multicolumn{1}{c}{$1/4$}
        & \multicolumn{1}{c}{$1/8$} \\ \midrule
        33.50 & 28.49 & 22.30 & 17.29 \\ 
        29.05 & 25.42 & 20.27 & 15.55 \\ 
        21.27 & 20.82 & 17.67 & 15.80 \\ 
        16.60 & 18.55 & 16.50 & 15.35 \\
      \end{tabular}
    \end{minipage}
    &
    \begin{minipage}[b]{0.26\textwidth}
      \FPeval{\MinNumber}{2.0}
      \FPeval{\MaxNumber}{17.0}
      \begin{tabular}{*{4}{X}}
        \multicolumn{1}{c}{$1/1$}
        & \multicolumn{1}{c}{$1/2$}
        & \multicolumn{1}{c}{$1/4$}
        & \multicolumn{1}{c}{$1/8$} \\ \midrule
        14.00 & 9.95 & 5.75 & 4.62 \\ 
        9.37 & 8.10 & 5.72 & 4.22 \\ 
        4.25 & 4.40 & 4.02 & 3.05 \\ 
        4.25 & 3.75 & 3.65 & 3.05 \\ 
      \end{tabular}
    \end{minipage}
    &
    \begin{minipage}[b]{0.24\textwidth}
      \FPeval{\MinNumber}{0.0}
      \FPeval{\MaxNumber}{7.0}
      \begin{tabular}{*{4}{X}}
        \multicolumn{1}{c}{$1/1$}
        & \multicolumn{1}{c}{$1/2$}
        & \multicolumn{1}{c}{$1/4$}
        & \multicolumn{1}{c}{$1/8$} \\ \midrule
        5.57 & 3.65 & 1.50 & 1.10 \\ 
        2.97 & 2.62 & 1.47 & 1.07 \\ 
        0.92 & 1.12 & 0.87 & 0.77 \\ 
        0.95 & 0.42 & 0.92 & 0.57 \\ 
      \end{tabular}
    \end{minipage}
    \\
    \midrule
    \rotatebox{90}{\hspace{-1em}RMIA}
    &
    \begin{minipage}[b]{3.15em}
    \begin{tabular}{@{}r@{}}
      $\ReferenceDatasetSize/\TrainDatasetSize =$ \\[4pt]
      $\ParameterBetaDistribution = 1$ \\
      $2$ \\
      $3$ \\
      $4$ 
    \end{tabular}
    \end{minipage}
    &
    \begin{subfigure}[b]{0.275\textwidth}
      \FPeval{\MinNumber}{13.0}
      \FPeval{\MaxNumber}{34.0}
      \begin{tabular}{*{4}{X}}
        \multicolumn{1}{c}{$1/1$}
        & \multicolumn{1}{c}{$1/2$}
        & \multicolumn{1}{c}{$1/4$}
        & \multicolumn{1}{c}{$1/8$} \\ \midrule
        31.62 & 30.02 & 24.22 & 14.14 \\ 
        28.95 & 29.37 & 22.40 & 15.52 \\ 
        23.15 & 23.15 & 18.67 & 13.55 \\ 
        18.05 & 18.95 & 15.05 & 15.27 \\ 
      \end{tabular}
    \end{subfigure}
    &
    \begin{minipage}[b]{0.26\textwidth}
      \FPeval{\MinNumber}{2.0}
      \FPeval{\MaxNumber}{17.0}
      \begin{tabular}{*{4}{X}}
        \multicolumn{1}{c}{$1/1$}
        & \multicolumn{1}{c}{$1/2$}
        & \multicolumn{1}{c}{$1/4$}
        & \multicolumn{1}{c}{$1/8$} \\ \midrule
        16.20 & 10.40 & 5.8 & 2.54 \\ 
        10.07 & 7.07 & 4.17 & 3.05 \\ 
        4.65 & 4.25 & 2.14 & 2.42 \\ 
        4.05 & 4.42 & 3.40 & 2.25 \\ 
      \end{tabular}
    \end{minipage}
    &
    \begin{minipage}[b]{0.24\textwidth}
      \FPeval{\MinNumber}{0.0}
      \FPeval{\MaxNumber}{7.0}
      \begin{tabular}{*{4}{X}}
        \multicolumn{1}{c}{$1/1$}
        & \multicolumn{1}{c}{$1/2$}
        & \multicolumn{1}{c}{$1/4$}
        & \multicolumn{1}{c}{$1/8$} \\ \midrule
        6.67 & 2.50 & 0.60 & 0.00 \\ 
        2.87 & 1.57 & 0.00 & 0.67 \\ 
        0.95 & 1.02 & 0.00 & 0.00 \\ 
        0.65 & 0.00 & 0.27 & 0.00 \\
      \end{tabular}
    \end{minipage}
    \\
    \bottomrule
  \end{tabular}
  \caption{Overall comparison of $\TrueDetectionRate (\%)$ ($\DatasetSize=1$) across
    Attack-P, Attack-R, LiRA, and RMIA on CIFAR-100.  Results are averaged over
    $250\times20$ detections. We trained $20$ WideResNet-28-2
    classifiers (WideResNet-28-2 is default model architecture in the previous works (e.g.,~\cite{carlini2022:membership})), 
    in each of which $250$ training samples of CIFAR-100
    were audited.  Lighter colors indicate larger improvement of our
    technique over these membership inference methods.}
  \label{fig:baselines:cifar100}
  \end{center}
\end{figure*}

\begin{figure*}[t]
    \begin{center}
    \begin{tabular}{rrccc}
      \toprule
      & $\FalseDetectionRate \le$ & $5\%$ & $1\%$ & $0.2\%$ \\
      \midrule
      \rotatebox{90}{\hspace{-0.5em}Ours} && 16.06 & 5.03 & 0.94 \\[6pt]
      \midrule
      \rotatebox{90}{\hspace{-1.0em}Attack-P} && 8.46 & 1.47 & 0.33 \\[6pt]
      \midrule
      \rotatebox{90}{\hspace{-1.35em}Attack-R}
      &
      \begin{minipage}[b]{3.15em}
      \begin{tabular}{@{}r@{}}
        $\ReferenceDatasetSize/\TrainDatasetSize =$ \\[4pt]
        $\ParameterBetaDistribution = 1$ \\
        $2$ \\
        $3$ \\
        $4$ 
      \end{tabular}
      \end{minipage}
      &
      \begin{minipage}[b]{0.275\textwidth}
        \FPeval{\MinNumber}{7.0}
        \FPeval{\MaxNumber}{14.0}
        \begin{tabular}{*{4}{X}}
          \multicolumn{1}{c}{$1/1$}
          & \multicolumn{1}{c}{$1/2$}
          & \multicolumn{1}{c}{$1/4$}
          & \multicolumn{1}{c}{$1/8$} \\ \midrule
          12.52 & 12.08 & 10.05 & 8.11 \\ 
          11.75 & 11.71 & 9.31 & 8.11 \\ 
          10.27 & 10.37 & 8.83 & 7.76 \\ 
          9.22 & 9.15 & 8.41 & 7.96 \\ 
        \end{tabular}
      \end{minipage}
      &
      \begin{minipage}[b]{0.26\textwidth}
        \FPeval{\MinNumber}{1.0}
        \FPeval{\MaxNumber}{5.0}
        \begin{tabular}{*{4}{X}}
          \multicolumn{1}{c}{$1/1$}
          & \multicolumn{1}{c}{$1/2$}
          & \multicolumn{1}{c}{$1/4$}
          & \multicolumn{1}{c}{$1/8$} \\ \midrule
          3.30 & 3.02 & 2.35 & 1.67 \\ 
          2.54 & 2.40 & 2.06 & 1.86 \\ 
          1.65 & 1.65 & 1.72 & 1.68 \\ 
          1.43 & 1.48 & 1.65 & 1.52 \\ 
        \end{tabular}
      \end{minipage}
      &
      \begin{minipage}[b]{0.24\textwidth}
        \FPeval{\MinNumber}{0.0}
        \FPeval{\MaxNumber}{1.5}
        \begin{tabular}{*{4}{X}}
          \multicolumn{1}{c}{$1/1$}
          & \multicolumn{1}{c}{$1/2$}
          & \multicolumn{1}{c}{$1/4$}
          & \multicolumn{1}{c}{$1/8$} \\ \midrule
          1.08 & 0.62 & 0.70 & 0.31 \\ 
          0.41 & 0.47 & 0.42 & 0.40 \\ 
          0.27 & 0.27 & 0.35 & 0.26 \\ 
          0.26 & 0.20 & 0.33 & 0.26 \\
        \end{tabular}
      \end{minipage}
      \\
      \midrule
      \rotatebox{90}{\hspace{-0.75em}LiRA}
      &
      \begin{minipage}[b]{3.15em}
      \begin{tabular}{@{}r@{}}
        $\ReferenceDatasetSize/\TrainDatasetSize =$ \\[4pt]
        $\ParameterBetaDistribution = 1$ \\
        $2$ \\
        $3$ \\
        $4$ 
      \end{tabular}
      \end{minipage}
      &
      \begin{minipage}[b]{0.275\textwidth}
        \FPeval{\MinNumber}{7.0}
        \FPeval{\MaxNumber}{14.0}
        \begin{tabular}{*{4}{X}}
          \multicolumn{1}{c}{$1/1$}
          & \multicolumn{1}{c}{$1/2$}
          & \multicolumn{1}{c}{$1/4$}
          & \multicolumn{1}{c}{$1/8$} \\ \midrule
          13.60 & 12.85 & 10.72 & 8.59 \\ 
          11.22 & 11.07 & 9.77 & 8.66 \\ 
          8.72 & 9.60 & 9.58 & 7.78 \\ 
          8.08 & 8.23 & 8.02 & 7.71 \\ 
        \end{tabular}
      \end{minipage}
      &
      \begin{minipage}[b]{0.26\textwidth}
        \FPeval{\MinNumber}{1.0}
        \FPeval{\MaxNumber}{5.0}
        \begin{tabular}{*{4}{X}}
          \multicolumn{1}{c}{$1/1$}
          & \multicolumn{1}{c}{$1/2$}
          & \multicolumn{1}{c}{$1/4$}
          & \multicolumn{1}{c}{$1/8$} \\ \midrule
          3.95 & 2.96 & 2.38 & 1.71 \\ 
          2.40 & 1.93 & 1.60 & 1.62 \\ 
          1.40 & 1.62 & 1.46 & 1.70 \\ 
          1.35 & 1.40 & 1.47 & 1.62 \\ 
        \end{tabular}
      \end{minipage}
      &
      \begin{minipage}[b]{0.24\textwidth}
        \FPeval{\MinNumber}{0.0}
        \FPeval{\MaxNumber}{1.5}
        \begin{tabular}{*{4}{X}}
          \multicolumn{1}{c}{$1/1$}
          & \multicolumn{1}{c}{$1/2$}
          & \multicolumn{1}{c}{$1/4$}
          & \multicolumn{1}{c}{$1/8$} \\ \midrule
          0.97 & 0.58 & 0.41 & 0.32 \\ 
          0.41 & 0.37 & 0.38 & 0.21 \\ 
          0.16 & 0.37 & 0.42 & 0.31 \\ 
          0.25 & 0.25 & 0.27 & 0.28 \\ 
        \end{tabular}
      \end{minipage}
      \\
      \midrule
      \rotatebox{90}{\hspace{-1em}RMIA}
      &
      \begin{minipage}[b]{3.15em}
      \begin{tabular}{@{}r@{}}
        $\ReferenceDatasetSize/\TrainDatasetSize =$ \\[4pt]
        $\ParameterBetaDistribution = 1$ \\
        $2$ \\
        $3$ \\
        $4$ 
      \end{tabular}
      \end{minipage}
      &
      \begin{subfigure}[b]{0.275\textwidth}
        \FPeval{\MinNumber}{7.0}
        \FPeval{\MaxNumber}{14.0}
        \begin{tabular}{*{4}{X}}
          \multicolumn{1}{c}{$1/1$}
          & \multicolumn{1}{c}{$1/2$}
          & \multicolumn{1}{c}{$1/4$}
          & \multicolumn{1}{c}{$1/8$} \\ \midrule
          13.56 & 13.41 & 10.67 & 9.96 \\ 
          13.30 & 11.40 & 10.50 & 8.96 \\ 
          10.28 & 10.78 & 8.66 & 8.72 \\ 
          9.47 & 9.42 & 8.69 & 8.56 \\ 
        \end{tabular}
      \end{subfigure}
      &
      \begin{minipage}[b]{0.26\textwidth}
        \FPeval{\MinNumber}{1.0}
        \FPeval{\MaxNumber}{5.0}
        \begin{tabular}{*{4}{X}}
          \multicolumn{1}{c}{$1/1$}
          & \multicolumn{1}{c}{$1/2$}
          & \multicolumn{1}{c}{$1/4$}
          & \multicolumn{1}{c}{$1/8$} \\ \midrule
          4.18 & 3.26 & 1.98 & 2.06 \\ 
          3.46 & 2.38 & 1.90 & 1.67 \\ 
          1.92 & 2.18 & 1.40 & 1.72 \\ 
          1.78 & 1.90 & 1.31 & 1.51 \\
        \end{tabular}
      \end{minipage}
      &
      \begin{minipage}[b]{0.24\textwidth}
        \FPeval{\MinNumber}{0.0}
        \FPeval{\MaxNumber}{1.5}
        \begin{tabular}{*{4}{X}}
          \multicolumn{1}{c}{$1/1$}
          & \multicolumn{1}{c}{$1/2$}
          & \multicolumn{1}{c}{$1/4$}
          & \multicolumn{1}{c}{$1/8$} \\ \midrule
          1.41 & 0.86 & 0.56 & 0.11 \\ 
          0.53 & 0.37 & 0.43 & 0.37 \\ 
          0.33 & 0.22 & 0.22 & 0.50 \\ 
          0.26 & 0.21 & 0.28 & 0.42 \\ 
        \end{tabular}
      \end{minipage}
      \\
      \bottomrule
    \end{tabular}

    \caption{Overall comparison of $\TrueDetectionRate (\%)$
      ($\DatasetSize=1$) across Attack-P, Attack-R, LiRA, and RMIA on
      TinyImageNet.  Results are averaged over $500\times20$
      detections. We trained $20$ ResNet-18 classifiers, in each of
      which $500$ training samples of TinyImageNet were audited.
      Lighter colors indicate larger improvement of our technique over
      these membership inference methods.}
      \label{fig:baselines:tinyimagenet}
    \end{center}
  \end{figure*}

\subsection{Robustness to Countermeatures}  \label{app:results:robustness}

We consider two image perturbation methods (namely adding Gaussian noise
and remarking), three image denoising methods (namely Gaussian smoothing,
median smoothing, and general smoothing), DPSGD, and strong regularization
as countermeatures.

\paragraph{Image perturbation}
We study the effectiveness of our data-use auditing method when the ML
practitioner perturbs the training samples he collected before their
use in training.  We consider two types of perturbation: (1) adding
Gaussian noise (parameterized by its standard deviation
$\GaussianNoiseStd$) and (2) applying our marking method (using the
default hyperparameters) to add additional perturbation, denoted as
``remarking''. Our results are shown in
\figref{fig:attack:cifar100} and \tblref{tbl:classifier:remark_smooth}.  
Both perturbation methods decreased our
$\TrueDetectionRate$: ``Remarking'' decreased $\TrueDetectionRate$ ($\DatasetSize=1$)
from $28.21\%$ to $12.60\%$ under $\FalseDetectionRate\leq5\%$ but the test accuracy on average dropped
by $3.90$ percentage points.  Adding Gaussian noise with a larger
$\GaussianNoiseStd$ made our $\TrueDetectionRate$ ($\DatasetSize=1$) closer to the
$\FalseDetectionRate$ level but it also sacrificed more model
utility. For example, adding Gaussian noise with
$\GaussianNoiseStd=25$ decreased $\TrueDetectionRate$ ($\DatasetSize=1$) to a level close
to $\FalseDetectionRate$ but at a cost of $10.11$ percentage points to
model utility.

When the data owner had more data instances (i.e., $\DatasetSize > 1$)
and all of them were used in training, our method became more
effective, i.e., our $\TrueDetectionRate$ increased with
$\DatasetSize$.  For example, when Gaussian noise
with $\GaussianNoiseStd=25$ was applied to perturb training images, our
$\TrueDetectionRate$ achieved $13.70\%$ ($\DatasetSize=8$) and
$47.21\%$ ($\DatasetSize=64$) under $\FalseDetectionRate\leq5\%$
(compared with $7.88\%$ for $\DatasetSize=1$).
When ``remarking'' was applied to perturb training images, our
$\TrueDetectionRate$ achieved $35.16\%$ ($\DatasetSize=8$) and
$97.66\%$ ($\DatasetSize=64$) under $\FalseDetectionRate\leq5\%$
(compared with $12.60\%$ for $\DatasetSize=1$).

\begin{figure}[ht!]
  \centering
  
\begin{subfigure}[b]{0.47\textwidth}
  \centering
  \resizebox{!}{3em}{\newenvironment{customlegend}[1][]{%
    \begingroup
    \csname pgfplots@init@cleared@structures\endcsname
    \pgfplotsset{#1}%
}{%
    \csname pgfplots@createlegend\endcsname
    \endgroup
}%

\def\addlegendimage{\csname pgfplots@addlegendimage\endcsname}

\begin{tikzpicture}

\definecolor{darkslategray38}{RGB}{38,38,38}
\definecolor{indianred1967882}{RGB}{196,78,82}
\definecolor{lightgray204}{RGB}{204,204,204}
\definecolor{mediumseagreen85168104}{RGB}{85,168,104}
\definecolor{peru22113282}{RGB}{221,132,82}
\definecolor{steelblue76114176}{RGB}{76,114,176}

\begin{customlegend}[
    legend style={{font={\small}},{draw=none}}, 
    legend columns=2,
    legend cell align={center},
    legend entries={{$\TrueDetectionRate (\DatasetSize=1)$}, {$\TrueDetectionRate (\DatasetSize=8)$}, {$\TrueDetectionRate (\DatasetSize=64)$}, {$\FalseDetectionRate$ bound}}]
\addlegendimage{very thick, color=steelblue76114176,mark=o}
\addlegendimage{very thick, peru22113282,mark=diamond,dashed,mark options={solid}}
\addlegendimage{very thick, mediumseagreen85168104,mark=triangle,dotted,mark options={solid}}
\addlegendimage{very thick, darkgray,dashdotted}
\addlegendimage{very thick, indianred1967882,mark=square,densely dashed,mark options={solid}}

\end{customlegend}

\end{tikzpicture}}
\end{subfigure}

  \begin{subfigure}[b]{0.235\textwidth}
      \centering
      \begin{tikzpicture}
          \begin{axis}[
              xlabel={$\GaussianNoiseStd$},
              xtick={1, 2, 3, 4, 5, 6},
              xticklabels={0, 5, 10, 15, 20, 25},
              ymode=log,
              log basis y=10, 
              ymin=0.07, ymax=140,
              ytick={0.1, 1, 10, 100},
              grid=major,
              width=1.0\textwidth,
              height=0.8\textwidth,
              ]
              \addplot[color=steelblue76114176,mark=o,line width=1.5] coordinates {
                (1, 28.21)
                (2, 20.058400000000002) 
                (3, 13.277599999999998) 
                (4, 9.968200000000005) 
                (5, 8.528500000000001) 
                (6, 7.889200000000002)  
              };
              \addplot[color=peru22113282,mark=diamond,dashed,mark options={solid},line width=1.5] coordinates {
                  (1,97.91)
                  (2,63.14)
                  (3,37.06)
                  (4,23.47)
                  (5,16.37)
                  (6,13.7)
              };
              \addplot[color=mediumseagreen85168104,mark=triangle,dotted,mark options={solid},line width=1.5] coordinates {
                  (1,100)
                  (2,100.0)
                  (3,97.86)
                  (4,82.98)
                  (5,61.56)
                  (6,47.21)
              };
              \addplot[color=darkgray,dashdotted,line width=1.5] coordinates {
                  (1, 5)
                  (2, 5)
                  (3, 5)
                  (4, 5)
                  (5, 5)
                  (6, 5)
              };
          \end{axis}
      \end{tikzpicture}
      \caption{$\TrueDetectionRate (\%) @ \FalseDetectionRate \leq 5\%$}
      \label{fig:attack:cifar100:5fdr}
  \end{subfigure}
  \hfill
  \begin{subfigure}[b]{0.235\textwidth}
      \centering
      \begin{tikzpicture}
          \begin{axis}[
              xlabel={$\GaussianNoiseStd$ },
              xtick={1, 2, 3, 4, 5, 6},
              xticklabels={0, 5, 10, 15, 20, 25},
              ymode=log,
              log basis y=10, 
              ymin=0.07, ymax=140,
              ytick={0.1, 1, 10, 100},
              grid=major,
              width=1.0\textwidth,
              height=0.8\textwidth,
              ]
              \addplot[color=steelblue76114176,mark=o,line width=1.5] coordinates {
                (1, 11.60)
                (2, 7.336400000000001) 
                (3, 3.975800000000001) 
                (4, 2.487800000000001) 
                (5, 1.9983000000000009) 
                (6, 1.5386000000000002)
              };
              \addplot[color=peru22113282,mark=diamond,dashed,mark options={solid},line width=1.5] coordinates {
                  (1, 91.2)
                  (2, 35.51)
                  (3, 14.6)
                  (4, 7.25)
                  (5, 4.8)
                  (6, 3.31)
              };
              \addplot[color=mediumseagreen85168104,mark=triangle,dotted,mark options={solid},line width=1.5] coordinates {
                    (1, 100)
                    (2, 99.95)
                    (3, 90.9)
                    (4, 59.15)
                    (5, 34.21)
                    (6, 21.59)
                };
                \addplot[color=darkgray,dashdotted,line width=1.5] coordinates {
                    (1, 1)
                    (2, 1)
                    (3, 1)
                    (4, 1)
                    (5, 1)
                    (6, 1)
                };
          \end{axis}
      \end{tikzpicture}
      \caption{$\TrueDetectionRate (\%) @ \FalseDetectionRate \leq 1\%$}
      \label{fig:attack:cifar100:1fdr}
  \end{subfigure}
  \\
  \begin{subfigure}[b]{0.235\textwidth}
      \centering
      \begin{tikzpicture}
          \begin{axis}[
              xlabel={$\GaussianNoiseStd$ },
              xtick={1, 2, 3, 4, 5, 6},
              xticklabels={0, 5, 10, 15, 20, 25},
              ymode=log,
              log basis y=10, 
              ymin=0.07, ymax=140,
              ytick={0.1, 1, 10, 100},
              grid=major,
              width=1.0\textwidth,
              height=0.8\textwidth,
              ]
              \addplot[color=steelblue76114176,mark=o,line width=1.5] coordinates {
                (1, 3.12)
                (2, 1.66) 
                (3, 0.5300000000000001) 
                (4, 0.3500000000000001) 
                (5, 0.26000000000000006) 
                (6, 0.16000000000000003)   
              };
              \addplot[color=peru22113282,mark=diamond,dashed,mark options={solid},line width=1.5] coordinates {
                  (1,73.48)
                  (2,13.75)
                  (3,3.76)
                  (4,1.41)
                  (5,0.85)
                  (6,0.4)
              };
              \addplot[color=mediumseagreen85168104,mark=triangle,dotted,mark options={solid},line width=1.5] coordinates {
                    (1,100)
                    (2,99.38)
                    (3,72.81)
                    (4,31.11)
                    (5,12.6)
                    (6,7.05)
                };
                \addplot[color=darkgray,dashdotted,line width=1.5] coordinates {
                    (1, 0.2)
                    (2, 0.2)
                    (3, 0.2)
                    (4, 0.2)
                    (5, 0.2)
                    (6, 0.2)
                };
          \end{axis}
      \end{tikzpicture}
      \caption{$\TrueDetectionRate (\%) @ \FalseDetectionRate \leq 0.2\%$}
      \label{fig:attack:cifar100:0.2fdr}
  \end{subfigure}
  \hfill
  \begin{subfigure}[b]{0.235\textwidth}
      \centering
      \begin{tikzpicture}
          \begin{axis}[
              xlabel={$\GaussianNoiseStd$ },
              xtick={1, 2, 3, 4, 5, 6},
              xticklabels={0, 5, 10, 15, 20, 25},
              ymin=60, ymax=80,
              grid=major,
              width=1.0\textwidth,
              height=0.8\textwidth,
              ]
              \addplot[color=indianred1967882,mark=square,densely dashed,mark options={solid},line width=1.5] coordinates {
                (1, 75.61)
                (2, 73.6245) 
                (3, 71.09299999999999) 
                (4, 68.807) 
                (5, 66.67500000000001) 
                (6, 64.501) 
              };
          \end{axis}
      \end{tikzpicture}
      \caption{Test accuracy $\Accuracy (\%)$}
      \label{fig:attack:cifar100:acc}
  \end{subfigure} 
  \caption{$\TrueDetectionRate (\%)$ of our
    auditing method for CIFAR-100 image classifiers under data
    perturbation using Gaussian noises (parameterized by standard deviation
    $\GaussianNoiseStd$)
    (\subfigsref{fig:attack:cifar100:5fdr}{fig:attack:cifar100:0.2fdr})
    and test accuracies of image classifiers
    (\figref{fig:attack:cifar100:acc}).}
  \label{fig:attack:cifar100}
\end{figure}

\paragraph{Image denoising}  \label{app:robustness:denoising}
We consider three image denoising methods, namely Gaussian smoothing,
median smoothing, and general smoothing. For Gaussian smoothing, we
set the kernel size to be $1$; For median smoothing, we set the
window size to be $3$. Our auditing results are shown in \tblref{tbl:classifier:remark_smooth}.
Image smoothing decreased our $\TrueDetectionRate$ but significantly
decreased the utility of the trained models. For example, general
smoothing reduced $\TrueDetectionRate$ ($\DatasetSize=1$) to $14.49\%$
under $\FalseDetectionRate\leq5\%$ but \Accuracy dropped by around
$11$ percentage points.

Again, when the data owner had more data instances (i.e., $\DatasetSize > 1$)
and all of them were used in training, our method became more
effective, i.e., our $\TrueDetectionRate$ increased with
$\DatasetSize$.

\begin{table}[ht!]
  \centering
  {\resizebox{0.47\textwidth}{!}{
  \begin{tabular}{@{}lc@{\hspace{0.5em}}cc@{\hspace{0.5em}}cc@{\hspace{0.5em}}cc@{\hspace{0.5em}}cc@{\hspace{0.5em}}c@{}}
  \toprule
   & \multicolumn{1}{c}{\multirow{2}{*}{$\Accuracy \%$}}& \multicolumn{1}{c}{\multirow{2}{*}{$\DatasetSize$}}& \multicolumn{3}{c}{$\FalseDetectionRate\leq$} \\ 
  & & & \multicolumn{1}{c}{$5\%$ }& \multicolumn{1}{c}{$1\%$ }& \multicolumn{1}{c}{$0.2\%$ }\\
  \midrule
  \multirow{3}{*}{No perturbation} & \multirow{3}{*}{$75.61$} & $1$ & $28.21$ &$11.59$ &$3.11$ \\
                                    & & $16$ & $97.91$ &$91.20$ &$73.48$ \\
                                    & & $64$ & $100.00$ &$100.00$ &$100.00$ \\
  \multirow{3}{*}{Remark} & \multirow{3}{*}{$71.70$} & $1$ & $12.60$ &$3.73$ &$0.58$ \\
                                    & & $16$ & $35.16$ &$13.75$ &$3.41$ \\
                                    & & $64$ & $97.66$ &$89.54$ &$70.27$ \\
  \multirow{3}{*}{Gaussian smooth} & \multirow{3}{*}{$39.51$} & $1$ & $7.57$ &$1.48$ &$0.19$ \\
                                            & & $16$ & $13.36$ &$3.33$ &$0.58$ \\
                                            & & $64$ & $46.45$ &$21.22$ &$6.51$ \\
  \multirow{3}{*}{Median smooth} & \multirow{3}{*}{$57.36$} & $1$ & $9.83$ &$2.39$ &$0.33$ \\
                                            & & $16$ & $24.72$ &$7.99$ &$1.41$ \\
                                            & & $64$ & $85.45$ &$63.04$ &$35.34$ \\
  \multirow{3}{*}{General smooth} & \multirow{3}{*}{$64.36$} & $1$ & $14.49$ &$3.78$ &$0.68$ \\
                                            & & $16$ & $44.55$ &$19.95$ &$5.71$ \\
                                            & & $64$ & $99.59$ &$97.59$ &$88.22$ \\
  \bottomrule 
  \end{tabular}
  }}
  \caption{$\TrueDetectionRate (\%)$ of our method
    after remarking or image smoothing when applied to audit the use of CIFAR-100 in 
    image classifier (ResNet-18). Results are
    averaged over $500\times20$ detections for CIFAR-100. We trained $20$
    classifiers, in each of which $500$ CIFAR-100 
    training samples were audited.}
  \label{tbl:classifier:remark_smooth}
\end{table}

\paragraph{DPSGD}
We study the effectiveness of our data-use auditing method on
classifiers trained using differentially private stochastic gradient
descent (DPSGD)~\cite{abadi2016:deep,hong2020:effectiveness}. DPSGD is
the state-of-the-art private learning algorithm reducing the
memorization and privacy leakage of training
samples~\cite{aerni2024:evaluations}, and thus it can be considered as
an attack to a data-use auditing method~\cite{huang2024:auditdata}.
It works by clipping the norm of the gradients and adding Gaussian
noise parameterized by a standard deviation (i.e., noise multiplier)
$\DPnoiseStd$ into gradients during training.  Our results on auditing
CIFAR-100 classifiers trained by DPSGD are presented in
\figref{fig:classifier:dpsgd}. As in \figref{fig:classifier:dpsgd},
when we set a larger $\DPnoiseStd$, the trained classifier memorized
its training samples less and thus our \TrueDetectionRate ($\DatasetSize=1$) decreased
under the same level of \FalseDetectionRate.  For example, under
$\FalseDetectionRate\leq5\%$, our \TrueDetectionRate ($\DatasetSize=1$) decreased from
$28.21\%$ to $9.21\%$ when we increased $\DPnoiseStd$ from $0$ to
$2\times10^{-3}$ ($\DPnoiseStd=0$ corresponds to the non-private
setting). However, the accuracy $\Accuracy$ of the trained classifiers
on the test samples decreased from $75.53\%$ to $65.82\%$ as
$\DPnoiseStd$ grew. We hypothesize that data-use auditing can be
applied to audit differential privacy in ML models; i.e., we could
audit the differential privacy guarantee of an ML model according to
our detection results.  We leave the exploration of the theoretical
relationship between differential privacy guarantees and our auditing
method as a direction for future work.

When the data owner had more data instances (i.e., $\DatasetSize > 1$)
and all of them were used in training, our method became more
effective, i.e., our $\TrueDetectionRate$ increased with
$\DatasetSize$, even when the ML practitioner applied a strong
privacy-preserving training method.  For example, when DPSGD with
$\DPnoiseStd=2\times10^{-3}$ was used to train the ML model, our
$\TrueDetectionRate$ achieved $24.34\%$ ($\DatasetSize=8$) and
$85.34\%$ ($\DatasetSize=64$) under $\FalseDetectionRate\leq5\%$
(compared with $9.21\%$ for $\DatasetSize=1$).

\begin{figure}[ht!]
  \centering

  \begin{subfigure}[b]{0.47\textwidth}
    \centering
    \resizebox{!}{3em}{\newenvironment{customlegend}[1][]{%
    \begingroup
    \csname pgfplots@init@cleared@structures\endcsname
    \pgfplotsset{#1}%
}{%
    \csname pgfplots@createlegend\endcsname
    \endgroup
}%

\def\addlegendimage{\csname pgfplots@addlegendimage\endcsname}

\begin{tikzpicture}

\definecolor{darkslategray38}{RGB}{38,38,38}
\definecolor{indianred1967882}{RGB}{196,78,82}
\definecolor{lightgray204}{RGB}{204,204,204}
\definecolor{mediumseagreen85168104}{RGB}{85,168,104}
\definecolor{peru22113282}{RGB}{221,132,82}
\definecolor{steelblue76114176}{RGB}{76,114,176}

\begin{customlegend}[
    legend style={{font={\small}},{draw=none}}, 
    legend columns=2,
    legend cell align={center},
    legend entries={{$\TrueDetectionRate (\DatasetSize=1)$}, {$\TrueDetectionRate (\DatasetSize=8)$}, {$\TrueDetectionRate (\DatasetSize=64)$}, {$\FalseDetectionRate$ bound}}]
\addlegendimage{very thick, color=steelblue76114176,mark=o}
\addlegendimage{very thick, peru22113282,mark=diamond,dashed,mark options={solid}}
\addlegendimage{very thick, mediumseagreen85168104,mark=triangle,dotted,mark options={solid}}
\addlegendimage{very thick, darkgray,dashdotted}
\addlegendimage{very thick, indianred1967882,mark=square,densely dashed,mark options={solid}}

\end{customlegend}

\end{tikzpicture}}
  \end{subfigure}

  \begin{subfigure}[b]{0.235\textwidth}
      \centering
      \begin{tikzpicture}
          \begin{axis}[
              xlabel={$\DPnoiseStd (\times 10^{-3})$},
              xtick={1, 2, 3, 4, 5},
              xticklabels={0, $0.5$, $1.0$, $1.5$, $2.0$},
              ymode=log,
              log basis y=10, 
              ymin=0.07, ymax=140,
              ytick={0.1, 1, 10, 100},
              grid=major,
              width=1.0\textwidth,
              height=0.8\textwidth,
              ]
              \addplot[color=steelblue76114176,mark=o,line width=1.5] coordinates {
                  (1, 28.21)
                  (2, 26.73)
                  (3, 16.65)
                  (4, 11.51)
                  (5, 9.21)
              };
              \addplot[color=peru22113282,mark=diamond,dashed,mark options={solid},line width=1.5] coordinates {
                  (1,83.7)
                  (2,83.02)
                  (3,61.12)
                  (4,36.18)
                  (5,24.34)
              };
              \addplot[color=mediumseagreen85168104,mark=triangle,dotted,mark options={solid},line width=1.5] coordinates {
                  (1,100.0)
                  (2,100.0)
                  (3,99.99)
                  (4,98.27)
                  (5,85.34)
              };
              \addplot[color=darkgray,dashdotted,line width=1.5] coordinates {
                    (1, 5)
                    (2, 5)
                    (3, 5)
                    (4, 5)
                    (5, 5)
                };
          \end{axis}
      \end{tikzpicture}
      \caption{$\TrueDetectionRate (\%) @ \FalseDetectionRate \leq 5\%$}
      \label{fig:classifier:dpsgd:5fdr}
  \end{subfigure}
  \hfill
  \begin{subfigure}[b]{0.235\textwidth}
      \centering
      \begin{tikzpicture}
          \begin{axis}[
              xlabel={$\DPnoiseStd (\times 10^{-3})$},
              xtick={1, 2, 3, 4, 5},
              xticklabels={0, $0.5$, $1.0$, $1.5$, $2.0$},
              ymode=log,
              log basis y=10, 
              ymin=0.07, ymax=140,
              ytick={0.1, 1, 10, 100},
              grid=major,
              width=1.0\textwidth,
              height=0.8\textwidth,
              ]
              \addplot[color=steelblue76114176,mark=o,line width=1.5] coordinates {
                  (1, 11.60)
                  (2, 10.03)
                  (3, 4.57)
                  (4, 2.68)
                  (5, 2.16)
              };
              \addplot[color=peru22113282,mark=diamond,dashed,mark options={solid},line width=1.5] coordinates {
                  (1,60.81)
                  (2,59.14)
                  (3,32.44)
                  (4,13.68)
                  (5,7.27)
              };
              \addplot[color=mediumseagreen85168104,mark=triangle,dotted,mark options={solid},line width=1.5] coordinates {
                  (1,100.0)
                  (2,100.0)
                  (3,99.83)
                  (4,91.88)
                  (5,63.53)
              };
              \addplot[color=darkgray,dashdotted,line width=1.5] coordinates {
                    (1, 1)
                    (2, 1)
                    (3, 1)
                    (4, 1)
                    (5, 1)
                };
          \end{axis}
      \end{tikzpicture}
      \caption{$\TrueDetectionRate (\%) @ \FalseDetectionRate \leq 1\%$}
      \label{fig:classifier:dpsgd:1fdr}
  \end{subfigure}
  \\
  \begin{subfigure}[b]{0.235\textwidth}
      \centering
      \begin{tikzpicture}
          \begin{axis}[
              xlabel={$\DPnoiseStd (\times 10^{-3})$},
              xtick={1, 2, 3, 4, 5},
              xticklabels={0, $0.5$, $1.0$, $1.5$, $2.0$},
              ymode=log,
              log basis y=10, 
              ymin=0.07, ymax=140,
              ytick={0.1, 1, 10, 100},
              grid=major,
              width=1.0\textwidth,
              height=0.8\textwidth,
              ]
              \addplot[color=steelblue76114176,mark=o,line width=1.5] coordinates {
                  (1, 3.12)
                  (2, 2.31)
                  (3, 0.67)
                  (4, 0.33)
                  (5, 0.32)
              };
              \addplot[color=peru22113282,mark=diamond,dashed,mark options={solid},line width=1.5] coordinates {
                  (1,33.1)
                  (2,30.34)
                  (3,11.02)
                  (4,3.34)
                  (5,1.48)
              };
              \addplot[color=mediumseagreen85168104,mark=triangle,dotted,mark options={solid},line width=1.5] coordinates {
                  (1,100.0)
                  (2,100.0)
                  (3,99.2)
                  (4,74.6)
                  (5,34.49)
              };
              \addplot[color=darkgray,dashdotted,line width=1.5] coordinates {
                    (1, 0.2)
                    (2, 0.2)
                    (3, 0.2)
                    (4, 0.2)
                    (5, 0.2)
                };
          \end{axis}
      \end{tikzpicture}
      \caption{$\TrueDetectionRate (\%) @ \FalseDetectionRate \leq 0.2\%$}
      \label{fig:classifier:dpsgd:0.2fdr}
  \end{subfigure}
  \hfill
  \begin{subfigure}[b]{0.235\textwidth}
      \centering
      \begin{tikzpicture}
          \begin{axis}[
              xlabel={$\DPnoiseStd (\times 10^{-3})$},
              xtick={1, 2, 3, 4, 5},
              xticklabels={0, $0.5$, $1.0$, $1.5$, $2.0$},
              ymin=60, ymax=80,
              grid=major,
              width=1.0\textwidth,
              height=0.8\textwidth,
              ]
              \addplot[color=indianred1967882,mark=square,densely dashed,mark options={solid},line width=1.5] coordinates {
                  (1, 75.53)
                  (2, 73.75)
                  (3, 71.20)
                  (4, 68.68)
                  (5, 65.82)
              };
          \end{axis}
      \end{tikzpicture}
      \caption{Test accuracy $\Accuracy (\%)$}
      \label{fig:classifier:dpsgd:acc}
  \end{subfigure}
  \caption{$\TrueDetectionRate (\%)$ of our
    auditing method for CIFAR-100 image classifiers trained by DPSGD
    (parameterized by noise multiplier $\DPnoiseStd$)
    (\subfigsref{fig:classifier:dpsgd:5fdr}{fig:classifier:dpsgd:0.2fdr})
    and test accuracies of image classifiers
    (\figref{fig:classifier:dpsgd:acc}).}
  \label{fig:classifier:dpsgd}
\end{figure}

\paragraph{Strong Regularization}

We study the effectiveness of our data-use audit-
ing method on classifiers trained under a stronger
regularization. We controled the regularization strength by
tuning the weight decay (denoted as $\WeightDecay$) parameter in the
SGD optimizer. Our auditing results are shown in \figref{fig:classifier:weightdecay}.
As shown in \figref{fig:classifier:weightdecay},
setting a large $\WeightDecay$ decreased our $\TrueDetectionRate$ ($\DatasetSize=1$)
but it decreased the accuracy of the trained classifier.
For example, when we set $\WeightDecay=8.0\times 10^{-3}$,
we achieved $\TrueDetectionRate$ of $7.02\%$ under $\FalseDetectionRate\leq5\%$
but the model accuracy dropped by $16.28$ percentage points.

Again, when the data owner had more data instances (i.e., $\DatasetSize > 1$)
and all of them were used in training, our method became more
effective, i.e., our $\TrueDetectionRate$ increased with
$\DatasetSize$.

\begin{figure}[ht!]
  \centering

  \begin{subfigure}[b]{0.47\textwidth}
    \centering
    \resizebox{!}{3em}{\newenvironment{customlegend}[1][]{%
    \begingroup
    \csname pgfplots@init@cleared@structures\endcsname
    \pgfplotsset{#1}%
}{%
    \csname pgfplots@createlegend\endcsname
    \endgroup
}%

\def\addlegendimage{\csname pgfplots@addlegendimage\endcsname}

\begin{tikzpicture}

\definecolor{darkslategray38}{RGB}{38,38,38}
\definecolor{indianred1967882}{RGB}{196,78,82}
\definecolor{lightgray204}{RGB}{204,204,204}
\definecolor{mediumseagreen85168104}{RGB}{85,168,104}
\definecolor{peru22113282}{RGB}{221,132,82}
\definecolor{steelblue76114176}{RGB}{76,114,176}

\begin{customlegend}[
    legend style={{font={\small}},{draw=none}}, 
    legend columns=2,
    legend cell align={center},
    legend entries={{$\TrueDetectionRate (\DatasetSize=1)$}, {$\TrueDetectionRate (\DatasetSize=8)$}, {$\TrueDetectionRate (\DatasetSize=64)$}, {$\FalseDetectionRate$ bound}}]
\addlegendimage{very thick, color=steelblue76114176,mark=o}
\addlegendimage{very thick, peru22113282,mark=diamond,dashed,mark options={solid}}
\addlegendimage{very thick, mediumseagreen85168104,mark=triangle,dotted,mark options={solid}}
\addlegendimage{very thick, darkgray,dashdotted}
\addlegendimage{very thick, indianred1967882,mark=square,densely dashed,mark options={solid}}

\end{customlegend}

\end{tikzpicture}}
  \end{subfigure}

  \begin{subfigure}[b]{0.235\textwidth}
      \centering
      \begin{tikzpicture}
          \begin{axis}[
              xlabel={$\WeightDecay (\times 10^{-3})$},
              xtick={1, 2, 3, 4, 5},
              xticklabels={$0.5$, $1.0$, $2.0$, $5.0$, $8.0$},
              ymode=log,
              log basis y=10, 
              ymin=0.07, ymax=140,
              ytick={0.1, 1, 10, 100},
              grid=major,
              width=1.0\textwidth,
              height=0.8\textwidth,
              ]
              \addplot[color=steelblue76114176,mark=o,line width=1.5] coordinates {
                  (1, 28.21)
                  (2, 28.919999999999995)
                  (3, 25.619999999999997)
                  (4, 10.760000000000002)
                  (5, 7.020000000000001)
              };
              \addplot[color=peru22113282,mark=diamond,dashed,mark options={solid},line width=1.5] coordinates {
                  (1,83.7)
                  (2,84.76)
                  (3,80.52)
                  (4,32.62)
                  (5,11.19)
              };
              \addplot[color=mediumseagreen85168104,mark=triangle,dotted,mark options={solid},line width=1.5] coordinates {
                  (1,100.0)
                  (2,100.0)
                  (3,100.0)
                  (4,96.42)
                  (5,35.68)
              };
              \addplot[color=darkgray,dashdotted,line width=1.5] coordinates {
                    (1, 5)
                    (2, 5)
                    (3, 5)
                    (4, 5)
                    (5, 5)
                };
          \end{axis}
      \end{tikzpicture}
      \caption{$\TrueDetectionRate (\%) @ \FalseDetectionRate \leq 5\%$}
      \label{fig:classifier:weightdecay:5fdr}
  \end{subfigure}
  \hfill
  \begin{subfigure}[b]{0.235\textwidth}
      \centering
      \begin{tikzpicture}
          \begin{axis}[
              xlabel={$\WeightDecay (\times 10^{-3})$},
              xtick={1, 2, 3, 4, 5},
              xticklabels={$0.5$, $1.0$, $2.0$, $5.0$, $8.0$},
              ymode=log,
              log basis y=10, 
              ymin=0.07, ymax=140,
              ytick={0.1, 1, 10, 100},
              grid=major,
              width=1.0\textwidth,
              height=0.8\textwidth,
              ]
              \addplot[color=steelblue76114176,mark=o,line width=1.5] coordinates {
                  (1, 11.60)
                  (2, 11.370000000000001)
                  (3, 9.600000000000003)
                  (4, 2.5800000000000005)
                  (5, 1.4500000000000004)
              };
              \addplot[color=peru22113282,mark=diamond,dashed,mark options={solid},line width=1.5] coordinates {
                  (1,60.81)
                  (2,61.8)
                  (3,55.78)
                  (4,11.9)
                  (5,2.62)
              };
              \addplot[color=mediumseagreen85168104,mark=triangle,dotted,mark options={solid},line width=1.5] coordinates {
                  (1,100.0)
                  (2,100.0)
                  (3,100.0)
                  (4,85.7)
                  (5,14.01)
              };
              \addplot[color=darkgray,dashdotted,line width=1.5] coordinates {
                    (1, 1)
                    (2, 1)
                    (3, 1)
                    (4, 1)
                    (5, 1)
                };
          \end{axis}
      \end{tikzpicture}
      \caption{$\TrueDetectionRate (\%) @ \FalseDetectionRate \leq 1\%$}
      \label{fig:classifier:weightdecay:1fdr}
  \end{subfigure}
  \\
  \begin{subfigure}[b]{0.235\textwidth}
      \centering
      \begin{tikzpicture}
          \begin{axis}[
              xlabel={$\WeightDecay (\times 10^{-3})$},
              xtick={1, 2, 3, 4, 5},
              xticklabels={$0.5$, $1.0$, $2.0$, $5.0$, $8.0$},
              ymode=log,
              log basis y=10, 
              ymin=0.07, ymax=140,
              ytick={0.1, 1, 10, 100},
              grid=major,
              width=1.0\textwidth,
              height=0.8\textwidth,
              ]
              \addplot[color=steelblue76114176,mark=o,line width=1.5] coordinates {
                  (1, 3.12)
                  (2, 3.120000000000000)
                  (3, 2.390000000000001)
                  (4, 0.36000000000000004)
                  (5, 0.25)
              };
              \addplot[color=peru22113282,mark=diamond,dashed,mark options={solid},line width=1.5] coordinates {
                  (1,33.1)
                  (2,33.65)
                  (3,28.39)
                  (4,2.89)
                  (5,0.35)
              };
              \addplot[color=mediumseagreen85168104,mark=triangle,dotted,mark options={solid},line width=1.5] coordinates {
                  (1,100.0)
                  (2,100.0)
                  (3,100.0)
                  (4,63.35)
                  (5,3.37)
              };
              \addplot[color=darkgray,dashdotted,line width=1.5] coordinates {
                    (1, 0.2)
                    (2, 0.2)
                    (3, 0.2)
                    (4, 0.2)
                    (5, 0.2)
                };
          \end{axis}
      \end{tikzpicture}
      \caption{$\TrueDetectionRate (\%) @ \FalseDetectionRate \leq 0.2\%$}
      \label{fig:classifier:weightdecay:0.2fdr}
  \end{subfigure}
  \hfill
  \begin{subfigure}[b]{0.235\textwidth}
      \centering
      \begin{tikzpicture}
          \begin{axis}[
              xlabel={$\WeightDecay (\times 10^{-3})$},
              xtick={1, 2, 3, 4, 5},
              xticklabels={$0.5$, $1.0$, $2.0$, $5.0$, $8.0$},
              ymin=55, ymax=80,
              grid=major,
              width=1.0\textwidth,
              height=0.8\textwidth,
              ]
              \addplot[color=indianred1967882,mark=square,densely dashed,mark options={solid},line width=1.5] coordinates {
                  (1, 75.53)
                  (2,75.88550000000002)
                  (3,75.277)
                  (4,69.27250000000001)
                  (5,59.2505)
              };
          \end{axis}
      \end{tikzpicture}
      \caption{Test accuracy $\Accuracy (\%)$}
      \label{fig:classifier:weightdecay:acc}
  \end{subfigure}
  \caption{$\TrueDetectionRate (\%)$ of our
    auditing method for CIFAR-100 image classifiers trained
    with varying weight decay
    (\subfigsref{fig:classifier:weightdecay:5fdr}{fig:classifier:weightdecay:0.2fdr})
    and test accuracies of image classifiers
    (\figref{fig:classifier:weightdecay:acc}).}
  \label{fig:classifier:weightdecay}
\end{figure}

\subsection{Additional Experimental Results of Auditing Image Classifier} \label{app:results:additional}

\paragraph{Across different model architectures}
We study the effectiveness of our data-use auditing method on auditing
data use ($\DatasetSize=1$) in image classifiers of different model
architectures, namely ResNet-18, ResNet-34~\cite{he2016:deep},
WideResNet-28-2~\cite{zagoruyko2016:wide},
VGG-16~\cite{simonyan2015:very}, and
ConvNetBN~\cite{ioffe2015:batch}. We only considered CIFAR-100 and
trained these classifiers using the same training algorithm (please
see \secref{sec:auditing_ML:classifier:setup}).  Our auditing results
are shown in \tblref{tbl:classifier:architecture}.  Our
$\TrueDetectionRate$ ranged from $20.80\%$ to $29.90\%$, from $7.20\%$
to $11.89\%$, and from $1.53\%$ to $3.25\%$ when the bounds on
$\FalseDetectionRate$ were set as $5\%$, $1\%$, and $0.2\%$,
respectively.

\begin{table}[ht!]
  \centering
  {\resizebox{0.47\textwidth}{!}{
  \begin{tabular}{@{}lc@{\hspace{0.5em}}cc@{\hspace{0.5em}}cc@{\hspace{0.5em}}cc@{\hspace{0.5em}}c@{}}
  \toprule
  \multicolumn{1}{c}{\multirow{2}{*}{}} & \multicolumn{3}{c}{$\FalseDetectionRate\leq$} \\ 
  & \multicolumn{1}{c}{$5\%$ }& \multicolumn{1}{c}{$1\%$ }& \multicolumn{1}{c}{$0.2\%$ } \\
  \midrule 
  ResNet-18 & $28.21(\pm1.60)$ &$11.59(\pm0.99)$ &$3.11(\pm0.69)$ \\
  ResNet-34 & $26.39(\pm1.63)$ &$10.52(\pm0.78)$ &$2.66(\pm0.41)$ \\ 
  WideResNet-28-2 & $29.10(\pm1.96)$ &$11.53(\pm1.43)$ &$2.99(\pm0.54)$  \\
  VGG-16 & $20.80(\pm1.09)$ &$7.20(\pm1.16)$ &$1.53(\pm0.45)$  \\
  ConvNetBN & $29.90(\pm1.33)$ &$11.89(\pm1.21)$ &$3.25(\pm0.69)$ \\ 
  \bottomrule 
  \end{tabular}
  }}
  \caption{$\TrueDetectionRate (\%)$ ($\DatasetSize=1$) of our data-use auditing method when
    applied to audit CIFAR-100 image instances in training image classifiers of
    different model architectures (ResNet-18 is the default). Results
    are averaged over $500\times20$ detections.  We trained $20$
    classifiers, in each of which $500$ training samples of CIFAR-100
    were audited. The numbers in the
    parenthesis are standard deviations among the 20 classifiers.}
  \label{tbl:classifier:architecture}
\end{table}

\paragraph{The impact of the utility-preservation parameter $\MarkBound$}
We study the impact of the utility-preservation parameter
$\MarkBound$ on the performance of our data auditing method
($\DatasetSize=1$), by changing $\MarkBound$ from $6$ to
$20$. $\MarkBound$ controls the visual quality (utility) of the marked
image. We present examples of marked CIFAR-100 images under different
$\MarkBound$ in \figref{fig:eps}. Our auditing results 
for varying $\MarkBound$ are presented in
\figref{fig:classifier:mark_bound}. As in \figref{fig:classifier:mark_bound}, 
a larger $\MarkBound$ led to a higher
$\TrueDetectionRate$ under the same $\FalseDetectionRate$ bound, which
shows a trade-off between utility-preservation and
$\TrueDetectionRate$.

\begin{figure}[ht!]
  \centering

\begin{subfigure}[b]{0.47\textwidth}
  \centering
  \resizebox{!}{1.8em}{\newenvironment{customlegend}[1][]{%
    \begingroup
    \csname pgfplots@init@cleared@structures\endcsname
    \pgfplotsset{#1}%
}{%
    \csname pgfplots@createlegend\endcsname
    \endgroup
}%

\def\addlegendimage{\csname pgfplots@addlegendimage\endcsname}

\begin{tikzpicture}

\definecolor{darkslategray38}{RGB}{38,38,38}
\definecolor{indianred1967882}{RGB}{196,78,82}
\definecolor{lightgray204}{RGB}{204,204,204}
\definecolor{mediumseagreen85168104}{RGB}{85,168,104}
\definecolor{peru22113282}{RGB}{221,132,82}
\definecolor{steelblue76114176}{RGB}{76,114,176}

\begin{customlegend}[
    legend style={{font={\small}},{draw=none}}, 
    legend columns=2,
    legend cell align={center},
    legend entries={{Our $\TrueDetectionRate$}, {$\FalseDetectionRate$ bound}}]
\addlegendimage{very thick, steelblue76114176, mark=o, solid}
\addlegendimage{very thick, darkgray,dashdotted}

\end{customlegend}

\end{tikzpicture}}
\end{subfigure}

  \begin{subfigure}[t]{0.32\textwidth}
    \centering
    \begin{tikzpicture}
      \begin{axis}[
        xlabel={$\MarkBound$},
        ylabel={\TrueDetectionRate (\%)},
        xtick={1, 2, 3, 4},
        xticklabels={6, 10, 16, 20},
        ymode=log,
      log basis y=10, 
      ymin=0.07, ymax=140,
      ytick={0.1, 1, 10, 100},
        grid=major,
        width=1.0\textwidth,
        height=0.5\textwidth,
        ]
        \addplot[color=steelblue76114176,mark=o,line width=1.5] coordinates {
          (1, 18.76)
          (2, 28.21)
          (3, 38.20)
          (4, 43.11)
        };
        \addplot[color=darkgray,dashdotted,line width=1.5] coordinates {
            (1, 5)
            (2, 5)
            (3, 5)
            (4, 5)
        };
      \end{axis}
    \end{tikzpicture}
    \caption[Short Caption]{
      \centering $\FalseDetectionRate \leq 5\%$}
  \end{subfigure}
  \hfill
  \begin{subfigure}[t]{0.32\textwidth}
    \centering
    \begin{tikzpicture}
      \begin{axis}[
        xlabel={$\MarkBound$},
        ylabel={\TrueDetectionRate (\%)},
        xtick={1, 2, 3, 4},
        xticklabels={6, 10, 16, 20},
        ymode=log,
      log basis y=10, 
      ymin=0.07, ymax=140,
      ytick={0.1, 1, 10, 100},
        grid=major,
        width=1.0\textwidth,
        height=0.5\textwidth,
        ]
        \addplot[color=steelblue76114176,mark=o,line width=1.5] coordinates {
          (1, 6.42)
          (2, 11.60)
          (3, 17.42)
          (4, 21.62)
        };
        \addplot[color=darkgray,dashdotted,line width=1.5] coordinates {
            (1, 1)
            (2, 1)
            (3, 1)
            (4, 1)
        };
      \end{axis}
    \end{tikzpicture}
    \caption[Short Caption]{
      \centering $\FalseDetectionRate \leq 1\%$}
  \end{subfigure}
  \hfill
  \begin{subfigure}[t]{0.32\textwidth}
    \centering
    \begin{tikzpicture}
      \begin{axis}[
        xlabel={$\MarkBound$},
        ylabel={\TrueDetectionRate (\%)},
        xtick={1, 2, 3, 4},
        xticklabels={6, 10, 16, 20},
        ymode=log,
      log basis y=10, 
      ymin=0.07, ymax=140,
      ytick={0.1, 1, 10, 100},
        grid=major,
        width=1.0\textwidth,
        height=0.5\textwidth,
        ]
        \addplot[color=steelblue76114176,mark=o,line width=1.5] coordinates {
          (1, 1.48)
          (2, 3.12)
          (3, 5.38)
          (4, 7.52)
        };
        \addplot[color=darkgray,dashdotted,line width=1.5] coordinates {
            (1, 0.2)
            (2, 0.2)
            (3, 0.2)
            (4, 0.2)
        };
      \end{axis}
    \end{tikzpicture}
    \caption[Short Caption]{
      \centering $\FalseDetectionRate \leq 0.2\%$}
  \end{subfigure}
  \caption{$\TrueDetectionRate(\%)$ ($\DatasetSize=1$) of our method for auditing CIFAR-100 instances 
          in image classifiers under varying $\MarkBound$ values, 
          plotted for three levels of $\FalseDetectionRate$. }
  \label{fig:classifier:mark_bound}
\end{figure}

\paragraph{Impact of using a different $\NumberMarkedData$}
We consider a setting where our marking algorithm generated
$\NumberMarkedData=200$,
$\NumberMarkedData=500$, or $\NumberMarkedData=5{,}000$ 
marked data per raw CIFAR-100 data
instance. Our results ($\DatasetSize=1$) are presented in
\tblref{tbl:classifier:large_n}. By comparing with the results of
$\NumberMarkedData=1{,}000$, we have the following observations: When
we set an $\FalseDetectionRate$ bound much larger than
$\frac{1}{\NumberMarkedData}$, e.g., $\FalseDetectionRate\leq5\%$ or
$\FalseDetectionRate\leq1\%$, $\TrueDetectionRate$ did not change much
when increasing $\NumberMarkedData$. However, when we considered
$\FalseDetectionRate\leq0.2\%$, $\TrueDetectionRate$ increased from
$3.11\%$ to $4.43\%$ if we increased $\NumberMarkedData$ from
$1{,}000$ to $5{,}000$.  In addition, setting a larger
$\NumberMarkedData$ allowed the data owner to detect her data-use
under a lower $\FalseDetectionRate$ bound (e.g.,
$\FalseDetectionRate\leq0.1\%$). But setting a larger
$\NumberMarkedData$ also brought a higher cost in marked data
generation and data-use detection, as shown in \figref{fig:large_n:overhead:marking}
and \figref{fig:large_n:overhead:detection}.

\begin{table}[ht!]
  \centering
  {\resizebox{0.47\textwidth}{!}{
  \begin{tabular}{@{}lc@{\hspace{0.5em}}cc@{\hspace{0.5em}}cc@{\hspace{0.5em}}cc@{\hspace{0.5em}}cc@{\hspace{0.5em}}c@{}}
  \toprule
   & \multicolumn{4}{c}{$\FalseDetectionRate\leq$} \\ 
   & \multicolumn{1}{c}{$5\%$ }& \multicolumn{1}{c}{$1\%$ }& \multicolumn{1}{c}{$0.2\%$ }& \multicolumn{1}{c}{$0.1\%$ } \\
  \midrule
  $\NumberMarkedData=200$ & $26.07(\pm1.61)$ &$6.96(\pm1.10)$ &$-$ &$-$ \\
  $\NumberMarkedData=500$ & $27.41(\pm1.72)$ &$10.19(\pm1.29)$ &$-$ &$-$ \\
  $\NumberMarkedData=1{,}000$ & $28.21(\pm1.60)$ &$11.59(\pm0.99)$ &$3.11(\pm0.69)$ &$-$ \\
  $\NumberMarkedData=5{,}000$ & $28.16(\pm2.01)$ &$11.79(\pm1.61)$ &$4.43(\pm0.96)$ &$2.01(\pm0.66)$ \\
  \bottomrule 
  \end{tabular}
  }}
  \caption{$\TrueDetectionRate (\%)$ ($\DatasetSize=1$) of our data-use auditing method when
    applied to audit the use of CIFAR-100 in image classifier (ResNet-18),
    when we set $\NumberMarkedData=200$,
    $\NumberMarkedData=500$, $\NumberMarkedData=1{,}000$, and $\NumberMarkedData=5{,}000$
    ($\NumberMarkedData=1{,}000$ is our default). Results are
    averaged over $500\times20$ detections for CIFAR-100. We trained $20$
    classifiers, in each of which $500$ CIFAR-100 
    training samples were audited. The numbers in the
    parenthesis are standard deviations among the 20 classifiers.} 
  \label{tbl:classifier:large_n}
\end{table}

\begin{figure}[ht!]
  \centering

  \begin{subfigure}[t]{0.32\textwidth}
    \centering
    \begin{tikzpicture}
      \begin{axis}[
        xlabel={$\NumberMarkedData$},
        ylabel={GPU time (min)},
        xtick={1, 2, 3, 4},
        xticklabels={200, 500, 1000, 5000},
        ymin=-2, ymax=62,
        grid=major,
        width=1.0\textwidth,
        height=0.5\textwidth,
        ]
        \addplot[color=steelblue76114176,mark=o,line width=1.5] coordinates {
          (1,1.6083197916666667)
          (2,4.004001302083333)
          (3,7.897980729166666)
          (4,39.082995833333335)
        };
      \end{axis}
    \end{tikzpicture}
  \end{subfigure}
  \hfill
  \begin{subfigure}[t]{0.32\textwidth}
    \centering
    \begin{tikzpicture}
      \begin{axis}[
        xlabel={$\NumberMarkedData$},
        ylabel={CPU time (min)},
        xtick={1, 2, 3, 4},
        xticklabels={200, 500, 1000, 5000},
        ymin=-2, ymax=62,
        grid=major,
        width=1.0\textwidth,
        height=0.5\textwidth,
        ]
        \addplot[color=steelblue76114176,mark=o,line width=1.5] coordinates {
          (1,1.7463630770166667)
          (2,4.368438106416667)
          (3,8.423586496433334)
          (4,40.891539929816666)
        };
      \end{axis}
    \end{tikzpicture}
  \end{subfigure}
  \hfill
  \begin{subfigure}[t]{0.32\textwidth}
    \centering
    \begin{tikzpicture}
      \begin{axis}[
        xlabel={$\NumberMarkedData$},
        ylabel={Peak memory usage (MB)},
        xtick={1, 2, 3, 4},
        xticklabels={200, 500, 1000, 5000},
        ymin=-100, ymax=3100,
        grid=major,
        width=1.0\textwidth,
        height=0.5\textwidth,
        ]
        \addplot[color=steelblue76114176,mark=o,line width=1.5] coordinates {
          (1,1297.63328)
          (2,1887.100928)
          (3,1937.793024)
          (4,1991.168)
        };
      \end{axis}
    \end{tikzpicture}
  \end{subfigure}
  \caption{Overhead of running our data-marking algorithm 
   using a varying $\NumberMarkedData$,
   for each audited data instance ($\DatasetSize=1$).}
  \label{fig:large_n:overhead:marking}
\end{figure}

\begin{figure}[ht!]
  \centering

  \begin{subfigure}[t]{0.32\textwidth}
    \centering
    \begin{tikzpicture}
      \begin{axis}[
        xlabel={$\NumberMarkedData$},
        ylabel={CPU time (sec)},
        xtick={1, 2, 3, 4},
        xticklabels={200, 500, 1000, 5000},
        ymin=-1, ymax=21,
        grid=major,
        width=1.0\textwidth,
        height=0.5\textwidth,
        ]
        \addplot[color=steelblue76114176,mark=o,line width=1.5] coordinates {
          (1,0.29823000099999997)
          (2,1.076791698)
          (3,2.526635149)
          (4,19.102947311999998)
        };
      \end{axis}
    \end{tikzpicture}
  \end{subfigure}
  \hfill
  \begin{subfigure}[t]{0.32\textwidth}
    \centering
    \begin{tikzpicture}
      \begin{axis}[
        xlabel={$\NumberMarkedData$},
        ylabel={Peak memory usage (MB)},
        xtick={1, 2, 3, 4},
        xticklabels={200, 500, 1000, 5000},
        ymin=-2, ymax=60,
        grid=major,
        width=1.0\textwidth,
        height=0.5\textwidth,
        ]
        \addplot[color=steelblue76114176,mark=o,line width=1.5] coordinates {
          (1,47.529984)
          (2,48.078848)
          (3,48.10752)
          (4,47.988736)
        };
      \end{axis}
    \end{tikzpicture}
  \end{subfigure}
  \caption{Overhead of applying prior-posterior-ratio martingale
    for confidence interval estimation in
    data-use detection for each audited data
    instance ($\DatasetSize=1$), under a varying $\NumberMarkedData$.}
  \label{fig:large_n:overhead:detection}
\end{figure}

\paragraph{Impact of using data augmentation in data-use detection}
We plot $\TrueDetectionRate$ ($\DatasetSize=1$) of our auditing method
using varying number $\NumAugment$ of data augmentations in
\figref{fig:classifier:augmentation}.  When we used a
larger number of data augmentations in the data-use detection
algorithm, our method achieved a higher $\TrueDetectionRate$ under the
same $\FalseDetectionRate$ bound. However, a larger number of data
augmentations requires more (black-box) queries to the ML model, and
thus it brings a higher query cost.

\begin{figure}[ht!]
  \centering
  
\begin{subfigure}[b]{0.47\textwidth}
  \centering
  \resizebox{!}{1.8em}{\newenvironment{customlegend}[1][]{%
    \begingroup
    \csname pgfplots@init@cleared@structures\endcsname
    \pgfplotsset{#1}%
}{%
    \csname pgfplots@createlegend\endcsname
    \endgroup
}%

\def\addlegendimage{\csname pgfplots@addlegendimage\endcsname}

\begin{tikzpicture}

\definecolor{darkslategray38}{RGB}{38,38,38}
\definecolor{indianred1967882}{RGB}{196,78,82}
\definecolor{lightgray204}{RGB}{204,204,204}
\definecolor{mediumseagreen85168104}{RGB}{85,168,104}
\definecolor{peru22113282}{RGB}{221,132,82}
\definecolor{steelblue76114176}{RGB}{76,114,176}

\begin{customlegend}[
    legend style={{font={\small}},{draw=none}}, 
    legend columns=2,
    legend cell align={center},
    legend entries={{Our $\TrueDetectionRate$}, {$\FalseDetectionRate$ bound}}]
\addlegendimage{very thick, steelblue76114176, mark=o, solid}
\addlegendimage{very thick, darkgray,dashdotted}

\end{customlegend}

\end{tikzpicture}}
\end{subfigure}

  \begin{subfigure}[t]{0.42\textwidth}
    \centering
    \begin{tikzpicture}
      \begin{axis}[
        xlabel={$\NumAugment$},
        ylabel={\TrueDetectionRate (\%)},
        xtick={1, 2, 3, 4, 5, 6, 7},
        xticklabels={1, 2, 4, 8, 16, 32, 64},
        ymode=log,
      log basis y=10, 
      ymin=0.07, ymax=140,
      ytick={0.1, 1, 10, 100},
        grid=major,
        width=1.0\textwidth,
        height=0.4\textwidth,
        ]
        \addplot[color=steelblue76114176,mark=o,line width=1.5] coordinates {
          (1, 21.55)
          (2, 23.52)
          (3, 24.91)
          (4, 26.30)
          (5, 28.21)
          (6, 29.04)
          (7, 29.09)
        };
        \addplot[color=darkgray,dashdotted,line width=1.5] coordinates {
            (1, 5)
            (2, 5)
            (3, 5)
            (4, 5)
            (5, 5)
            (6, 5)
            (7, 5)
        };
      \end{axis}
    \end{tikzpicture}
    \caption[Short Caption]{
      \centering $\FalseDetectionRate \leq 5\%$}
  \end{subfigure}
  \hfill
  \begin{subfigure}[t]{0.42\textwidth}
    \centering
    \begin{tikzpicture}
      \begin{axis}[
        xlabel={$\NumAugment$},
        ylabel={\TrueDetectionRate (\%)},
        xtick={1, 2, 3, 4, 5, 6, 7},
        xticklabels={1, 2, 4, 8, 16, 32, 64},
        ymode=log,
      log basis y=10, 
      ymin=0.07, ymax=140,
      ytick={0.1, 1, 10, 100},
        grid=major,
        width=1.0\textwidth,
        height=0.4\textwidth,
        ]
        \addplot[color=steelblue76114176,mark=o,line width=1.5] coordinates {
          (1, 7.30)
          (2, 8.79)
          (3, 9.45)
          (4, 10.39)
          (5, 11.60)
          (6, 12.34)
          (7, 12.78)
        };
        \addplot[color=darkgray,dashdotted,line width=1.5] coordinates {
            (1, 1)
            (2, 1)
            (3, 1)
            (4, 1)
            (5, 1)
            (6, 1)
            (7, 1)
        };
      \end{axis}
    \end{tikzpicture}
    \caption[Short Caption]{
      \centering $\FalseDetectionRate \leq 1\%$}
  \end{subfigure}
  \hfill
  \begin{subfigure}[t]{0.42\textwidth}
    \centering
    \begin{tikzpicture}
      \begin{axis}[
        xlabel={$\NumAugment$},
        ylabel={\TrueDetectionRate (\%)},
        xtick={1, 2, 3, 4, 5, 6, 7},
        xticklabels={1, 2, 4, 8, 16, 32, 64},
        ymode=log,
      log basis y=10, 
      ymin=0.07, ymax=140,
      ytick={0.1, 1, 10, 100},
        grid=major,
        width=1.0\textwidth,
        height=0.4\textwidth,
        ]
        \addplot[color=steelblue76114176,mark=o,line width=1.5] coordinates {
          (1, 1.46)
          (2, 1.89)
          (3, 2.36)
          (4, 2.82)
          (5, 3.12)
          (6, 3.34)
          (7, 3.75)
        };
        \addplot[color=darkgray,dashdotted,line width=1.5] coordinates {
            (1, 0.2)
            (2, 0.2)
            (3, 0.2)
            (4, 0.2)
            (5, 0.2)
            (6, 0.2)
            (7, 0.2)
        };
      \end{axis}
    \end{tikzpicture}
    \caption[Short Caption]{
      \centering $\FalseDetectionRate \leq 0.2\%$}
  \end{subfigure}
  \caption{$\TrueDetectionRate(\%)$ ($\DatasetSize=1$) of our auditing method for auditing CIFAR-100 instances 
    in image classifiers, using varying $\NumAugment$ values ($\NumAugment$ is
    the number of data augmentations) in data-use detection, 
    plotted for three levels of $\FalseDetectionRate$.}
  \label{fig:classifier:augmentation}
\end{figure}

\paragraph{Factor impacting auditability} 
We study a factor that might impact the auditability of an image.
Given an ML model $\MLModel$ trained using a (marked) image
$\MarkedData$ ($\DatasetSize = 1$) and membership inference method
$\MIAlg{\MLModel}(\cdot)$, we measured the auditability of an image
instance $\MarkedData$ by its
$\Rank{\MarkedData,\HiddenInfo,\MIAlg{\MLModel}}$, i.e., a higher
$\Rank{\MarkedData,\HiddenInfo,\MIAlg{\MLModel}}$ indicates more
auditability.  We considered the difference between the loss of a
model that is not trained on the audited data instance $\MarkedData$
and the loss of the model $\MLModel$ that is trained on $\MarkedData$,
denoted as
$\Loss(\MLModel{-\MarkedData},\MarkedData)-\Loss(\MLModel,\MarkedData)$
where $\Loss$ denotes the loss function and $\MLModel{-\MarkedData}$
denotes the model that is not trained on $\MarkedData$.  The loss
difference indicates how a marked data instance is vulnerable to a
membership inference attack~\cite{carlini2022:membership}, i.e., a
data instance with larger loss difference is more vulnerable to
membership inference.  We plot the distribution of
$(\Rank{\MarkedData,\HiddenInfo,\MIAlg{\MLModel}},
\Loss(\MLModel{-\MarkedData},\MarkedData)-\Loss(\MLModel,\MarkedData))$
in \figref{fig:auditability_factors}. We calculated the Pearson
correlation coefficients~\cite{cohen2009:pearson} between
$\Rank{\MarkedData,\HiddenInfo,\MIAlg{\MLModel}}$ and
$\Loss(\MLModel{-\MarkedData},\MarkedData)-\Loss(\MLModel,\MarkedData)$.
The resulted coefficients for CIFAR-100 and TinyImageNet are $0.331$
and $0.295$ respectively. These results indicate a weak positive
linear correlation between
$\Rank{\MarkedData,\HiddenInfo,\MIAlg{\MLModel}}$ and
$\Loss(\MLModel{-\MarkedData},\MarkedData)-\Loss(\MLModel,\MarkedData)$. In
other words, it would be easier to audit the use of a data instance
with a larger loss difference. This observation is consistent with
that from previous works on privacy
attacks~\cite{carlini2022:membership}.  We leave exploring other
factors impacting auditability of a data instance as a direction for
future work.

\begin{figure}
  \centering
  \begin{subfigure}[b]{0.23\textwidth}
      \setlength\figureheight{1.9in}
      \centering
      \resizebox{\textwidth}{!}{\input{figures/loss_rank_cifar100.tex}}
      \caption{CIFAR-100}
    \end{subfigure}
    \hfill
    \begin{subfigure}[b]{0.23\textwidth}
      \setlength\figureheight{1.9in}
      \centering
      \resizebox{\textwidth}{!}{\input{figures/loss_rank_tinyimagenet.tex}}
      \caption{TinyImageNet}
    \end{subfigure}
  \caption[Short caption without TikZ commands]{$\Rank{\MarkedData,\HiddenInfo,\MIAlg{\MLModel}}$ vs.\ $\Loss(\MLModel{-\MarkedData},\MarkedData)-\Loss(\MLModel,\MarkedData)$}
  \label{fig:auditability_factors}
\end{figure}

\paragraph{Ablation study on data-marking algorithm}
We study the impact of two components of our data-marking algorithm:
(1) optimizing the $\NumberMarkedData$ unit vectors such that their
minimum pairwise distance is maximized (compared to generating them by
sampling randomly), and (2) optimizing the added marks such that the
distance between any pair of generated marked data is maximized
(compared to generating added marks randomly).  We denote the method
of generating the added marks randomly (i.e., each pixel of a mark is
uniformly at random sampled from $\{-\MarkBound, \MarkBound\}$) as
\texttt{RM}.  We denote the method of generating marks by generating
random unit vectors and optimizing the marks to maximize the distance
between any pair of marked data, as \texttt{RUV+OM} (only component
(2) included).  We denote the method of generating marks by both
optimizing the unit vectors and the marks, as \texttt{OUV+OM} (both
component (1) and (2) included).  The results ($\DatasetSize=1$) are
presented in \tblref{tbl:classifier:ablation_study}.  As in
\tblref{tbl:classifier:ablation_study}, \texttt{OUV+OM} achieved the
highest \TrueDetectionRate under the same level of
\FalseDetectionRate. Compared with \texttt{RM}, \texttt{OUV+OM}
improved by $5.44$ percentage points, $3.36$ percentage points, and
$1.01$ percentage points, under the false-detection bounds of $5\%$,
$1\%$, and $0.2\%$, respectively. Such improvement demonstrates that a
useful feature extractor helps in generating ``effective'' marked
data. This is because using a feature extractor can embed marks into
the high level features of an image such that the high level features
of $\NumberMarkedData$ marked data are maximally different and so it
is easier for a membership inference method to distinguish a model
trained on one but not the others. Compared with \texttt{RUV+OM},
\texttt{OUV+OM} achieved a slightly higher \TrueDetectionRate. Random
sampling can generate unit vectors whose minimum pairwise distance is
large enough and thus optimizing unit vectors only made marginal
improvement.

\begin{table}[ht!]
  \centering
  \begin{tabular}{@{}lc@{\hspace{0.5em}}cc@{\hspace{0.5em}}cc@{\hspace{0.5em}}cc@{\hspace{0.5em}}c@{}}
  \toprule
  \multicolumn{1}{c}{\multirow{2}{*}{}} & \multicolumn{3}{c}{$\FalseDetectionRate\leq$} \\ 
  & \multicolumn{1}{c}{$5\%$ }& \multicolumn{1}{c}{$1\%$ }& \multicolumn{1}{c}{$0.2\%$ } \\
  \midrule 
  \texttt{RM} & $22.77(\pm1.58)$ &$8.23(\pm1.02)$ &$2.10(\pm0.68)$ \\
  \texttt{RUV+OM} & $27.79(\pm1.47)$ &$11.08(\pm1.19)$ &$2.97(\pm0.95)$ \\
  \texttt{OUV+OM} & $28.21(\pm1.60)$ &$11.59(\pm0.99)$ &$3.11(\pm0.69)$ \\
  \bottomrule 
  \end{tabular}
  \caption{$\TrueDetectionRate (\%)$ ($\DatasetSize=1$) of auditing CIFAR-100 instances in image
    classifiers under different choices of data-marking methods
    (\texttt{OUV+OM} is the default).  Results are averaged over
    $500\times20$ detections. We trained $20$ classifiers, in each of
    which $500$ training samples of CIFAR-100 were audited. The numbers in the
    parenthesis are standard deviations among the 20 classifiers.}
    \label{tbl:classifier:ablation_study}
\end{table}

\paragraph{Ablation study on data-use detection algorithm}
We study the effectiveness of using the rank of the added mark of the
published data in detecting data use. In other words, we measured the
``memorization'' scores of the added marks from $\NumberMarkedData$
generated marked data, then estimated the rank of the added mark of
$\MarkedData$, and detected data use based on the estimated rank.
This is the adaption of the idea from Carlini et al.'s
work~\cite{carlini2019:secret} to our setting.  Our results
($\DatasetSize=1$) in \tblref{tbl:classifier:using_mark} show that the
rank of the added mark did not provide strong evidence of data-use in
visual models, i.e., $\TrueDetectionRate$ values were low and close to
the level of $\FalseDetectionRate$.  This is because the visual model
memorizes the whole marked data instance in its training rather than
the added mark.

\begin{table}[ht!]
  \centering
  {\resizebox{0.47\textwidth}{!}{
  \begin{tabular}{@{}lc@{\hspace{0.5em}}cc@{\hspace{0.5em}}cc@{\hspace{0.5em}}cc@{\hspace{0.5em}}c@{}}
  \toprule
   & \multicolumn{3}{c}{$\FalseDetectionRate\leq$} \\ 
  & \multicolumn{1}{c}{$5\%$ }& \multicolumn{1}{c}{$1\%$ }& \multicolumn{1}{c}{$0.2\%$ }\\
  \midrule 
  Our method & $28.21(\pm1.60)$ &$11.59(\pm0.99)$ &$3.11(\pm0.69)$ \\
  Use rank of mark & $6.21(\pm4.58)$ &$1.57(\pm2.28)$ &$0.32(\pm0.94)$ \\
  \bottomrule 
  \end{tabular}
  }}
  \caption{$\TrueDetectionRate (\%)$ ($\DatasetSize=1$) of the method using the rank of the added mark and our method when
    applied to audit the use of CIFAR-100 in image classifier (ResNet-18). Results are
    averaged over $500\times20$ detections for CIFAR-100. We trained $20$
    classifiers, in each of which $500$ CIFAR-100 
    training samples were audited.}
  \label{tbl:classifier:using_mark}
\end{table}

\subsection{Additional Experimental Results of Auditing CLIP and BLIP} \label{app:results:additional_clip_blip}

\paragraph{CLIP}
Our results on auditing the data-use in the fine-tuned CLIP 
by more than one epochs are shown
in \figref{fig:clip_additional}.  With $\DatasetSize=1$, when we fine-tuned the CLIP
model by more epochs, its \Accuracy slightly increased from $85.44\%$
to $86.57\%$ and \TrueDetectionRate also increased, e.g., from
$9.60\%$ to $16.38\%$ under $\FalseDetectionRate\leq5\%$. Fine-tuning
CLIP by more epochs makes it memorize its training samples more and
thus it is easier to audit data-use (i.e., a higher
\TrueDetectionRate), which was also observed by previous works on
dataset-level data-use auditing (e.g.,~\cite{huang2024:auditdata}).
When the data owner had more data instances (i.e.,
$\DatasetSize>1$) and all of them were used in training, our method
became significantly more effective, i.e., our $\TrueDetectionRate$
increased with $\DatasetSize$.

\begin{figure}[ht!]
  \centering

  \begin{subfigure}[b]{0.47\textwidth}
    \centering
    \resizebox{!}{3em}{\newenvironment{customlegend}[1][]{%
    \begingroup
    \csname pgfplots@init@cleared@structures\endcsname
    \pgfplotsset{#1}%
}{%
    \csname pgfplots@createlegend\endcsname
    \endgroup
}%

\def\addlegendimage{\csname pgfplots@addlegendimage\endcsname}

\begin{tikzpicture}

\definecolor{darkslategray38}{RGB}{38,38,38}
\definecolor{indianred1967882}{RGB}{196,78,82}
\definecolor{lightgray204}{RGB}{204,204,204}
\definecolor{mediumseagreen85168104}{RGB}{85,168,104}
\definecolor{peru22113282}{RGB}{221,132,82}
\definecolor{steelblue76114176}{RGB}{76,114,176}

\begin{customlegend}[
    legend style={{font={\small}},{draw=none}}, 
    legend columns=2,
    legend cell align={center},
    legend entries={{$\TrueDetectionRate (\DatasetSize=1)$}, {$\TrueDetectionRate (\DatasetSize=8)$}, {$\TrueDetectionRate (\DatasetSize=64)$}, {$\FalseDetectionRate$ bound}}]
\addlegendimage{very thick, color=steelblue76114176,mark=o}
\addlegendimage{very thick, peru22113282,mark=diamond,dashed,mark options={solid}}
\addlegendimage{very thick, mediumseagreen85168104,mark=triangle,dotted,mark options={solid}}
\addlegendimage{very thick, darkgray,dashdotted}
\addlegendimage{very thick, indianred1967882,mark=square,densely dashed,mark options={solid}}

\end{customlegend}

\end{tikzpicture}}
  \end{subfigure}

  \begin{subfigure}[b]{0.235\textwidth}
      \centering
      \begin{tikzpicture}
          \begin{axis}[
              xlabel={Epochs},
              xtick={1, 2, 3, 4, 5, 6},
              xticklabels={0, 1, 2, 3, 4, 5},
              ymode=log,
              log basis y=10, 
              ymin=0.07, ymax=140,
              ytick={0.1, 1, 10, 100},
              grid=major,
              width=1.0\textwidth,
              height=0.8\textwidth,
              ]
              \addplot[color=steelblue76114176,mark=o,line width=1.5] coordinates {
                  (1, 4.48)
                  (2, 9.60)
                  (3, 12.48)
                  (4, 14.66)
                  (5, 15.56)
                  (6, 16.38)
              };
              \addplot[color=peru22113282,mark=diamond,dashed,mark options={solid},line width=1.5] coordinates {
                  (1, 4.33)
                  (2,18.41)
                  (3,27.81)
                  (4,33.76)
                  (5,37.95)
                  (6,39.32)
              };
              \addplot[color=mediumseagreen85168104,mark=triangle,dotted,mark options={solid},line width=1.5] coordinates {
                  (1,3.73)
                  (2,67.06)
                  (3,90.37)
                  (4,95.73)
                  (5,98.13)
                  (6,98.88)
              };
              \addplot[color=darkgray,dashdotted,line width=1.5] coordinates {
                (1, 5)
                (2, 5)
                (3, 5)
                (4, 5)
                (5, 5)
                (6, 5)
              };
          \end{axis}
      \end{tikzpicture}
      \caption{$\TrueDetectionRate (\%) @ \FalseDetectionRate \leq 5\%$}
      \label{fig:clip_additional:5fdr}
  \end{subfigure}
  \hfill
  \begin{subfigure}[b]{0.235\textwidth}
      \centering
      \begin{tikzpicture}
          \begin{axis}[
              xlabel={Epochs},
              xtick={1, 2, 3, 4, 5, 6},
              xticklabels={0, 1, 2, 3, 4, 5},
              ymode=log,
              log basis y=10, 
              ymin=0.07, ymax=140,
              ytick={0.1, 1, 10, 100},
              grid=major,
              width=1.0\textwidth,
              height=0.8\textwidth,
              ]
              \addplot[color=steelblue76114176,mark=o,line width=1.5] coordinates {
                  (1, 0.80)
                  (2, 2.64)
                  (3, 3.96)
                  (4, 4.48)
                  (5, 5.16)
                  (6, 5.52)
              };
              \addplot[color=peru22113282,mark=diamond,dashed,mark options={solid},line width=1.5] coordinates {
                  (1,0.73)
                  (2,5.6)
                  (3,9.68)
                  (4,13.23)
                  (5,16.28)
                  (6,17.26)
              };
              \addplot[color=mediumseagreen85168104,mark=triangle,dotted,mark options={solid},line width=1.5] coordinates {
                  (1,0.68)
                  (2,40.13)
                  (3,72.7)
                  (4,85.28)
                  (5,91.26)
                  (6,94.49)
              };
              \addplot[color=darkgray,dashdotted,line width=1.5] coordinates {
                (1, 1)
                (2, 1)
                (3, 1)
                (4, 1)
                (5, 1)
                (6, 1)
              };
          \end{axis}
      \end{tikzpicture}
      \caption{$\TrueDetectionRate (\%) @ \FalseDetectionRate \leq 1\%$}
      \label{fig:clip_additional:1fdr}
  \end{subfigure}
  \\
  \begin{subfigure}[b]{0.235\textwidth}
      \centering
      \begin{tikzpicture}
          \begin{axis}[
              xlabel={Epochs},
              xtick={1, 2, 3, 4, 5, 6},
              xticklabels={0, 1, 2, 3, 4, 5},
              ymode=log,
              log basis y=10, 
              ymin=0.07, ymax=140,
              ytick={0.1, 1, 10, 100},
              grid=major,
              width=1.0\textwidth,
              height=0.8\textwidth,
              ]
              \addplot[color=steelblue76114176,mark=o,line width=1.5] coordinates {
                  (1, 0.1)
                  (2, 0.38)
                  (3, 0.84)
                  (4, 0.96)
                  (5, 1.06)
                  (6, 1.18)
              };
              \addplot[color=peru22113282,mark=diamond,dashed,mark options={solid},line width=1.5] coordinates {
                  (1, 0.1)
                  (2,1.21)
                  (3,2.45)
                  (4,3.46)
                  (5,4.51)
                  (6,5.35)
              };
              \addplot[color=mediumseagreen85168104,mark=triangle,dotted,mark options={solid},line width=1.5] coordinates {
                (1, 0.1)
                (2,16.88)
                (3,46.23)
                (4,62.99)
                (5,74.13)
                (6,81.14)
              };
              \addplot[color=darkgray,dashdotted,line width=1.5] coordinates {
                (1, 0.2)
                (2, 0.2)
                (3, 0.2)
                (4, 0.2)
                (5, 0.2)
                (6, 0.2)
              };
          \end{axis}
      \end{tikzpicture}
      \caption{$\TrueDetectionRate (\%) @ \FalseDetectionRate \leq 0.2\%$}
      \label{fig:clip_additional:0.2fdr}
  \end{subfigure}
  \hfill
  \begin{subfigure}[b]{0.235\textwidth}
      \centering
      \begin{tikzpicture}
          \begin{axis}[
              xlabel={Epochs},
              xtick={1, 2, 3, 4, 5, 6},
              xticklabels={0, 1, 2, 3, 4, 5},
              ymin=79, ymax=87,
              grid=major,
              width=1.0\textwidth,
              height=0.8\textwidth,
              ]
              \addplot[color=indianred1967882,mark=square,densely dashed,mark options={solid},line width=1.5] coordinates {
                  (1, 80.06)
                  (2, 85.44)
                  (3, 85.98)
                  (4, 86.28)
                  (5, 86.44)
                  (6, 86.57)
              };
          \end{axis}
      \end{tikzpicture}
      \caption{Test accuracy $\Accuracy (\%)$}
      \label{fig:clip_additional:acc}
  \end{subfigure}
  \caption{$\TrueDetectionRate (\%)$ of our auditing method for CLIP
    fine-tuned on Flickr30k
    (\subfigsref{fig:clip_additional:5fdr}{fig:clip_additional:0.2fdr}) and test accuracies
    of fine-tuned CLIP (\figref{fig:clip_additional:acc}). Results
    are averaged over $250\times20$ detections. We fine-tuned $20$
    CLIP models, in each of which $250$ training samples were
    audited.}
  \label{fig:clip_additional}
\end{figure}

\paragraph{BLIP}
Our results on auditing the data-use in the fine-tuned BLIP 
by more than one epochs are shown
in \figref{fig:blip_additional}.  With $\DatasetSize=1$, when we fine-tuned the BLIP
model by more epochs, its BLEU score decreased due to overfitting, 
but \TrueDetectionRate increased, e.g., from
$13.07\%$ to $24.42\%$ under $\FalseDetectionRate\leq5\%$. Fine-tuning
BLIP by more epochs makes it memorize its training samples more and
thus it is easier to audit data-use (i.e., a higher
\TrueDetectionRate). Again, when the data owner had more data instances (i.e.,
$\DatasetSize>1$) and all of them were used in training, our method
became significantly more effective, i.e., our $\TrueDetectionRate$
increased with $\DatasetSize$.

\begin{figure}[ht!]
  \centering

  \begin{subfigure}[b]{0.47\textwidth}
    \centering
    \resizebox{!}{3em}{\newenvironment{customlegend}[1][]{%
    \begingroup
    \csname pgfplots@init@cleared@structures\endcsname
    \pgfplotsset{#1}%
}{%
    \csname pgfplots@createlegend\endcsname
    \endgroup
}%

\def\addlegendimage{\csname pgfplots@addlegendimage\endcsname}

\begin{tikzpicture}

\definecolor{darkslategray38}{RGB}{38,38,38}
\definecolor{indianred1967882}{RGB}{196,78,82}
\definecolor{lightgray204}{RGB}{204,204,204}
\definecolor{mediumseagreen85168104}{RGB}{85,168,104}
\definecolor{peru22113282}{RGB}{221,132,82}
\definecolor{steelblue76114176}{RGB}{76,114,176}

\begin{customlegend}[
    legend style={{font={\small}},{draw=none}}, 
    legend columns=2,
    legend cell align={center},
    legend entries={{$\TrueDetectionRate (\DatasetSize=1)$}, {$\TrueDetectionRate (\DatasetSize=8)$}, {$\TrueDetectionRate (\DatasetSize=64)$}, {$\FalseDetectionRate$ bound}}]
\addlegendimage{very thick, color=steelblue76114176,mark=o}
\addlegendimage{very thick, peru22113282,mark=diamond,dashed,mark options={solid}}
\addlegendimage{very thick, mediumseagreen85168104,mark=triangle,dotted,mark options={solid}}
\addlegendimage{very thick, darkgray,dashdotted}
\addlegendimage{very thick, indianred1967882,mark=square,densely dashed,mark options={solid}}

\end{customlegend}

\end{tikzpicture}}
  \end{subfigure}

  \begin{subfigure}[b]{0.235\textwidth}
      \centering
      \begin{tikzpicture}
          \begin{axis}[
              xlabel={Epochs},
              xtick={1, 2, 3, 4},
              xticklabels={0, 1, 2, 3},
              ymode=log,
              log basis y=10, 
              ymin=0.07, ymax=140,
              ytick={0.1, 1, 10, 100},
              grid=major,
              width=1.0\textwidth,
              height=0.8\textwidth,
              ]
              \addplot[color=steelblue76114176,mark=o,line width=1.5] coordinates {
                  (1,5.13)
                  (2,13.07)
                  (3,20.54)
                  (4,24.42)
              };
              \addplot[color=peru22113282,mark=diamond,dashed,mark options={solid},line width=1.5] coordinates {
                  (1,5.21)
                  (2,31.31)
                  (3,56.4)
                  (4,65.64)
              };
              \addplot[color=mediumseagreen85168104,mark=triangle,dotted,mark options={solid},line width=1.5] coordinates {
                  (1,5.19)
                  (2,94.58)
                  (3,99.91)
                  (4,99.99)
              };
              \addplot[color=darkgray,dashdotted,line width=1.5] coordinates {
                (1, 5)
                (2, 5)
                (3, 5)
                (4, 5)
              };
          \end{axis}
      \end{tikzpicture}
      \caption{$\TrueDetectionRate (\%) @ \FalseDetectionRate \leq 5\%$}
      \label{fig:blip_additional:5fdr}
  \end{subfigure}
  \hfill
  \begin{subfigure}[b]{0.235\textwidth}
      \centering
      \begin{tikzpicture}
          \begin{axis}[
              xlabel={Epochs},
              xtick={1, 2, 3, 4},
              xticklabels={0, 1, 2, 3},
              ymode=log,
              log basis y=10, 
              ymin=0.07, ymax=140,
              ytick={0.1, 1, 10, 100},
              grid=major,
              width=1.0\textwidth,
              height=0.8\textwidth,
              ]
              \addplot[color=steelblue76114176,mark=o,line width=1.5] coordinates {
                  (1,0.97)
                  (2,3.43)
                  (3,7.6)
                  (4,9.64)
              };
              \addplot[color=peru22113282,mark=diamond,dashed,mark options={solid},line width=1.5] coordinates {
                  (1,1.13)
                  (2,12.0)
                  (3,30.6)
                  (4,39.33)
              };
              \addplot[color=mediumseagreen85168104,mark=triangle,dotted,mark options={solid},line width=1.5] coordinates {
                  (1,0.92)
                  (2,82.1)
                  (3,99.51)
                  (4,99.97)
              };
              \addplot[color=darkgray,dashdotted,line width=1.5] coordinates {
                (1, 1)
                (2, 1)
                (3, 1)
                (4, 1)
              };
          \end{axis}
      \end{tikzpicture}
      \caption{$\TrueDetectionRate (\%) @ \FalseDetectionRate \leq 1\%$}
      \label{fig:blip_additional:1fdr}
  \end{subfigure}
  \\
  \begin{subfigure}[b]{0.235\textwidth}
      \centering
      \begin{tikzpicture}
          \begin{axis}[
              xlabel={Epochs},
              xtick={1, 2, 3, 4},
              xticklabels={0, 1, 2, 3},
              ymode=log,
              log basis y=10, 
              ymin=0.07, ymax=140,
              ytick={0.1, 1, 10, 100},
              grid=major,
              width=1.0\textwidth,
              height=0.8\textwidth,
              ]
              \addplot[color=steelblue76114176,mark=o,line width=1.5] coordinates {
                  (1,0.14)
                  (2,0.55)
                  (3,1.78)
                  (4,2.71)
              };
              \addplot[color=peru22113282,mark=diamond,dashed,mark options={solid},line width=1.5] coordinates {
                  (1,0.06)
                  (2,2.77)
                  (3,11.22)
                  (4,16.85)
              };
              \addplot[color=mediumseagreen85168104,mark=triangle,dotted,mark options={solid},line width=1.5] coordinates {
                (1,0.13)
                (2,57.81)
                (3,97.02)
                (4,99.53)
              };
              \addplot[color=darkgray,dashdotted,line width=1.5] coordinates {
                (1, 0.2)
                (2, 0.2)
                (3, 0.2)
                (4, 0.2)
              };
          \end{axis}
      \end{tikzpicture}
      \caption{$\TrueDetectionRate (\%) @ \FalseDetectionRate \leq 0.2\%$}
      \label{fig:blip_additional:0.2fdr}
  \end{subfigure}
  \hfill
  \begin{subfigure}[b]{0.235\textwidth}
      \centering
      \begin{tikzpicture}
          \begin{axis}[
              xlabel={Epochs},
              xtick={1, 2, 3, 4},
              xticklabels={0, 1, 2, 3},
              ymin=18, ymax=24,
              grid=major,
              width=1.0\textwidth,
              height=0.8\textwidth,
              ]
              \addplot[color=indianred1967882,mark=square,densely dashed,mark options={solid},line width=1.5] coordinates {
                  (1, 19.30)
                  (2, 22.77)
                  (3, 21.20)
                  (4, 19.64)
              };
          \end{axis}
      \end{tikzpicture}
      \caption{BLEU $(\%)$}
      \label{fig:blip_additional:acc}
  \end{subfigure}
  \caption{$\TrueDetectionRate (\%)$ of our auditing method for BLIP
    fine-tuned on Flickr30k
    (\subfigsref{fig:blip_additional:5fdr}{fig:blip_additional:0.2fdr}) and test accuracies
    of fine-tuned BLIP (\figref{fig:blip_additional:acc}). Results
    are averaged over $250\times20$ detections. We fine-tuned $20$
    BLIP models, in each of which $250$ training samples were
    audited.}
  \label{fig:blip_additional}
\end{figure}

\section{Verification on Machine Unlearning} \label{app:unlearning}

\subsection{Approximate Unlearning Methods and Their Implementations} \label{app:unlearning:introduction}
\paragraph{Warnecke et al.'s gradient-based method~\cite{warnecke2021:machine}}
In Warnecke et al.'s work, they formulate machine unlearning as an optimization problem:
\begin{align*}
  \MLModel \leftarrow \argmin_{\MLModelAlt} & \frac{1}{\setSize{\TrainDataset}} \sum_{\TrainDataSample \in \TrainDataset} \Loss(\MLModelAlt, \TrainDataSample)  \\
  & + \UnlearnWeight \sum_{\PerturbedData \in \PerturbedDataset} \Loss(\MLModelAlt, \PerturbedData) - \UnlearnWeight \sum_{\UnlearntData \in \UnlearntDataset} \Loss(\MLModelAlt, \UnlearntData),
\end{align*}
where $\TrainDataset$ is its original training dataset, $\UnlearnWeight\in[0,1]$ is a small value controlling how much the influence of data instances is removed,
$\UnlearntDataset$ is the set of data instances to be unlearned, and $\PerturbedDataset$ is the perturbed version of $\UnlearntDataset$.
Instead of solving the optimization exactly, they approximately solve it by applying a gradient-based update:
\begin{align*}
  \MLModel \leftarrow \OriginalMLModel - \UnlearningRate \big(\sum_{\PerturbedData \in \PerturbedDataset} \gradient{\OriginalMLModel}\Loss(\OriginalMLModel, \PerturbedData) - \sum_{\UnlearntData \in \UnlearntDataset} \gradient{\OriginalMLModel}\Loss(\OriginalMLModel, \UnlearntData)\big),
\end{align*}
where $\OriginalMLModel$ denotes the ML model before unlearning and $\UnlearningRate$ is the unlearning rate.\footnote{Warnecke et al also propose a second-order method
but it requires the computation of inverse Hessian, which makes it unsuitable for large deep neural network model.}

In the implementation of Warnecke et al.'s gradient-based method to unlearn data instances from an image classifier,
we set $\UnlearntDataset = \{(\MarkedData, \DataLabel)\}$ where $\DataLabel$ is the label of $\MarkedData$; 
We created a random image with label of $\DataLabel$ and used this labeled random image as $\PerturbedDataset$. 
In its implementation of unlearning data instances from a fine-tuned CLIP model,
we used a batch of audited data instances with their text descriptions as $\UnlearntDataset$; 
We created a set of random images of the same size as that of $\UnlearntDataset$,
assigned them with same text descriptions as those in $\UnlearntDataset$, and used this set of random images with text descriptions
as $\PerturbedDataset$.

\paragraph{Fine-tuning-based method}
The basic idea of the fine-tuning-based method is to fine-tune the ML model by one epoch on
the set of unlearned data instances assigned with randomly selected incorrect labels/text descriptions.
In other words,
\begin{align*}
  \MLModel \leftarrow \OriginalMLModel - \UnlearningRate \sum_{\PerturbedData \in \PerturbedDataset} \gradient{\OriginalMLModel}\Loss(\OriginalMLModel, \PerturbedData),
\end{align*}
where $\OriginalMLModel$ is the ML model before unlearning, $\UnlearningRate$ is the unlearning rate, and $\PerturbedDataset$ denotes a set of unlearned data instances assigned
with randomly selected incorrect labels or text descriptions.

In the implementation of the fine-tuning-based method to unlearn data instances from an image classifier,
we set $\PerturbedDataset = \{(\MarkedData, \IncorrectLabel)\}$ where $\IncorrectLabel$ is a 
randomly chosen incorrect label for $\MarkedData$;
In its implementation of unlearning data instances from a fine-tuned CLIP model,
we used a batch of audited data instances with their text descriptions randomly mismatched as $\PerturbedDataset$.

\subsection{Additional Experimental Results} \label{app:unlearning:results}

\figref{fig:unlearning:tinyimagenet} presents our auditing results after applying approximate unlearning methods to remove
a TinyImageNet data instance from a classifier. \figref{fig:unlearning:clip:gradient}
and \figref{fig:unlearning:clip:finetune} present auditing results after applying approximate unlearning
to remove data instances from a finetuned CLIP model.
As shown there, when we set a larger unlearning rate
$\UnlearningRate$, our $\TrueDetectionRate$ decreased, indicating that
more information of the unlearned data instance was removed from the
updated model.  However, a larger $\UnlearningRate$ led to a lower
model utility as measured by $\Accuracy$, the average fraction of test
data samples correctly predicted by the updated model. 
In contrast, to maintain good model utility, a small
$\UnlearningRate$ could be used, but neither the gradient-based nor
fine-tuning-based unlearning methods with a small $\UnlearningRate$
could sufficiently remove the unlearned data, since our
$\TrueDetectionRate$ was significantly larger than its
$\FalseDetectionRate$ bound. From our results, both the
gradient-based and fine-tuning-based unlearning methods failed to
decrease $\TrueDetectionRate$ to the level of $\FalseDetectionRate$
even after diminishing model utility by $10\%$, and both methods
failed to remove the influence of the unlearned data when maintaining
model utility.

\begin{figure}[ht!]
  \centering
  
\begin{subfigure}[b]{0.47\textwidth}
  \centering
  \resizebox{!}{3em}{\newenvironment{customlegend}[1][]{%
    \begingroup
    \csname pgfplots@init@cleared@structures\endcsname
    \pgfplotsset{#1}%
}{%
    \csname pgfplots@createlegend\endcsname
    \endgroup
}%

\def\addlegendimage{\csname pgfplots@addlegendimage\endcsname}

\begin{tikzpicture}

\definecolor{darkslategray38}{RGB}{38,38,38}
\definecolor{indianred1967882}{RGB}{196,78,82}
\definecolor{lightgray204}{RGB}{204,204,204}
\definecolor{mediumseagreen85168104}{RGB}{85,168,104}
\definecolor{peru22113282}{RGB}{221,132,82}
\definecolor{steelblue76114176}{RGB}{76,114,176}

\begin{customlegend}[
    legend style={{font={\small}},{draw=none}}, 
    legend columns=2,
    legend cell align={center},
    legend entries={{Before unlearning}, {Gradient-based}, {Fine-tuning-based}, {$\FalseDetectionRate$ bound}}]
\addlegendimage{very thick, red, dotted}
\addlegendimage{very thick, steelblue76114176, mark=o, solid}
\addlegendimage{very thick, peru22113282,mark=diamond,dashed,mark options={solid}}
\addlegendimage{very thick, darkgray,dashdotted}

\end{customlegend}

\end{tikzpicture}}
\end{subfigure}

  \begin{subfigure}[b]{0.235\textwidth}
      \centering
      \begin{tikzpicture}
          \begin{axis}[
              xlabel={$\UnlearningRate (\times 0.01)$ },
              xtick={1, 2, 3, 4, 5},
              xticklabels={2, 3, 4, 5, 6},
              ymode=log,
              log basis y=10, 
              ymin=0.07, ymax=140,
              ytick={0.1, 1, 10, 100},
              grid=major,
              width=1.0\textwidth,
              height=0.8\textwidth,
              ]
              \addplot[color=steelblue76114176,mark=o,line width=1.5] coordinates {
                  (1, 14.083) 
                  (2, 11.203750000000001) 
                  (3, 9.5436) 
                  (4, 8.248799999999997) 
                  (5, 7.183649999999999)
              };
              \addplot[color=peru22113282,mark=diamond,dashed,mark options={solid},line width=1.5] coordinates {
                  (1, 16.6982) 
                  (2, 12.993450000000005) 
                  (3, 9.878849999999998) 
                  (4, 7.8189) 
                  (5, 6.57335) 
              };
              \addplot[color=red,dotted,line width=1.5] coordinates {
                  (1, 18.10)
                  (2, 18.10)
                  (3, 18.10)
                  (4, 18.10)
                  (5, 18.10)
              };
              \addplot[color=darkgray,dashdotted,line width=1.5] coordinates {
                  (1, 5)
                  (2, 5)
                  (3, 5)
                  (4, 5)
                  (5, 5)
              };
          \end{axis}
      \end{tikzpicture}
      \caption{$\TrueDetectionRate (\%) @ \FalseDetectionRate \leq 5\%$}
      \label{fig:unlearning:tinyimagenet:5fdr}
  \end{subfigure}
  \hfill
  \begin{subfigure}[b]{0.235\textwidth}
      \centering
      \begin{tikzpicture}
          \begin{axis}[
              xlabel={$\UnlearningRate (\times 0.01)$ },
              xtick={1, 2, 3, 4, 5},
              xticklabels={2, 3, 4, 5, 6},
              ymode=log,
              log basis y=10, 
              ymin=0.07, ymax=140,
              ytick={0.1, 1, 10, 100},
              grid=major,
              width=1.0\textwidth,
              height=0.8\textwidth,
              ]
              \addplot[color=steelblue76114176,mark=o,line width=1.5] coordinates {
                  (1, 4.2566000000000015) 
                  (2, 3.162000000000001) 
                  (3, 2.3982500000000004) 
                  (4, 2.113150000000001) 
                  (5, 1.5038500000000004)  
              };
              \addplot[color=peru22113282,mark=diamond,dashed,mark options={solid},line width=1.5] coordinates {
                  (1, 5.41135) 
                  (2, 3.898150000000001) 
                  (3, 2.632950000000001) 
                  (4, 1.8981500000000007) 
                  (5, 1.6384500000000006)
              };
              \addplot[color=red,dotted,line width=1.5] coordinates {
                  (1, 6.03)
                  (2, 6.03)
                  (3, 6.03)
                  (4, 6.03)
                  (5, 6.03)
              };
              \addplot[color=darkgray,dashdotted,line width=1.5] coordinates {
                  (1, 1)
                  (2, 1)
                  (3, 1)
                  (4, 1)
                  (5, 1)
              };
          \end{axis}
      \end{tikzpicture}
      \caption{$\TrueDetectionRate (\%) @ \FalseDetectionRate \leq 1\%$}
      \label{fig:unlearning:tinyimagenet:1fdr}
  \end{subfigure}
  \\
  \begin{subfigure}[b]{0.235\textwidth}
      \centering
      \begin{tikzpicture}
          \begin{axis}[
              xlabel={$\UnlearningRate (\times 0.01)$ },
              xtick={1, 2, 3, 4, 5},
              xticklabels={2, 3, 4, 5, 6},
              ymode=log,
              log basis y=10, 
              ymin=0.07, ymax=140,
              ytick={0.1, 1, 10, 100},
              grid=major,
              width=1.0\textwidth,
              height=0.8\textwidth,
              ]
              \addplot[color=steelblue76114176,mark=o,line width=1.5] coordinates {
                  (1, 0.66) 
                  (2, 0.4500000000000001) 
                  (3, 0.27000000000000013) 
                  (4, 0.29500000000000004) 
                  (5, 0.17)  
              };
              \addplot[color=peru22113282,mark=diamond,dashed,mark options={solid},line width=1.5] coordinates {
                  (1, 1.0000000000000004) 
                  (2, 0.675) 
                  (3, 0.335) 
                  (4, 0.25500000000000006) 
                  (5, 0.2700000000000001) 
              };
              \addplot[color=red,dotted,line width=1.5] coordinates {
                  (1, 1.39)
                  (2, 1.39)
                  (3, 1.39)
                  (4, 1.39)
                  (5, 1.39)
              };
              \addplot[color=darkgray,dashdotted,line width=1.5] coordinates {
                  (1, 0.2)
                  (2, 0.2)
                  (3, 0.2)
                  (4, 0.2)
                  (5, 0.2)
              };
          \end{axis}
      \end{tikzpicture}
      \caption{$\TrueDetectionRate (\%) @ \FalseDetectionRate \leq 0.2\%$}
      \label{fig:unlearning:tinyimagenet:0.2fdr}
  \end{subfigure}
  \hfill
  \begin{subfigure}[b]{0.235\textwidth}
      \centering
      \begin{tikzpicture}
          \begin{axis}[
              xlabel={$\UnlearningRate (\times 0.01)$ },
              xtick={1, 2, 3, 4, 5},
              xticklabels={2, 3, 4, 5, 6},
              ymin=40, ymax=65,
              grid=major,
              width=1.0\textwidth,
              height=0.8\textwidth,
              ]
              \addplot[color=steelblue76114176,mark=o,line width=1.5] coordinates {
                  (1, 56.537028500000005) 
                  (2, 55.179012499999985) 
                  (3, 53.22460699999999) 
                  (4, 50.608312) 
                  (5, 47.39055700000001)  
              };
              \addplot[color=peru22113282,mark=diamond,dashed,mark options={solid},line width=1.5] coordinates {
                  (1, 58.56222150000001) 
                  (2, 57.56700300000002) 
                  (3, 55.79305300000003) 
                  (4, 52.771501500000014) 
                  (5, 48.145481499999995) 
              };
              \addplot[color=red,dotted,line width=1.5] coordinates {
                  (1, 59.86)
                  (2, 59.86)
                  (3, 59.86)
                  (4, 59.86)
                  (5, 59.86)
              };
          \end{axis}
      \end{tikzpicture}
      \caption{Test accuracy $\Accuracy (\%)$}
      \label{fig:unlearning:tinyimagenet:acc}
  \end{subfigure}
  \caption{$\TrueDetectionRate(\%)$ of our auditing method for
    TinyImageNet image classifiers after applying approximate
    unlearning
    (\subfigsref{fig:unlearning:tinyimagenet:5fdr}{fig:unlearning:tinyimagenet:0.2fdr})
    and test accuracies of image classifiers
    (\figref{fig:unlearning:tinyimagenet:acc}).}
  \label{fig:unlearning:tinyimagenet}
\end{figure}

\begin{figure}[ht!]
  \centering

\begin{subfigure}[b]{0.47\textwidth}
  \centering
  \resizebox{!}{1.6em}{\newenvironment{customlegend}[1][]{%
    \begingroup
    \csname pgfplots@init@cleared@structures\endcsname
    \pgfplotsset{#1}%
}{%
    \csname pgfplots@createlegend\endcsname
    \endgroup
}%

\def\addlegendimage{\csname pgfplots@addlegendimage\endcsname}

\begin{tikzpicture}

\definecolor{darkslategray38}{RGB}{38,38,38}
\definecolor{indianred1967882}{RGB}{196,78,82}
\definecolor{lightgray204}{RGB}{204,204,204}
\definecolor{mediumseagreen85168104}{RGB}{85,168,104}
\definecolor{peru22113282}{RGB}{221,132,82}
\definecolor{steelblue76114176}{RGB}{76,114,176}

\begin{customlegend}[
    legend style={{font={\small}},{draw=none}}, 
    legend columns=3,
    legend cell align={center},
    legend entries={{Before unlearning}, {After unlearning}, {$\FalseDetectionRate$ bound}}]
\addlegendimage{very thick, red, dotted}
\addlegendimage{very thick, steelblue76114176, mark=o, solid}
\addlegendimage{very thick, darkgray,dashdotted}

\end{customlegend}

\end{tikzpicture}}
\end{subfigure}

  \begin{subfigure}[b]{0.235\textwidth}
      \centering
      \begin{tikzpicture}
          \begin{axis}[
              xlabel={$\UnlearningRate (\times 10^{-4})$ },
              xtick={1, 2, 3, 4, 5, 6},
              xticklabels={2, 3, 4, 5, 6, 7},
              ymode=log,
              log basis y=10, 
              ymin=0.07, ymax=140,
              ytick={0.1, 1, 10, 100},
              grid=major,
              width=1.0\textwidth,
              height=0.8\textwidth,
              ]
              \addplot[color=steelblue76114176,mark=o,line width=1.5] coordinates {
                  (1, 14.238000000000001) 
                  (2, 13.618200000000005) 
                  (3, 12.738600000000002) 
                  (4, 12.096800000000004) 
                  (5, 11.578200000000004) 
                  (6, 10.898400000000006)
              };
              \addplot[color=red,dotted,line width=1.5] coordinates {
                  (1, 16.38)
                  (2, 16.38)
                  (3, 16.38)
                  (4, 16.38)
                  (5, 16.38)
                  (6, 16.38)
              };
              \addplot[color=darkgray,dashdotted,line width=1.5] coordinates {
                  (1, 5)
                  (2, 5)
                  (3, 5)
                  (4, 5)
                  (5, 5)
                  (6, 5)
              };
          \end{axis}
      \end{tikzpicture}
      \caption{$\TrueDetectionRate (\%) @ \FalseDetectionRate \leq 5\%$}
      \label{fig:unlearning:clip:gradient:5fdr}
  \end{subfigure}
  \hfill
  \begin{subfigure}[b]{0.235\textwidth}
      \centering
      \begin{tikzpicture}
          \begin{axis}[
              xlabel={$\UnlearningRate (\times 10^{-4})$ },
              xtick={1, 2, 3, 4, 5, 6},
              xticklabels={2, 3, 4, 5, 6, 7},
              ymode=log,
              log basis y=10, 
              ymin=0.07, ymax=140,
              ytick={0.1, 1, 10, 100},
              grid=major,
              width=1.0\textwidth,
              height=0.8\textwidth,
              ]
              \addplot[color=steelblue76114176,mark=o,line width=1.5] coordinates {
                  (1, 4.456600000000002) 
                  (2, 4.077400000000002) 
                  (3, 3.8968000000000007) 
                  (4, 3.4382) 
                  (5, 3.0778000000000008) 
                  (6, 3.0774) 
              };
              \addplot[color=red,dotted,line width=1.5] coordinates {
                  (1, 5.52)
                  (2, 5.52)
                  (3, 5.52)
                  (4, 5.52)
                  (5, 5.52)
                  (6, 5.52)
              };
              \addplot[color=darkgray,dashdotted,line width=1.5] coordinates {
                  (1, 1)
                  (2, 1)
                  (3, 1)
                  (4, 1)
                  (5, 1)
                  (6, 1)
              };
          \end{axis}
      \end{tikzpicture}

      \caption{$\TrueDetectionRate (\%) @ \FalseDetectionRate \leq 1\%$}
      \label{fig:unlearning:clip:gradient:1fdr}
  \end{subfigure}
  \\
  \begin{subfigure}[b]{0.235\textwidth}
      \centering
      \begin{tikzpicture}
          \begin{axis}[
              xlabel={$\UnlearningRate (\times 10^{-4})$ },
              xtick={1, 2, 3, 4, 5, 6},
              xticklabels={2, 3, 4, 5, 6, 7},
              ymode=log,
              log basis y=10, 
              ymin=0.07, ymax=140,
              ytick={0.1, 1, 10, 100},
              grid=major,
              width=1.0\textwidth,
              height=0.8\textwidth,
              ]
              \addplot[color=steelblue76114176,mark=o,line width=1.5] coordinates {
                  (1, 0.9200000000000004) 
                  (2, 0.7800000000000001) 
                  (3, 0.7000000000000001) 
                  (4, 0.6400000000000001) 
                  (5, 0.42000000000000004) 
                  (6, 0.4000000000000001)  
              };
              \addplot[color=red,dotted,line width=1.5] coordinates {
                (1, 1.18)
                (2, 1.18)
                (3, 1.18)
                (4, 1.18)
                (5, 1.18)
                (6, 1.18)
            };
            \addplot[color=darkgray,dashdotted,line width=1.5] coordinates {
                  (1, 0.2)
                  (2, 0.2)
                  (3, 0.2)
                  (4, 0.2)
                  (5, 0.2)
                  (6, 0.2)
              };
          \end{axis}
      \end{tikzpicture}
      \caption{$\TrueDetectionRate (\%) @ \FalseDetectionRate \leq 0.2\%$}
      \label{fig:unlearning:clip:gradient:0.2fdr}
  \end{subfigure}
  \hfill
  \begin{subfigure}[b]{0.235\textwidth}
      \centering
      \begin{tikzpicture}
          \begin{axis}[
              xlabel={$\UnlearningRate (\times 10^{-4})$ },
              xtick={1, 2, 3, 4, 5, 6},
              xticklabels={2, 3, 4, 5, 6, 7},
              ymin=70, ymax=90,
              grid=major,
              width=1.0\textwidth,
              height=0.8\textwidth,
              ]
              \addplot[color=steelblue76114176,mark=o,line width=1.5] coordinates {
                  (1, 85.72915834640818) 
                  (2, 85.27615292263881) 
                  (3, 84.52035618546822) 
                  (4, 83.3510435944689) 
                  (5, 79.91696725998607) 
                  (6, 77.62224468336042)   
              };
              \addplot[color=red,dotted,line width=1.5] coordinates {
                (1, 86.57)
                (2, 86.57)
                (3, 86.57)
                (4, 86.57)
                (5, 86.57)
                (6, 86.57)
            };
          \end{axis}
      \end{tikzpicture}
      \caption{Test accuracy $\Accuracy (\%)$}
      \label{fig:unlearning:clip:gradient:acc}
  \end{subfigure}
  \caption{$\TrueDetectionRate(\%)$ of our auditing method for CLIP
    after applying gradient-based approximate unlearning
    (\subfigsref{fig:unlearning:clip:gradient:5fdr}{fig:unlearning:clip:gradient:0.2fdr})
    and test accuracies of image classifiers
    (\figref{fig:unlearning:clip:gradient:acc}).}
  \label{fig:unlearning:clip:gradient}
\end{figure}

\begin{figure}[ht!]
  \centering

\begin{subfigure}[b]{0.47\textwidth}
  \centering
  \resizebox{!}{1.6em}{\newenvironment{customlegend}[1][]{%
    \begingroup
    \csname pgfplots@init@cleared@structures\endcsname
    \pgfplotsset{#1}%
}{%
    \csname pgfplots@createlegend\endcsname
    \endgroup
}%

\def\addlegendimage{\csname pgfplots@addlegendimage\endcsname}

\begin{tikzpicture}

\definecolor{darkslategray38}{RGB}{38,38,38}
\definecolor{indianred1967882}{RGB}{196,78,82}
\definecolor{lightgray204}{RGB}{204,204,204}
\definecolor{mediumseagreen85168104}{RGB}{85,168,104}
\definecolor{peru22113282}{RGB}{221,132,82}
\definecolor{steelblue76114176}{RGB}{76,114,176}

\begin{customlegend}[
    legend style={{font={\small}},{draw=none}}, 
    legend columns=3,
    legend cell align={center},
    legend entries={{Before unlearning}, {After unlearning}, {$\FalseDetectionRate$ bound}}]
\addlegendimage{very thick, red, dotted}
\addlegendimage{very thick, steelblue76114176, mark=o, solid}
\addlegendimage{very thick, darkgray,dashdotted}

\end{customlegend}

\end{tikzpicture}}
\end{subfigure}

  \begin{subfigure}[b]{0.235\textwidth}
      \centering
      \begin{tikzpicture}
          \begin{axis}[
              xlabel={$\UnlearningRate (\times 10^{-5})$ },
              xtick={1, 2, 3, 4, 5, 6},
              xticklabels={1, 2, 5, 10, 12, 15},
              ymode=log,
              log basis y=10, 
              ymin=0.07, ymax=140,
              ytick={0.1, 1, 10, 100},
              grid=major,
              width=1.0\textwidth,
              height=0.8\textwidth,
              ]
              \addplot[color=steelblue76114176,mark=o,line width=1.5] coordinates {
                  (1, 15.018000000000002) 
                  (2, 14.357600000000001) 
                  (3, 12.758400000000002) 
                  (4, 10.957800000000006) 
                  (5, 9.6386) 
                  (6, 9.138400000000004)
              };
              \addplot[color=red,dotted,line width=1.5] coordinates {
                  (1, 16.38)
                  (2, 16.38)
                  (3, 16.38)
                  (4, 16.38)
                  (5, 16.38)
                  (6, 16.38)
              };
              \addplot[color=darkgray,dashdotted,line width=1.5] coordinates {
                  (1, 5)
                  (2, 5)
                  (3, 5)
                  (4, 5)
                  (5, 5)
                  (6, 5)
              };
          \end{axis}
      \end{tikzpicture}
      \caption{$\TrueDetectionRate (\%) @ \FalseDetectionRate \leq 5\%$}
      \label{fig:unlearning:clip:finetune:5fdr}
  \end{subfigure}
  \hfill
  \begin{subfigure}[b]{0.235\textwidth}
      \centering
      \begin{tikzpicture}
          \begin{axis}[
              xlabel={$\UnlearningRate (\times 10^{-5})$ },
              xtick={1, 2, 3, 4, 5, 6},
              xticklabels={1, 2, 5, 10, 12, 15},
              ymode=log,
              log basis y=10, 
              ymin=0.07, ymax=140,
              ytick={0.1, 1, 10, 100},
              grid=major,
              width=1.0\textwidth,
              height=0.8\textwidth,
              ]
              \addplot[color=steelblue76114176,mark=o,line width=1.5] coordinates {
                  (1, 4.937800000000002) 
                  (2, 4.796400000000003) 
                  (3, 3.996600000000001) 
                  (4, 3.1966000000000006) 
                  (5, 2.838) 
                  (6, 2.2992000000000012)
              };
              \addplot[color=red,dotted,line width=1.5] coordinates {
                  (1, 5.52)
                  (2, 5.52)
                  (3, 5.52)
                  (4, 5.52)
                  (5, 5.52)
                  (6, 5.52)
              };
              \addplot[color=darkgray,dashdotted,line width=1.5] coordinates {
                  (1, 1)
                  (2, 1)
                  (3, 1)
                  (4, 1)
                  (5, 1)
                  (6, 1)
              };
          \end{axis}
      \end{tikzpicture}
      \caption{$\TrueDetectionRate (\%) @ \FalseDetectionRate \leq 1\%$}
      \label{fig:unlearning:clip:finetune:1fdr}
  \end{subfigure}
  \\
  \begin{subfigure}[b]{0.235\textwidth}
      \centering
      \begin{tikzpicture}
          \begin{axis}[
              xlabel={$\UnlearningRate (\times 10^{-5})$ },
              xtick={1, 2, 3, 4, 5, 6},
              xticklabels={1, 2, 5, 10, 12, 15},
              ymode=log,
              log basis y=10, 
              ymin=0.07, ymax=140,
              ytick={0.1, 1, 10, 100},
              grid=major,
              width=1.0\textwidth,
              height=0.8\textwidth,
              ]
              \addplot[color=steelblue76114176,mark=o,line width=1.5] coordinates {
                  (1, 1.1400000000000003) 
                  (2, 1.1400000000000003) 
                  (3, 0.9800000000000003) 
                  (4, 0.49999999999999994) 
                  (5, 0.54) 
                  (6, 0.48000000000000015) 
              };
              \addplot[color=red,dotted,line width=1.5] coordinates {
                (1, 1.18)
                (2, 1.18)
                (3, 1.18)
                (4, 1.18)
                (5, 1.18)
                (6, 1.18)
            };
            \addplot[color=darkgray,dashdotted,line width=1.5] coordinates {
                  (1, 0.2)
                  (2, 0.2)
                  (3, 0.2)
                  (4, 0.2)
                  (5, 0.2)
                  (6, 0.2)
              };
          \end{axis}
      \end{tikzpicture}
      \caption{$\TrueDetectionRate (\%) @ \FalseDetectionRate \leq 0.2\%$}
      \label{fig:unlearning:clip:finetune:0.2fdr}
  \end{subfigure}
  \hfill
  \begin{subfigure}[b]{0.235\textwidth}
      \centering
      \begin{tikzpicture}
          \begin{axis}[
              xlabel={$\UnlearningRate (\times 10^{-5})$ },
              xtick={1, 2, 3, 4, 5, 6},
              xticklabels={1, 2, 5, 10, 12, 15},
              ymin=70, ymax=90,
              grid=major,
              width=1.0\textwidth,
              height=0.8\textwidth,
              ]
              \addplot[color=steelblue76114176,mark=o,line width=1.5] coordinates {
                  (1, 86.01633498179065) 
                  (2, 85.98080647311039) 
                  (3, 85.59132744558156) 
                  (4, 84.04455956204663) 
                  (5, 79.59192568702356) 
                  (6, 76.89465401760702)
              };
              \addplot[color=red,dotted,line width=1.5] coordinates {
                (1, 86.57)
                (2, 86.57)
                (3, 86.57)
                (4, 86.57)
                (5, 86.57)
                (6, 86.57)
            };
          \end{axis}
      \end{tikzpicture}
      \caption{Test accuracy $\Accuracy (\%)$}
      \label{fig:unlearning:clip:finetune:acc}
  \end{subfigure}
  \caption{$\TrueDetectionRate(\%)$ of our auditing method for CLIP
    after applying fine-tuning-based approximate unlearning
    (\subfigsref{fig:unlearning:clip:finetune:5fdr}{fig:unlearning:clip:finetune:0.2fdr})
    and test accuracies of image classifiers
    (\figref{fig:unlearning:clip:finetune:acc}).}
  \label{fig:unlearning:clip:finetune}
\end{figure}

\end{document}